# Evolution of the IGG concept at IGF from 2004 to 2007

G. Modanese
Free University of Bolzano
Faculty of Science and Technology
Piazza Università, 5
39100 Bolzano – Italy
E-mail: giovanni.modanese@unibz.it

**Abstract -** IGG is the acronym for Impulse Gravity Generator, a device developed by E. Podkletnov in 1997-2003 for generating high-voltage discharges through YBCO electrodes. According to Podkletnov, an anomalous force beam is generated at the discharge, which acts on distant material target of any composition with a small repulsive force proportional to the target mass. An independent replication of this device was started in 2004 at IGF, Germany (Institut für Gravitationsforschung, Göde Foundation). The author was involved as theoretical consultant and his first assignment was to study a possible scaled-down version of the device. This required a thorough analysis of the physical working principles of the apparatus, which was documented in several internal reports from 2004 to 2009. The whole content of those from 2004 to 2007 is given here. Several parts are outdated, but useful for an understanding of the phenomenon. In particular, the discharge mechanism was eventually found to be different, namely a vacuum spark discharge instead of a low pressure discharge with cascade gas ionization [1]. Also outdated is the theoretical model developed in Ch. 1 on the "acceleration of Cooper pairs through the superconductor" as a possible basis for the anomalous emission. This naïve representation, however, is useful to show that such an acceleration does not make sense in a superconductor like YBCO and that the correct picture is that of (low-voltage) tunnelling of pairs through intrinsic Josephson junctions [2,3]. The present paper should therefore be regarded as an "historical summary", mainly valuable as a reference for further developments.

Recent references cited in the abstract:

# CONTENTS

*Chapter 3*   **Evaluation of the pulse parameters, target velocity and beam energy for the IU1045 and IU1080 Marx generators (2005)**

# Chapter 1

# Scaling down/(up) of the impulse gravity generator experiment
# Part 1 (2004)

**Introduction**

The apparatus described by E. Podkletnov and myself in Ref. [1] is composed of a high-Tc superconductor cathode made from $YBa_2Cu_3O_{7-x}$ which sends a brief, high voltage electrical discharge to a copper anode through a low pressure gas. Under certain conditions, this discharge was accompanied by a beam of anomalous forces that was reported to propagate in the same direction as the electrical discharge. This beam had approximately the same diameter as the superconducting emitter (10 cm) and exhibited no detectable spread over a distance of 150 m.

The aim of this work is to find and explain in detail the relationships between the main experimental parameters of the device: duration of the discharge, pressure in the chamber, applied voltage, voltage-current relation, ion mobility in the discharge, threshold potential for the flat discharge, temperature, frequency and wavelength of the emitted anomalous radiation, spacing of the microscopic crystals in the emitter, current density, area and magnetic flux in the emitter.

Based on the relationships mentioned above, I will examine in a forthcoming report the possibility of scaling down some of the crucial parameters, in order to have a device simpler to build and operate.

Please be advised that the figures and graphs of this report just have an explication purpose.

**A. Objective phenomenology of the gas discharge**

In this Part A we analyse the phenomena which occur in the discharge chamber, without any specific reference to the emission of anomalous radiation (which we suppose to occur in the superconducting emitter; see Part B). For this scope, the entire device can be schematically represented, from the electric point of view, like a sort of RC circuit. Special attention will be devoted to the problem of the definition of the breakdown voltage and the optimal gap length.

*A.1 Total capacitance, charge and electrostatic energy of the Marx generator. Duration of the discharge*

The Marx generator is an electric circuit which allows to charge a parallel capacitors array to a voltage relatively simple to achieve (of the order of 100 kV), and then change the connection from parallel to serial in a very short time. In this way, a high voltage pulse can be generated. The fast switching from the parallel to the serial configuration is obtained by inducing discharges in air between certain nodes of the circuit, which are therefore temporarily short-circuited. The rise time of the voltage pulse is sometimes called "erecting time" and depends on the resistors inserted in the circuit.

The value of the voltage rise-time was not given in [1]. As we shall see, this value is important, especially in comparison with the formation time of the discharge $\Delta t_1$ (see Fig. 2). If the rise-time of the voltage pulse is of the same order as $\Delta t_1$, the breakdown will occur at a voltage which is definitely smaller than the peak voltage, and the discharge current will consequently be smaller, too. In private communications, E. Podkletnov indicated that reduction of the rise-time in his device led to a remarkable increase (up to a factor 2) of the peak current and therefore of the anomalous emission. This seems to imply that the rise-time is not much shorter than its critical value $\Delta t_1$, i.e. of the order of $10^{-8}$ to $10^{-7}$ s. For several reasons, we deem it desirable to reduce this time by one or two magnitude orders. We shall discuss this possibility in our 2$^{nd}$ report.

In our Marx generator there are 20 capacitors, each with a capacitance of 25 nF, initially charged at a voltage $V_P$ between 50 and 100 kV. Let us for instance consider $V_P$=100 kV. The total capacitance in the parallel configuration is $C_P = 20 \cdot 25 \cdot 10^{-9}$ F = $5 \cdot 10^{-7}$ F. The total charge is $Q_P = C_P V_P = 5 \cdot 10^{-2}$ C. The total electrostatic energy is given by $U_P = ½ C_P V_P^2 = 2.5 \cdot 10^3$ J (not $10^6$ J as erroneously typed in [1]). After switching to the serial configuration, the voltage is multiplied by 20, so $V_S$=2·10$^6$ V. The effective charge is that of one single capacitor, i.e. $Q_S$=2.5·10$^{-3}$ C. The total equivalent capacity is given by $C_S = [20(25 \cdot 10^{-9})^{-1}]^{-1} = 1.25 \cdot 10^{-9}$ F. One easily checks that $Q_S = C_S V_S$ and that $U_S = ½ C_S V_S^2$ is equal to $U_P$; this means that in the ideal circuit there is no energy dissipation at the switching between the serial and parallel configuration.

When a high voltage pulse with $V_S$ > 500 kV obtained through the Marx generator is sent to the electrodes of the discharge chamber, the electric field between the electrodes is sufficient to cause bulk ionization of the gas and a "flat" discharge (see Section A.3). The discharge current $I$ has a peak value of ~$10^4$ A. The shape of the voltage and current pulses was not observed/specified in [1]. The duration of the gas discharge as deduced from the photodiode signal is between $10^{-5}$ and $10^{-4}$ s. It is easy to show, however, that the duration $\Delta t_2$ of the current pulse must be shorter. Namely, the electric power is given, independently from any detail of the conduction mechanism, by $P(t)=I(t)V(t)$. Its time integral must be equal to the energy $U_S$ stored in the capacitors of the Marx generator, which we can rewrite as $U_S = ½ Q_S V_S$. Thus as magnitude order we have

$$IV_S \Delta t \approx Q_S V_S \Rightarrow I \Delta t \approx Q_S \approx 10^{-3} \, \mathrm{C} \tag{A1}$$

So with $I$~$10^4$ A we find $\Delta t$~$10^{-7}$ s. The discharge is much shorter than the light signal. Its duration is relevant for all the subsequent analysis.

*A.2 Equivalent electric scheme. Effective resistance of the emitter*

After electric breakdown has occurred in the discharge chamber, the device behaves as a *RC* circuit, in which *C* is the total serial capacity of the Marx generator and *R* is the total effective resistance of the superconducting electrode plus the normal layer. We expect indeed that the superconducting electrode opposes a resistance to the flow of high-frequency AC current (see below). The discharge time is $\Delta t = RC$, and since $\Delta t$ ~ $10^{-7}$ s, *R* should be of the order of $10^2$ Ω. This is confirmed by the *I/V* ratio: with a peak value of the current $I = 10^4$ A under a voltage $V = 10^6$ V, one can again guess an effective resistance $R$ ~ $V/I$ ~ $10^2$ Ω. Note however that only the magnitude order of *I* for maximum voltage is given in [1], not the *I(V)* characteristics. We cannot state that the *I/V* ratio is constant, although it certainly is constant as magnitude order. Knowledge of the *I(V)* graph would be relevant for other purposes as well. See Section B.5.

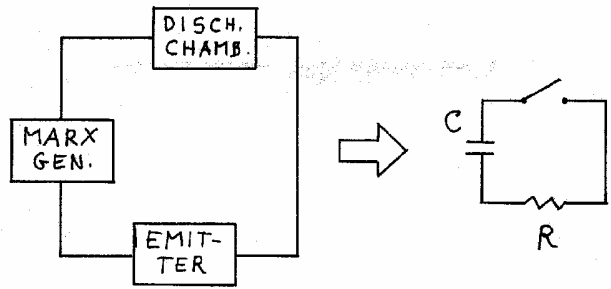

Fig. 1 - Simplified equivalent electric scheme of the device, represented by three blocks: Marx generator, discharge chamber, superconducting emitter. After breakdown has occurred, the discharge chamber behaves essentially as a short-circuit. The switching time is $\Delta t_1$ (compare Fig. 2).

Below its irreversibility temperature $T_{irrev}$ (see our final notes, Part C) the superconducting layer of the emitter has zero resistance with respect to direct current; but all type-II superconductors exhibit in general resistance to AC currents. From the microscopic point of view, it is known that conduction in the *c* direction in YBCO implies a tunnelling of the Cooper pairs wave-function between the *ab* planes. If forced to happen in a very short time, this tunnelling process can be accompanied by an effective resistance and by dissipation. It follows that across the superconducting layer of the emitter there will be a voltage drop, which in fact pushes the Cooper pairs from one plane to the other. In the following (Part B) we shall discuss if this picture is consistent, and how the dissipation would occur.

According to [1], the normal layer of the cathode has a small resistance (about 1 Ω) which cannot play any role. This figure refers, however, to DC conductivity, while in our case there is a very short pulse with high-frequency AC components. Furthermore, it is known that in the ceramic cuprate superconductors the conductivity can exhibit large anisotropy (up to a factor $10^6$) between the *ab* and the *c* directions; such anisotropy depends on several poorly known factors. Therefore, it is well possible that a part of the effective resistance of the emitter is due to the normal-conducting layer. This point is related to the problem of the still unclear role played by the N-layer in the discharge and in the whole anomalous emission phenomenon. A better understanding of this role would be important, because the presence of the N-layer considerably complicates the fabrication of the emitter. We shall discuss this issue in our 2[nd] report. According to the theoretical model presented in Part B, the minimum voltage drop on the emitter needed for the anomalous emission is of the order of several tens of kilovolts, depending on the dominant radiating transition (eq. (B4) and table). This implies that if the N-layer has an effective resistance, this can account only for a part of the total effective resistance of the emitter.

*A.3 Breakdown voltage, optimum gap length.*

As described in [1], the generation of anomalous radiation is only observed when the voltage is sufficiently high to produce a flat discharge, i.e. a discharge in which the whole mass of the gas is involved and becomes ionized. The threshold value of the voltage which allows to obtain a flat discharge is about 500 kV. Below such value, spark discharges are observed, either with single or multiple sparks. In spark discharges the ionization regions are long and narrow, and the current density is not uniform in the gas; the current only flows along the ionized paths of the sparks. This implies for a low-pressure gas, like in our case ($P \sim 1$ Pa), that the total current can not be very large. Spark discharges usually occur in long gaps, where by "gap" the space between the electrodes is meant; flat discharges are usually observed in short gaps, of no more than few centimetres. In our case, the situation is intermediate, with gaps between 15 and 40 cm. Another important distinction is the following: in spark discharges, hot-spots with thermoionic emission usually appear on the cathode, because the positive ion current hits small areas of the cathode; this does not happen in the flat discharges. In our case there is no local over-heating of the cathode and the emission of secondary electrons occurs almost certainly by field effect, due to the strong applied electric field (see below).

The electrical breakdown between two electrodes containing a low pressure gas obeys in general the laws by Townsend and Paschen [3,4]. According to Townsend's theory, the discharge starts in the bulk of the gas when the mechanism for primary ionization (collisions of ions and electrons in the gas) generates an amount of secondary ionization (expulsion of electrons from the cathode by impact or other processes) sufficient to initiate a self-amplification process. This happens when the applied voltage reaches a value, called breakdown voltage $V_B$, which is sharply defined, such that for values smaller than $V_B$ by only a few volts, breakdown does not yet occur. The values of the breakdown voltage experimentally found for several gases are given in the literature and are referred to the static case, i.e. to the case when the gas is irradiated with UV radiation or subjected to other ionization causes, and the voltage is gradually increased until the breakdown occurs. At breakdown, the current increases exponentially, and this leads to the collapse of the applied voltage in a very short time, thus turning the insulating gap into a short-circuit. The actual collapsing time of the voltage depends on the other components of the circuit ($\Delta t \sim RC \sim 10^{-7}$ s in our case).

If a voltage pulse of short duration is present, like in our case, an over-voltage is generally needed to start the discharge, typically of the order of few percent.

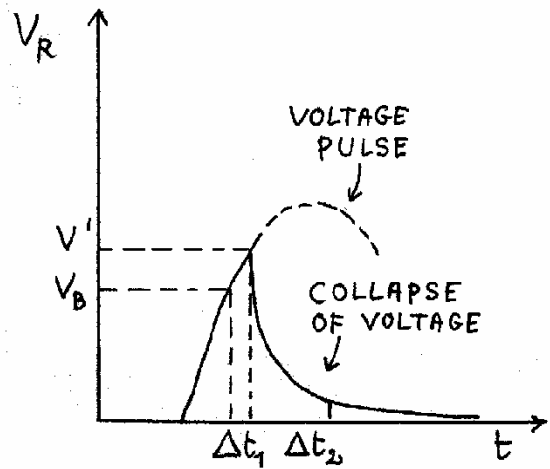

Fig. 2 - Time dependence of the voltage $V_R$ applied on the external load (in our case, on the emitter) before and after the original pulse produced by the generator has caused breakdown of a gas gap. The wave form of the voltage (sinusoidal in this case) is not much relevant. The time for formation of the discharge is $\Delta t_1$. Breakdown starts when the voltage has reached the static (Townsend) value $V_B$; the potential reaches an "over-voltage" $V'$ before the gap is turned into a short-circuit. The load voltage then collapses in a time $\Delta t_2$ defined by the load; in our case $\Delta t_2 \sim RC$ (Fig. 1).

The over-voltage depends on the rise time of the pulse and on the formation time of the breakdown current. The analysis of discharges produced by short voltage pulses should be based on the knowledge of the coefficients of primary and secondary ionization measured in static conditions. Such measurements were not done by Podkletnov, however. In order to apply Paschen's law, we will refer to data available in the literature, for instance for air. Paschen's law states that the breakdown voltage, in volts, is given by the equation

$$V_B = \frac{Bx}{C + \ln x} \tag{A2}$$

where $x$ is the product of the pressure $p$ by the gap length $d$. The voltage is therefore not a function of $p$ and $d$ independently, but of their product. The units employed are usually Tor·cm for $p$ and cm for $d$. The $B$ coefficient depends on the gas and describes the efficiency of primary ionization by electrons collisions; the $C$ coefficient accounts for the efficiency of secondary electron emission at the cathode. Ref. [1] does not give any precise information on which kind of gas is present in the discharge chamber. The use of nitrogen and helium in the cooling system might imply that some amount of these gases is present in the chamber. A further, even larger uncertainty concerns the $C$ coefficient, which depends on the material of the cathode and on the microscopic features of its surface. Even for metallic cathodes, a different treatment of the surface leads to different secondary emission. Again, this coefficient should be measured in static conditions before one can predict the behaviour of the system under pulsed voltage. Nevertheless, since we will be interested into the value of the $pd$ product which gives a breakdown voltage tending to infinity (see below), our analysis can proceed without a detailed knowledge of the $B$ and $C$ coefficients. Let us plot the breakdown voltage $V_B$ as given by the equation above with $B=577$ (air) and $C=1.57$ (standard metallic cathode).

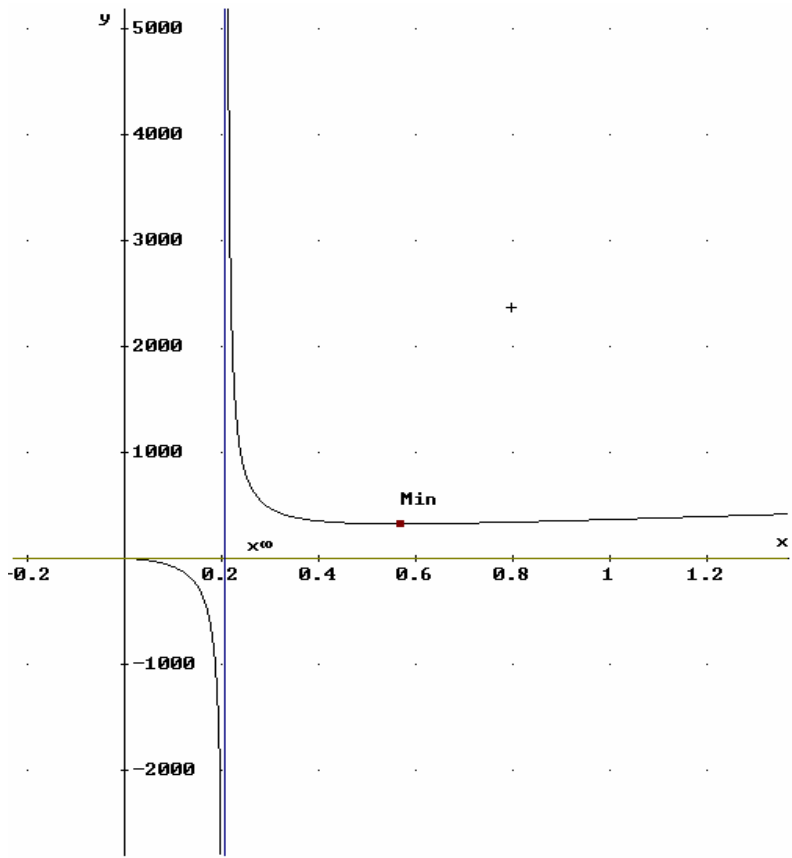

Fig. 3 - Paschen Law (eq. A2): breakdown voltage of an air gap with tungsten electrodes, pressure $p$ and length $d$. On the y-axis is the voltage in volts; on the horizontal axis, $x=pd$, in units Tor·cm. The negative branch of the curve does not have physical meaning. The minimum value of the breakdown voltage (~327 V) is obtained for $x=x_{min}$~0.567 Tor·cm. $x_{inf}$ ~ 0.2080 is the minimum possible value of the $pd$ product in these conditions; when $x$ approaches $x_{inf}$ from the right, the breakdown voltage tends to infinity.

The minimum sparking potential $V_{B,min}=327$ V is obtained for a $pd$ product $x_{min}=0.567$. (In practice, often $V_{B,min}$ and $x_{min}$ are measured, and from them one computes $B$ and $C$, according to the relations $B= V_{B,min}/x_{min}$ and $C=1-\ln(x_{min})$.) Note that this value of the potential is very low in comparison to those in our device; our value of $x$ instead is only slightly smaller than $x_{min}$, of the order of 0.1-0.2 tor·cm (1 Pa~ 1/132 Tor). In ref. [1] the "optimal gap lengths" are not specified, nor it is explained how these are determined; it is only stated that "the distance between the electrodes can vary from 15 to 40 cm in order to find the optimum length for each type of emitter". In view of all the above, we then conclude by formulating the following work hypothesis:

The optimal gap length, empirically determined as the one giving good flat discharges, is the length $d$ at the given pressure such that the breakdown voltage predicted by Paschen's law tends to infinity. The corresponding value of $x$ is denoted as $x_{inf}$ in the graph and is equal to exp(-$C$). Namely, only by regulating the gap in this way, one can prevent breakdown from occurring when the potential is still to low, of the order of a few kV.

In order to obtain the emission of anomalous radiation it is necessary, on the contrary, that the gap is not short-circuited until the voltage has reached the threshold value of 500 kV, and this for three reasons:

1. The electric field must be strong enough to cause copious field emission of secondary electrons from the whole surface of the cathode, leading to a flat discharge and not to isolated sparks.

2. Since the peak value of the current at the breakdown is given by the ratio $V/R$ ($R$ is the effective resistance of the emitter), the breakdown voltage $V$ must be high enough to give a large current; this is in turn crucial, as we shall see, to cause an intense anomalous emission.

3. There is a minimum voltage value required by the microscopic mechanisms of inter-plane tunnelling which is at the basis of the anomalous emission (see B). This value is of the order of several tens of kilovolts, depending on the dominant radiating transition (eq. (B4) and table). In the conditions described in [1] this minimum value is smaller than the threshold for the flat discharge, but it could be relevant under different conditions.

Note that the condition $x=x_{inf}$ (tuning the gap length in order to obtain $V_B$ infinite) can only be satisfied with limited precision. With the above $B$ and $C$ coefficients, for instance, one finds that $x_{inf}$ =0.2080; when $x$ approaches this value from the right, the breakdown voltage increases, but not very fast (for instance, for $x$=0.2090, $V_B$=26 kV). $V_B$ reaches the 450 kV threshold for $x$=0.2081, which means a tuning within 0.05%. Clearly it is difficult to achieve such a precision, both for the gap length and for pressure. A possible remedy is to use voltage pulses with very short rise time; this is a crucial point, also in view of the possible scaling and simplification of the device. I will discuss it in more detail in my 2$^{nd}$ report.

Until now, we have regarded the pressure as a constant, supposing that only the gap length *d* is varied. This corresponds to the conditions of the original experiment, in which we are interested in this first part of our study. A spontaneous question occurs: what happens if we change *p* and *d* while keeping their product constant, for instance if we make *d* 10 times smaller and *p* 10 times larger? From the practical point of view, this would lead to a simplification of the device. Furthermore, it is known from the general theory of gaseous discharges [3] that the features of the discharge at the microscopic level depend on the ratio *E*/*p* between pressure and electric field. In particular, the Townsend coefficient for primary ionization $\alpha$ depends on this ratio and not on *E* and *p* separately. The average energy of ions in the gas also depends only on *E*/*p*. If the gap length *d* decreases and the voltage is constant (which is necessary, for the three reasons above) then the electric field *E*=*V*/*d* increases in proportion to *p* and the ratio *E*/*p* is constant. This would be an additional motivation in favour of a reduction of the gap. As I will discuss in my 2$^{nd}$ report, however, a stronger electric field on the cathode could have the negative effect of starting the breakdown when the voltage is still too low, unless the pulse rise time is very short.

*A.4 Electron drift velocity. Ionization rate in the gas*

Usually, in a flat Townsend discharge the mean free path of electrons between an inelastic (ionising) collision with a gas molecule and the next collision is given by ~$1/\alpha$ where $\alpha$ is the Townsend coefficient of primary ionization. The theory predicts that the ratio $\alpha$/*P* depends only on the ratio *E*/*P* between electric field and pressure. See for instance in Fig. 4 the typical dependence for hydrogen and nitrogen. In our case the *E*/*P* ratio is very large (~$10^6$ V/(tor cm)), because *E* is large (~$10^4$ V/cm) and *P* is small (~$10^{-2}$ tor).

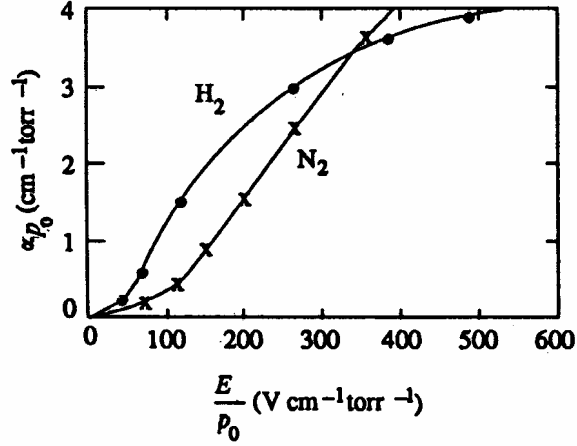

Fig. 4 – (From Naidu, ref. [4]) The Townsend α-coefficient over pressure *P* as a function of the ratio *E/P* between electric field and pressure, for molecular hydrogen and nitrogen. Pressure is reduced to its value at zero centigrades.

The average energy given by the field to the electrons is $U_e \sim eEl$, where $l$ is the electronic mean free path. This path is inversely proportional to the pressure, therefore large values of *E/P* imply large values of the average electron energy. The ionization cross section for electrons reaches a maximum at an energy of the order of $10^2$ eV and then tends to decrease at higher energies (outside the range of Fig. 4). In the absence of more precise data, we can guess that in our conditions (1) there is a ionization at each collision; (2) the electronic mean free path is that obtained from a cross section σ of the order of the square of the atomic radius, i.e. $\sigma \sim 10^{-19}$ m$^2$, $l = RT/(P\sigma N_A) \sim 10^{-3}$ m. We conclude that the average electron energy is $U_e \sim 10^3$ eV and the electronic drift velocity $v_d \sim \sqrt{(elE/m_e)} \sim 10^7$ m/s. This implies in turn that the formation time $\Delta t_1$ of the discharge is less than $10^{-7}$ s.

From this we can also estimate the average ionization rate of the gas. First, in order to compute the electronic current density in the chamber, let us consider a pressure $P \sim 1$ Pa as stated in [1] and $T \sim 70$ K. The number $n$ of gas moles per cubic meter is

$$n = \frac{PV}{RT} = \frac{1}{8.31 \cdot 70} = 1.7 \cdot 10^{-3} \qquad (A4)$$

If the gas was entirely ionized, with ions of charge $1^\pm$, the charge density would be $e\rho = enN_A = 1.6 \cdot 10^2$ C/m$^3$. The electronic current density is given by $j = e\rho v_d$, where $v_d$ is the electronic drift velocity. We then would have $j \sim 3 \cdot 10^9$ A/m$^2$ for a completely ionized gas. Since the current density in the emitter at the discharge is of the order of $10^6$ A/m$^2$ and the sectional area of the emitter is the same as for the active gas column, this implies that the ionization rate needed is of the order of $10^{-3}$.

*Data summary for Part A*

1. Data available or directly computable

| Total capacitance of the Marx generator in serial configuration | $C_S$ | $1.25 \cdot 10^{-9}$ F |
|---|---|---|
| Maximum voltage output of the Marx generator | $V$ | $2 \cdot 10^6$ V |
| Electrostatic energy in the capacitors at maximum voltage | $U_S$ | $2.5 \cdot 10^3$ J |
| Peak value of the discharge current | $I$ | $10^4$ A |
| Current density | $J$ | $10^6$ A/m$^2$ |
| Pressure in the discharge chamber | $p$ | 1 Pa |
| Gap length | $d$ | 15-40 cm |
| Paschen *pd* product | $pd$ | 0.1 - 0.2 Tor·cm |

2. Data computable with straightforward assumptions

| Duration of the current pulse | $\Delta t_2$ | $10^{-7}$ s |
|---|---|---|
| Effective resistance of the emitter | $R$ | 100 Ω |
| Paschen *C* coefficient of secondary emission | $C$ | 1.6 – 2.3 |
| Free mean path of electrons in the discharge | $l$ | $10^{-3}$ m |
| Average drift velocity of electrons in the discharge | $v_d$ | $10^7$ m/s |
| Ionization rate of atoms/molecules in the discharge | | $10^{-3}$ |
| Formation time | $\Delta t_1$ | $10^{-8}$ - $10^{-7}$ s |

3. Data not known

| Rise time of the voltage pulse | Probably not less than $10^{-8}$ s |
|---|---|
| Kind of gases present in the discharge chamber | Probably O$_2$, N$_2$, He |
| $I(V)$ relation | Probably almost linear, with saturation at high $V$ |
| Optimal gap length $d$ | Probably defined by the relation $pd=\exp(-C)$ |

## B. Working upon the hypothesis that all crystal planes radiate in phase

In this Part B we focus our attention on the superconducting emitter. A model will be proposed, which allows to describe the emission of anomalous radiation and to find some relationships between the wavelength and frequency of the radiation, the microscopic features of the emitter and the electric features analysed in Part A. This model successfully predicts the mechanical effects of the anomalous radiation, i.e. the value of the velocity imparted by the pulse to the targets and the maximum energy available in the beam.

*B.1 Explanation and justification of this hypothesis*

As discussed in [1], the radiation emitted by the superconducting cathode at the discharge should actually be called a "virtual" radiation, because it does not obey the usual relation $\lambda=c/f$, but one finds instead $\lambda<<c/f$. This implies that the momentum $p=h\lambda^{-1}$ carried by one radiation quantum is much larger than its energy $E=hf$ divided by $c$, while for free photons or gravitons $p=E/c$. We believe that such radiation only exists as an intermediate state of a quantum process, which begins with the emission of the radiation from the cathode and ends with its absorption in the target. This virtual radiation would not be able to propagate freely to infinity like a real radiation.

At the microscopic level, we suppose that the quanta of virtual radiation are emitted when the Cooper pairs pass by tunnelling from one superconducting *ab* plane to the other, in the *c* direction, under the action of the strong electric field at the discharge. There are definite theoretical motivations to this hypothesis; in particular, we have shown that the only way to obtain a large negative vacuum-like energy density in a superconductor is to produce sharp maxima in the Cooper pairs density, and this is exactly what happens when the supercurrent, driven by the electric field, "pushes" on the borders of the 2D interplane potential barriers to tunnel through. Without going into the details of how this condition triggers an anomalously strong gravitational-like emission, let us work out now the consequences of this simple hypothesis, namely that each *ab* plane emits in phase when crossed by the supercurrent. This is in any case reasonable, since the Cooper pairs wave-function is spatially coherent, if not over the whole area of the cathode, at least over macroscopic portions.

*B.2 Computation of the voltage-per-plane required to impart to Cooper pairs the correct value of momentum* p *(p = $\lambda$/h).*

In order to simplify the analysis at this initial stage, let us suppose that the spacing between superconducting planes in the *c* direction is constant. (In fact, the stacking of *ab* planes in YBCO is more complicated, with interleaved alternated non-SC planes at distances of 2 and 4 lattice spacings; see below.) Let us denote this interplane spacing as $s_c$. Its magnitude order is 1 nm. It is clear that in order to have an emission from all the bulk that interferes constructively, the wavelength $\lambda$ of the virtual radiation needs to be equal to $s_c$. More precisely, we can say that among the infinite possible modes of virtual radiation, only this mode has a high excitation probability.

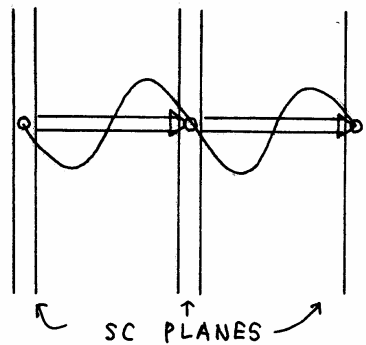

Fig. 5 – Condition for the coherence of radiation emitted by Cooper pairs tunnelling between the *ab* planes, in a hypothetical cuprate superconductor where all planes are superconducting. For the real situation in YBCO see Fig.s 6 and 7.

The momentum $p=h\lambda^{-1}$ carried by each radiation quantum must be supplied by the electric field. We suppose that each quantum is emitted (with a certain probability) in an elementary process in which one Cooper pair crosses the interplane barrier. One could consider multi-particle processes in which more Cooper pairs concur to the emission of a single radiation quantum, but such processes should be far less probable.

So we need to compute the momentum given by the electric field to a Cooper pair while it accelerates between one lattice plane and the next one. (Later we shall also estimate the minimum energy needed, related to the frequency *f* of the radiation quanta; it turns out, however, to be very small.) Let $E_P$ be the kinetic energy imparted by the electric field to a pair over a distance $s_c$. We have $p_P = \sqrt{2m_P E_P} = \sqrt{4m_e E_P}$. Expressing $E_P$ in eV, we find

$$p_P \cong 7.64 \cdot 10^{-25} \sqrt{E_P} \quad (E_P \text{ in eV}) \tag{B1}$$

Inserting $p_P = h\lambda^{-1}$, $\lambda = s_c$, it follows

$$E_P \cong \left(\frac{8.67 \cdot 10^{-10}}{s_c}\right)^2 \quad (E_P \text{ in eV}) \tag{B2}$$

For instance, taking $s_c = 1$ nm, one finds $E_P \sim 0.75$ eV. Now consider the emitter with thickness $\delta = 4$ mm. It contains $4 \cdot 10^6$ planes, and the total voltage needed is $1.5 \cdot 10^6$ V. It is easy to write a general formula for the total voltage needed as a function of $\delta$ and $s_c$:

$$V_{tot} \cong \left(\frac{8.67 \cdot 10^{-10}}{s_c}\right)^2 \cdot \frac{1}{2} \cdot \frac{\delta}{s_c} \cong 3.76 \cdot 10^{-19} \frac{\delta}{s_c^3} \tag{B3}$$

If the total voltage drop through the superconducting part of the cathode is less than this value, then the emission of coherent radiation will not be possible. This is true independently from the nature of the radiation, i.e. whether it is gravitational-like or not. Note that if the radiation is actually emitted, the supercurrent will be subject to a kind of dissipation while crossing the cathode, as if the cathode would have an effective resistance. This dissipation has some peculiar features, because a large voltage drop is necessary in order to give momentum to the Cooper pairs and the radiation quanta, but only a small part of the pairs energy is transferred to the radiation quanta (see table below). The rest of the energy is dissipated in other "parasitic" processes, the most important probably being an electromagnetic emission in direction opposite to the discharge. This "back-radiation" on the back of the emitter has been mentioned by E. Podkletnov in a communication separated from Ref. [1] and must be properly taken into account also for safety reasons. From the theoretical point of view, this electromagnetic emission has not been analysed yet. Anyway, remembering that the effective potential converted into radiation energy amounts to just a few volts, while the applied voltage is of the order of hundreds of kilovolts, one can conclude that the energetic efficiency of the anomalous radiation generation is very low; instead, the device is more effective for the transmission of momentum, and therefore for exerting a force on low-velocity targets.

Is there any relation between the 500 kV threshold for the flat discharge and this microscopic threshold for the emission of anomalous radiation? If it were so, the threshold value of $V$ should depend on the thickness of the superconducting layer of the emitter, which is not true: for "Emitter 2" of thickness 8 mm the threshold voltage is 500 kV, like for "Emitter 1" of thickness 4 mm. Therefore the threshold is probably set by the condition for the onset of secondary ionization at the emitter through field emission (Part A).

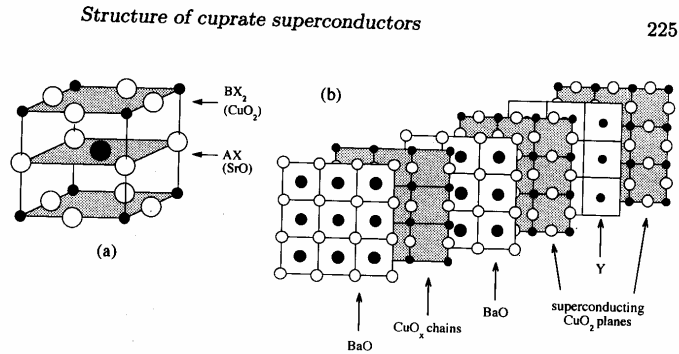

Fig. 6 – (From Waldram, Ref. [2]) Stacking of *ab* planes in YBCO. (a) Partial lateral view showing the crystal cell. (b) Top view of the whole cycle of 6 planes.

Let us now look at the crystal structure in detail and compute the exact value of the total voltage drop needed. The Cooper pairs are accelerated over a distance of 2, 4, 6 … lattice-spacings before emitting a radiation quantum (Fig. 7). We shall call this distance "acceleration space" $s_{accel}$. The interplane spacing in the $c$ direction is denoted by $s_c$ and in the samples employed by Podkletnov is approximately equal to 1.17 nm. The simplest possibilities, in order to have coherent emission, are the following:

(a) $s_{accel}=2s_c$ (the acceleration space is twice the lattice-spacing in the $c$ direction; the voltage drop on the subsequent 4 lattice-spacings is not completely exploited). We can have coherent emission for $\lambda=s_c$ or $\lambda=2s_c$. In this latter case, the potential needed must be computed by replacing $s_c$ by $2s_c$ in eq. (B2).

(b) $s_{accel}=6s_c$ (no intermediate emission). We can have coherent emission for $\lambda=s_c$, $\lambda=2s_c$, $\lambda=3s_c$ or $\lambda=6s_c$.

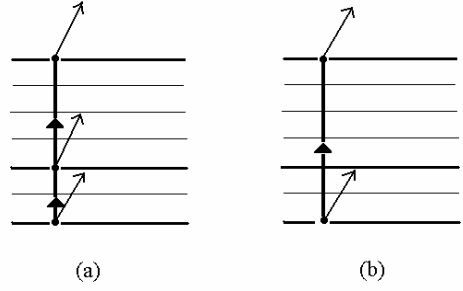

(a)            (b)

Fig. 7 - Schematic representation of the two basic ways of emission of coherent anomalous radiation by Cooper pairs tunnelling between *ab* superconducting planes in the emitter at the discharge. The horizontal lines represent the *ab* planes; the bold horizontal lines are the $CuO_2$ superconducting planes. In (a) the pairs emit two quanta when crossing a six planes unit, in (b) only one quantum. Emitted quanta are represented as thin arrows, pointing sideways only for graphic reasons (in fact, their momentum is parallel to that of the pairs).

Actually, several processes can occur at the same time, with different probabilities. Let us introduce the integers $n$ and $k$, to denote respectively the acceleration space and the wavelength in units of $s_c$: $s_{accel}=ns_c$, $\lambda=ks_c$. Eq. (B3) for the total voltage needed must be generalized as follows:

$$V_{tot} \cong 3.76 \cdot 10^{-19} \frac{\delta}{s_c^3} \frac{1}{k^2 n} \tag{B4}$$

and for the processes depicted in Fig. 7 we find the following values:

| $n$ (acceleration space in lattice units $s_c$) | $k$ (wavelength) | $k^2 n$ | $V_{tot}$ (kV) |
|---|---|---|---|
| 2 | 1 | 2 | 470 |
| 2 | 2 | 8 | 117 |
| 6 | 1 | 6 | 157 |
| 6 | 2 | 24 | 39 |
| 6 | 3 | 54 | 17 |
| 6 | 6 | 216 | 4 |

*B.3 Frequency f of the emitted radiation as a function of current density, lattice spacing and carriers density*

In order to estimate the frequency *f* of the anomalous radiation, consider the time interval between two collective tunnelling events of Cooper pairs from plane to plane. We recall that all pairs are described by a single macroscopic wave-function $\Psi$, of which one can consider the spatial dependence at a given instant and also the temporal dependence.

This temporal dependence of the wave function is characterized by one or more dominant frequencies, and the frequency of the emitted radiation is necessarily defined by these dominant frequencies. The dominant frequencies include the reciprocal of the duration $\Delta t_2$ ($f \sim 10^7$ Hz) of the discharge and the reciprocal of the average time the Cooper pairs take for tunnelling between superconducting planes. Let us estimate this latter time.

The current through the emitter is of the order of $10^4$ A and its surface of the order of $10^{-2}$ m$^2$, so the current density $j = \rho v_P$ is of the order of $10^6$ A/m$^2$. For an estimate of $\rho$ we can consider the London penetration length $\Lambda$ in the *c* direction, $\Lambda_c \sim 890$ nm. The pairs density is given by $\rho = m_e/(2\mu_0 e^2 \Lambda^2) \sim 1.8 \cdot 10^{25}$ m$^{-3}$. This value should possibly be corrected for two reasons: (1) We are not at $T=0$, so $\rho$ is actually smaller. (2) The penetration length $\Lambda$ is well defined for direct currents, not for a high-frequency pulse as in our case.

With this value of $\rho$ we find for the average pair velocity $v_P = j/(2e\rho) = 0.17$ m/s. Cooper pairs cross the planes at the average frequency $f = v_P/s_{accel} = (k/n)v_P/\lambda$. When we know *f* we can compute the energy emitted in each elementary process, the fraction of the total voltage effectively converted into radiation energy and the maximum energy available in the beam. The effective total voltage is computed taking into account the total number of emitting planes. For the processes with $n=2$ and $n=6$ considered above, we find the following values:

| Acceleration space $n$ in lattice units | Average frequency $f$ of emission (Hz) | Average energy $E_f=hf$ of radiation quanta (J) | "Effective" voltage-per-emitting-plane $V' = E_f/2e$ converted into radiation energy (V) | "Effective" total voltage $V'_{tot}$ for the 4 mm emitter (V) | Maximum energy in the beam $U_{max}=IV'_{tot}\Delta t_2$ for the 4 mm emitter (J) |
|---|---|---|---|---|---|
| 2 | $7.26 \cdot 10^7$ | $4.81 \cdot 10^{-26}$ | $1.50 \cdot 10^{-7}$ | 0.171 | $1.71 \cdot 10^{-4}$ |
| 6 | $2.42 \cdot 10^7$ | $1.60 \cdot 10^{-26}$ | $0.50 \cdot 10^{-7}$ | 0.0285 | $2.85 \cdot 10^{-5}$ |

Note that the estimated maximum energy available in the beam is 5-10 times smaller than the targets energy deduced from the experimental data (see the next section). This could imply that certain transitions (those with large $k^2n$ factor and low $V_{tot}$, see table after eq. (B4)) actually produce multiple emissions.

*B.4 The λ-f relation compared with the ratio of energy and momentum absorbed by the targets*

Using the values of λ and $f$ found in the previous sections, we are now able to compute the ratio $E/p$ for the radiation quanta:

$$\frac{E_f}{p_\lambda} = \frac{hf}{h\lambda^{-1}} = f\lambda \tag{B5}$$

According to our model, when the radiation pulse hits the target it must be completely absorbed, because the virtual radiation exists only as intermediate state and all its energy and momentum are transferred to the target. If the target, having mass $m$, is initially at rest, and reaches a velocity $v_t$ (much smaller than $c$) after absorbing the pulse, the ratio between its energy and its momentum is

$$\frac{E_t}{p_t} = \frac{\frac{1}{2}mv_t^2}{mv_t} = \frac{v_t}{2} \tag{B6}$$

The two ratios (B5) and (B6) must be equal, so we find the very useful formula for the target velocity

$$v_t = 2f\lambda\frac{n}{k} = 2v_P\frac{n}{k} = \frac{I}{e\rho S}\frac{n}{k} \tag{B7}$$

With the value above for $v_P$ and $n/k=1$ we have $v_t \approx 0.34$ m/s. This agrees well with the target velocity observed in the experiment. The use of ballistic pendulums as targets allows to deduce $v_t$ easily. For small oscillation angles, the relation between the height $\Delta h$ reached by the pendulum and its swing half-amplitude $\Delta x$ is $\Delta h \approx \Delta x^2/(2L)$, where $L$ is the length of the wire. From this one computes the velocity with which the pendulum has left its rest position after absorbing the brief radiation pulse:

$$v_t = \sqrt{\frac{2\Delta U}{m}} = \sqrt{2g\Delta h} = \sqrt{\frac{g}{L}} \cdot \Delta x \qquad (B8)$$

For instance, with $V$=1500 kV, the 4 mm emitter gives a deflection (see graph in Fig. 3 of [1], reported below) $\Delta x$ =0.1 m; this amounts (with $m$=18 g) to a pendulum energy $\Delta U$=1.1·10$^{-3}$ J and a target velocity $v_t$ = 0.35 m/s. The complete table of values for Emitter 1 is the following:

| Voltage (kV) | Pendulum deflection (mm) | Target velocity (m/s) | Target energy for $m$=18 g (mJ) |
|---|---|---|---|
| 500 | 40 | 0.14 | 0.2 |
| 750 | 70 | 0.25 | 0.5 |
| 1000 | 85 | 0.30 | 0.8 |
| 1250 | 95 | 0.33 | 1.0 |
| 1500 | 100 | 0.35 | 1.1 |
| 1750 | 105 | 0.37 | 1.2 |
| 2000 | 110 | 0.39 | 1.3 |

The agreement between the experimental data and the predicted $E_t/p_t$ ratio is remarkable and supports our theoretical picture. It should be stressed, however, that small adjustments of the theoretical predictions might still be necessary. The exact value of the effective electron mass used to estimate the pairs density $\rho$ and the value of the current $I$ can require the insertion of a numerical factor of order 1.

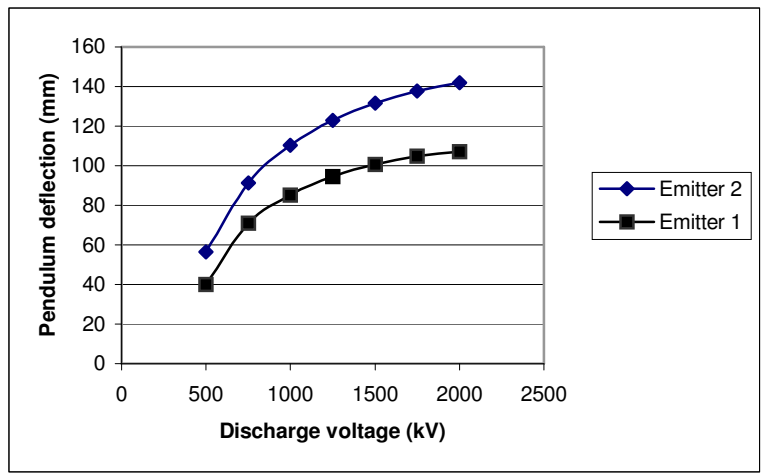

Fig. 8 - Effect of the anomalous radiation beam on a ballistic pendulum of mass 18 g (from Ref. [1]). If our model is correct, the *I(V)* graph should have exactly the same shape.

*B.5 Relation between the frequency* f *of the virtual radiation, the total current* I *and the emitter surface* S

While the wavelength $\lambda$ of the emission is fixed by the microscopic structure of the emitter, the frequency *f* depends on the drift velocity $v_P$ of the pairs in the current pulse: $f=v_P/\lambda$. The velocity of targets hit by the radiation is proportional to *f* and is indeed equal to twice the pairs drift velocity. The pairs drift velocity is given by $v_P = j/2e\rho = I/2e\rho S$ (eq. (B7)). Let us analyse this expression. For a given material at a given *T*, $\rho$ is fixed, and well below $T_c$ (*T*<70 K in our case) the dependence of $\rho$ on *T* is weak. From the relation $v_P = j/2e\rho$ we see that $v_P$ is limited in principle by $j_c/2e\rho$; in our case, however, $j \sim 10^6$ A/m$^2$, while $j_c \sim 10^8$ A/m$^2$; we are therefore far from that limit.

We also see that $v_P$ increases when the emitter surface *S* decreases, provided *I* is constant. However, a smaller emitter cannot collect enough positive ions from the gas, and *I* cannot remain constant. This explains why the application of a magnetic field to the emitter allows to decrease the effective emitter section for the pairs current (because magnetic flux penetration reduces the SC regions), while still leaving a large surface for ions collection. This reportedly improves the effect up to 25% [1].

This 25% increase factor is not entirely clear, however. From the deflection data one deduces that the difference between the performance of Emitter 2 (thickness 8 mm, high trapped field) and that of Emitter 1 (thickness 4 mm, lower trapped field) is of 35%; but the values of the trapped field are not known and so we can not estimate the difference in the effective cross section *S*. Furthermore, we do not known if the current is the same in the two cases; this can depend on the ratio between the effective resistance of the normal layer (which is roughly the same for the two emitters, because the thickness is approximately the same) and that of the superconducting layer.

Finally, we observe that since the pendulum deflection $d$ is proportional to $v_t$, and in our picture $v_t$ is proportional to $I$, the relation $\Delta x(V)$ of the figure above should also represent, up to a fixed factor, the $I(V)$ relation. Experimental determination of the $I(V)$ relation would then allow to check the validity of our picture. We see that the slope of the curve decreases for high $V$. If our picture is correct, this might imply that for high $V$ the conduction in the gas saturates, as ionization rate and ion collection at the emitter are not able to grow further.

*Data summary for Part B*

1. Data independent from the theoretical model

| Density of "Cooper" pairs in the emitter | $\rho$ | $1.8 \cdot 10^{25}$ m$^{-3}$ |
|---|---|---|
| Drift velocity of the pairs for peak current | $v_P$ | 0.17 m/s |
| Targets velocity for peak current | $v_t$ | 0.35 – 0.40 m/s |
| Maximum targets energy for peak current | $U_{max}$ | $10^{-3}$ J |
| Maximum targets mass | $m$ | 18.5 g |
| Energetic efficiency ($U_{max}$/energy in the capacitors) | $U_{max}/U_S$ | $10^{-6}$ |
| Targets acceleration during the pulse | $v_t/\Delta t_2$ | $10^6$ m/s$^2$ |
| Lattice spacing in the *c* direction | $s_c$ | 1.17 nm |

2. Model-dependent data

| Wavelength of anomalous radiation | $\lambda$ | $k \cdot s_c$, $k$=1,2,3,6… |
|---|---|---|
| Frequency of anomalous radiation | $f$ | $10^7 – 10^8$ Hz |
| Acceleration space of the pairs | $s_{accel}$ | $n \cdot s_c$, $n$=2,6 |
| Minimum emitter voltage | $V_{tot}$ | $840/(k^2 n)$ kV |
| Effective voltage per emitting plane converted into radiation energy | | $10^{-7}$ V |
| Estimated maximum energy in the beam | $U_{max}$ | $10^{-4} - 10^{-3}$ J |

3. Data not known

| Trapped magnetic field in the emitter | Probably not less than 0.5 T |
|---|---|

**C. Final notes. Open topics.**

In conclusion I have been able to give, for the first time, quantitative estimates for several macroscopic and microscopic parameters which are crucial to the understanding of the working principles of the device. I am satisfied of this achievement, which has also been possible because I shifted the focus of my investigation from an explanation of the anomalous emission at a quantum gravity level [5], towards a more phenomenological analysis.

The agreement between theory and experimental data is rather good. The picture emerging from all the analysis is that of a coherent, laser-like emission, though with very low energetic efficiency. Some open points are the following:

1. Rise time of the voltage pulse of the Marx generator. This has a great influence on the discharge parameters, in particular on the breakdown over-voltage and the optimal gap length and pressure. These discharge parameters, in turn, define the peak value of the current through the emitter and therefore the frequency of the emitted radiation. In general, it would be useful to make the rise time shorter. I think this is possible even with commercially available Marx generators, still it must be checked if a shorter rise time is compatible with the required power, maximum voltage etc. This issue will have to be considered in detail in the framework of the scaling analysis in my 2$^{nd}$ report.

2. "Dominant" transitions. We have seen that several elementary emission processes are possible, when Cooper pairs tunnel between superconducting planes. Such processes can be labelled by two integers $n$ and $k$ (see table after eq. (B4)). For each elementary process, a minimum "activation" value of the total voltage can be computed, which also depends on $k$ and $n$. Multiple emission processes are also possible. At the present stage, it is still unknown which (if any) are the dominant transitions.

3. The features of the electromagnetic back radiation which dissipates a large part of the energy of the discharge (only a fraction of the order of $10^{-6}$ is carried by the anomalous radiation beam) are still unknown, both theoretically and experimentally.

4. Irreversibility temperature. The anomalous emission reportedly occurs only at temperatures below ~70 K. It is possible that this value corresponds to the irreversibility temperature of the material composing the emitter. Podkletnov did not report any DC or AC measurements of the irreversibility temperature of his samples in the magnetic field present on the emitter. Since the phase coupling between *ab*-planes diminishes abruptly above the irreversibility temperature, one can expect that a lower temperature is necessary to ensure the coherence of the emission.

5. Role of the normal layer of the emitter. A better understanding of this role is imperative, because the presence of the N-layer considerably complicates the fabrication of the emitter. We have seen that the N-layer can contribute to the effective AC resistance of the emitter. If this were its only effect, it would be easy to replace it with another resistive element. One should first explain, however, how a larger resistance can lead to the observed "stabilization" of the discharge. Another possible role of the N-layer can be to avoid a direct junction between the S layer of the emitter and a metal. In any case, it is easy to show that given the brief duration of the current pulse, only a very small fraction of the charge carriers involved in the discharge has to pass through the junctions.

The following is an (incomplete) list of issues to be addressed in my 2$^{nd}$ report:

- scaling properties of the device, in size as well as in pressure, voltage, temperature, magnetic field;
- coherence of the emission, as compared to laser coherence;
- connection between the coherence of the emitted radiation and the microscopic structure of the ceramic emitter;
- comparison between the impulse gravity generator and a free electron laser;
- conceivable simplifications in the structure and fabrication method of the emitter;
- recommended features of the beam detector.

**Symbols**

*E*: electric field

*V*: voltage

$\delta$: thickness of the superconducting layer of the emitter

*S*: emitter surface

*h*: Planck constant

$\rho$: pairs density in the emitter

*j*: current density in the emitter

*P*: pressure in the discharge chamber

*d*: gap length (distance of the electrodes)

*x*: product *Pd*

$V_B$: Paschen breakdown voltage

*B*, *C*: Paschen coefficients

$\alpha$ : Townsend coefficient of primary ionization

*l*: free mean path of ions or electrons in the gas

$s_c$: lattice spacing in the *c* direction

$\Delta x$, $\Delta h$: horizontal and vertical displacement of the ballistic pendulum

$\Delta U$: mechanical energy of the ballistic pendulum

*L*, *m*: length and mass of the ballistic pendulum

$\Delta t_1$: formation time of the discharge

$\Delta t_2$: duration of the current pulse

$\lambda$, *f*: wavelength and frequency of virtual radiation

$v_d$: drift velocity of electrons in the gas at the discharge

$v_p$: average velocity of Cooper pairs in the emitter at the discharge

$v_t$: velocity of the target after it has been hit by the radiation pulse

$p_\lambda$, $E_f$: momentum and energy of a virtual radiation quantum

$p_P$, $E_P$: momentum and energy of a pair in the emitter at the discharge

$p_t$, $E_t$: momentum and energy of the target

$s_{accel}$: acceleration space of pairs

*n*: acceleration space of pairs in lattice units

*k*: radiation wavelength in lattice units

# Chapter 2

# Scaling down/(up) of the impulse gravity generator experiment. Part 2 (2004)

**1. Introduction. Parameters which is desirable to scale down**

The original IGG experiment is complex and requires bulky and expensive equipment. I believe that the reproduction of this experiment in a western laboratory depends, in the medium and long term, on our ability to clarify and explain its working principles and to work out some key simplifications in the design and operation of the device. The prototype was developed and gradually improved on an empirical basis and according to the accessible equipment. The results of the theoretical analysis contained in my previous report, besides being interesting in themselves, give us more confidence in the phenomenon. By establishing relationships between the experimental conditions and the performance of the device, these results can guide us in taking decisions on the scaling-down procedure. Such decisions can be grouped in two categories:

· Reasonably safe, reliable modifications, to be included in any scaling procedure from the beginning.
· Modifications which are desirable, but whose effects are not entirely predictable. Such modifications should be introduced in a reversible way, one after the other, each followed by performance tests.

We shall discuss both kinds of modifications in detail and give a summary of our recommendations in Section 8. It is of course necessary to be very careful, because the idea of a demonstration of concept at small scale has been unsuccessful up to now for the rotating-disk experiments and has generated some scepticism.

Let us start with a list of parameters for which scaling is desirable, in order to make the construction and operation of the device simpler and less expensive.

- *Voltage*. The generation of the required voltage is not particularly difficult, thanks to the multiplication occurring in the Marx generator. The main problems come from the need for electric insulation of the whole chamber. Safety measures are also implied. Furthermore, the high voltage discharge produces parasitic electromagnetic emissions in the radio to infrared range, which can damage electronic equipment and biological tissues.

- *Size*. The construction of a superconducting emitter with 10 cm diameter and uniform alignment of crystal planes is a non trivial task. The construction method is not standard, and Podkletnov's "recipe" appears to be unknown to the commercial suppliers. Even admitting that the construction method is just as described by Podkletnov and yields samples with the required features, it is clear that a reduction in the size of the emitter leads to a simplification. In order to fix the ideas, we shall consider in the following a diameter reduction of 50%. Such a scaling will be considerably helpful in the construction and operation of the discharge chamber, which must support high vacuum and high voltage and must be cooled below 70 K. The large solenoid surrounding the chamber will also benefit much from a 50% size reduction. Actually, under certain conditions it might be possible to eliminate it altogether (see Section 4).

- *Vacuum level*. The required pressure of 1 Pa is not prohibitive in itself, but the low gas density could limit the discharge current and seems to impose unnecessarily long gaps, which in turn make lateral deviations of the discharge more likely and call for a wider chamber and a strong focussing magnetic field.

- *Operation temperature*. The requisite of temperatures below 70 K makes the use of helium mandatory. The use of nitrogen (possible for $T > 77$ K) would be much more convenient.

An area where further improvement of the experiment is possible is the detection of the anomalous radiation and the overall monitoring of the process. More sensitive detectors can compensate the reduction in performance consequent to size scaling. A more sophisticated monitoring than described in [1] would allow to record the *I/V* relationship, the shape of the pulses and the timing between the signals from the discharge and those from the detectors.

## 2. Scaling relations for the current density $j$, $I$, $U_{max}$ (maximum energy available in the beam)

We have seen that the target velocity $v_t$ is twice the average pairs velocity $v_P$. Since $v_P=j/\rho$, where $j$ is the current density and $\rho$ is the pairs density in the emitter, and $\rho$ is essentially constant, we see that $v_t$ is proportional to $j$. Therefore, in order to obtain the same target velocity, it is sufficient that $j$ stays constant upon size scaling. If, for instance, the emitter diameter is reduced by 50%, then its surface becomes 1/4 of the original surface and, provided $j$ is constant, the total current also becomes 1/4. The requirement of a constant $j$ is very natural, because if the structure of the emitter is unchanged upon scaling, then the current density it can support will also be essentially unchanged. The same holds for the current density in the gas, provided the field/pressure ratio and the ionization rate are unchanged. In other words, looking at a portion of the cross-section of the discharge region and of the emitter, all physical processes are still the same.

Also we have seen that the $I/V$ relation in the emitter is approximately ohmic, i.e. $I=V/R$, with $R \sim 100$ Ohms. This holds for the whole range of $V$, as magnitude order, though for the highest values of the voltage a saturation occurs, and the current increases slower than $V$. Since $V$ and the emitter thickness are supposed to be unaffected by the scaling (see below), the ohmic $I/V$ relation is equivalent, for the intensive quantities, to $j=\sigma E$, where $\sigma$ is the conductivity, $\sigma \sim 0.004$ in SI units. Therefore $j$ is constant under scaling, and so is $v_t$.

The maximum energy available in the beam $U_{max}$ is proportional to $I$, not to $j$. If $I \to I/4$, then $U_{max} \to U_{max}/4$, too. This implies that the maximum target mass becomes 4 times smaller (from $\sim 18$ g to $\sim 4.5$ g, if the emitter as the same efficiency as those by Podkletnov). The application of a magnetic field on the emitter increases $v_t$ (by 25%, according to Podkletnov), but it should not increase $U_{max}$.

## 3. Rise time of the Marx generator

The rise time of the voltage pulses produced by Podkletnov's Marx generator can be estimated to be $\sim 10^{-8} - 10^{-7}$ s. This parameter is very important, because it affects several other aspects of the design.

(Direct measurements of the rise time with a high voltage probe plus oscilloscope were not given by Podkletnov. The figure above is deduced from two facts: (1) When the anomalous radiation beam is brought to interact with a laser beam, it causes an intensity reduction of the laser beam with a rise time of that order of magnitude. (2) Personal communications with Podkletnov indicated that any small improvement, i.e. reduction, of the pulses rise time had the effect of increasing the discharge current. This implies that the rise time is close to the lower limit for production of an optimal current, which we denoted earlier as $\Delta t_1$ and estimated to be ~ $10^{-7}$ s.)

It is remarkable that the given figure for the rise time coincides with the rise time of commercially available Marx generators having the same voltage and current output (and therefore the same power; for such a pulse power, the only pulse generators available are just the Marx generators - see Section 7). A quick Internet search yielded the following data.

Compact Marx Generators by Samtech (http://www.samtech.co.uk):

| CMG-01 Specifications | CMG-02 Specifications | CMG-03 Specifications |
|---|---|---|
| Output voltage 1.2 MV | Output voltage 0.7 MV | Output voltage 0.3 MV |
| Rise time 50 ns | Rise time 30 ns | Rise time 20 ns |
| Peak current 20 kA | Peak current 20 kA | Peak current 20 kA |
| Number of stages 15 | Number of stages 10 | Number of stages 6 |
| Pulse repetition rate 1 Hz | Pulse repetition rate 1 Hz | Pulse repetition rate 1 Hz |
| Approx. dimensions 1 m$^3$ | Approx. dimensions 0.75 m$^3$ | Approx. dimensions 0.5 m$^3$ |
| Comments: Oil insulated | Comments: Air insulated | Comments: Air insulated |

Magnavolt (http://www.magnavolt.com): Identical models

North Star (http://www.northstar-research.com): "Marx generators have been supplied from 50 - 1,500 kV (1.5 MV) in both conventional and Marx/PFN configurations. Please specify voltage, current, pulse risetime and pulse duration to receive a quotation."

So there seems to be little space for improvement, i.e. for a reduction of the rise time, which would be necessary in order to work at higher pressure and avoid lateral deviations of the discharge (see Section 4). Still, it is necessary to contact the producers specifying the exact requirements. A diminution of the rise time might well be possible at the expenses of other performance parameters which are less relevant in our case, like the repetition rate (quoted as 1 Hz above; 1 repetition per minute could be enough for us, since the emitter must be cooled down a bit after each pulse).

On the other hand, a rapid search through research papers about Marx generators reveals that faster generators have actually been built; for instance, Ref. [3] mentions a generator with rise time 0.4 ns. Costs and overall performance are less easy to evaluate in this case.

**4. "Safe" scaling hypotheses**

Voltage - I think it should not be reduced, for two reasons:

· Only with high voltage can the high-frequency transitions with small $k$ and $n$ be excited (see table ...).
· The total current $I$ and the current density $j$ are both proportional to $V$, and so are the target velocity and the maximum energy available in the beam.

In principle, one could reduce voltage and thickness in the same proportion, thus maintaining the same voltage-per-plane. Since, however, stimulated emission is very likely to play an important role (see below), cutting the emitter thickness could reduce the emission esponentially, and should be avoided.

Size - I will consider a 50% reduction in emitter diameter (but not thickness) and in the size of the discharge chamber. The thickness of the emitter should stay at 4 mm for the superconducting layer plus at least 5 mm for the normal layer. On this basis, there are further distinct possibilities.

*Hypothesis 1*: minimal pressure increase ($p\rightarrow 2p$). In order to leave the ionization processes in the discharge unchanged, the ratio of the electric field to the pressure must stay the same. If the gap between the electrodes is reduced by 50% and $V$ is unchanged, then the field in the discharge chamber before breakdown is doubled (unlike the field in the emitter after breakdown, which stays the same). So the pressure must be doubled, too.

In this way, the Paschen *pd* product and the breakdown potential $V_B$ also stay the same. So the scaling is consistent. The only difference in the discharge process, with respect to the original full-size device, is that the formation time of the discharge is shorter, for two reasons:

· the electric field is stronger, and the emission of the initiatory electrons from the cathode starts earlier, on the rising voltage front;
· since the gap is smaller and the drift velocity $v_d$ is the same, the secondary electrons cascade takes half the time to reach the anode.

As a consequence, the gap is in principle short-circuited before the maximum voltage is reached. This can be avoided, however, providing the rise-time is at least 2 times shorter, which is not difficult to achieve.

*Hypothesis 2*: more substantial pressure increase (e.g. $p \rightarrow 10p$). Pressure is increased tenfold, and the gap becomes ten times smaller. The *E/p* ratio and the *pd* product stay the same, and so $V_B$ and $v_d$ are constant.

This configuration has important advantages:

· The narrow gap (ca. 1.5 cm instead of ca. 15 cm, compared to an emitter of 5 cm) makes a lateral deviation of the discharge far less likely. This allows for a discharge chamber with less void space on the side of the electrodes (compare Fig. 3) and allows perhaps even to eliminate the focussing magnet.
· The higher gas density implies that a larger current can flow in the gas, and the saturation for large *V* (the drop in the *I/V* ratio) can be possibly avoided.

On the other hand, higher pressure and shorter gap can cause problems:

· The discharge formation time becomes even shorter. Combining the effect of the stronger *E* on the emitter and the shorter gap, the formation time can decrease by a factor between 10 and 100. As above, this will require a Marx generator with shorter rise time (ca. less then $10^{-9}$ s).
· Condensation of vapour on the emitter is more likely. The original IGG design is described as using a rotary pump to create a rough vacuum, and then a cryo-pump to get the pressure down to 1 Pa. Podkletnov suggests that this is needed to prevent condensation on the emitter. However, if all that is needed is to get the partial pressure of water vapor and, perhaps, nitrogen, down to a minimal level to prevent condensation, then that might be accomplished through other means.

I think these problems can be solved with reasonable efforts, and so in view of the advantages, I would recommend operating at higher pressure.

Temperature - Temperatures about 70 K are quite typical as irreversibility temperatures in ceramic superconductors. Therefore the need for temperatures below 70 probably amounts to a need to stay in the irreversibility region of the *B-T* (magnetic field-temperature) diagram. In this region the pinning is effective enough as to prevent the breakdown of flux lines into a "pancake" made of flux rings localized in the *ab* planes; it is well known that the critical current supported by the pancake is much smaller. The identification of 70 K as the irreversibility temperature $T_{irrev}$ in our samples is only an hypothesis, because Podkletnov did not report any measurement of $T_{irrev}$, neither under DC nor under AC. It is a plausible hypothesis, however, from the theoretical point of view, because it is known that above $T_{irrev}$ the coupling between *ab* planes drops abruptly, while just such a coupling is what is needed in order to maintain the phase coherence of the emission.

Apart from this, I do not see other strong reasons to stay below 70 K. When *T* decreases, the superconducting carriers density increases, but at 70 K the density has already reached about the 50% of its zero-temperature value, and at 50 K the 75%. The target velocity is inversely proportional to the carriers density.

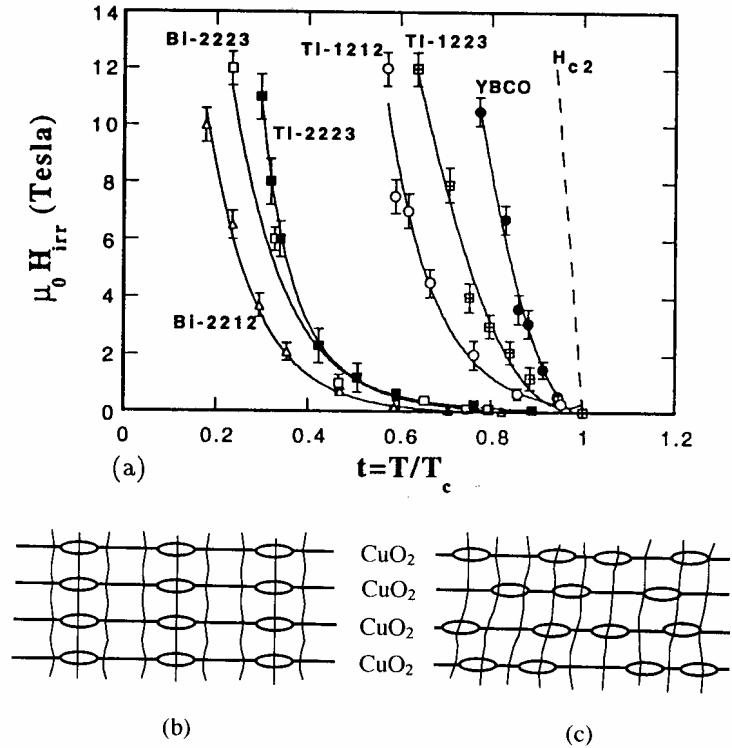

Fig. 1 – (a) The irreversibility lines in a temperature/magnetic field graph for several ceramic superconductors. (b) Typical flux lines pattern below the irreversibility line. (c) Above the irreversibility line, flux lines break to form a "pancake" of flux rings, as the *ab* crystal planes (horizontal) become decoupled. [From Waldram, Ref. 2]

It is known that $T_{irrev}$ depends on the applied magnetic field. Diagrams for YBCO are available in the literature, but depend stronlgy on the specific doping. In general, $T_{irrev}$ decreases when $B$ increases. For pure YBCO, however, the increase is not very large, unless $B$ is very strong (Fig. 1). The magnetic field on our emitter does not exceed 1 T, so for pure YBCO we would expect that $T_{irrev}$ is quite larger then 70 K. It is possible that the special fabrication method employed by Podkletnov acts in the sense of decreasing $T_{irrev}$, instead of increasing it as usually sought for in applications to levitation and high-$j_c$ devices.

In conclusion, it seems impossible to operate the IGG above 77 K, and so avoid the use of helium. It might be possible to increase somewhat the operation temperature, and so to work in the upper part of the 50-70 K range indicated by Podkletnov, provided the magnetic field on the emitter is eliminated or much reduced. This field comes from two sources: (1) The small coil around the emitter, which has the specific purpose of letting in some flux in order to increase the target velocity by approx. 25%. I believe this moderate increase is not of special interest in a demonstration of principle; the main problem in a demonstration is that of not depressing $U_{max}$ too much, and actually any increase in the target velocity which comes without a corresponding increase in the total current has the effect to diminish the maximum admissible value of the target mass.

Another source of magnetic field on the emitter is the large external solenoid for the concentration of the discharge. We have seen that it is conceivable to eliminate this, too, thus obtaining a considerable simplification, provided the optimal gap can be made much narrower than in the original device, of the order of 1 cm.

**5. Scaled figures with some practical remarks**

<u>*B* field and discharge deviation</u>

Fig. 2 below gives the measures of a 50% reduced version. It was made just by copying the original figure of Ref. [1] and adding the new measures, all scaled in proportion to a 5 cm emitter. Note that in his description of the discharge chamber, Podkletnov mentioned a chamber diameter of 100 cm, while in the figure it appeared to be 50 cm (5 times larger than the original 10 cm emitter). This discrepancy was introduced for graphical reasons, i.e. to avoid a drawing with too much "void" all around the emitter. According to personal communications with Podkletnov, however, the void was really needed, because if the chamber walls were too close, then the discharge tended to deviate from its central trajectory and run to the walls. This problem also led to the addition of the big focussing solenoid around the chamber. Another reason for using a large chamber was originally the need to try electrodes of different sizes and to vary their distance in a substantial way.

The large solenoid generates a field of ~ 1 T. Remembering that the drift velocity of the electrons in the discharge is ~ $10^7$ m/s, and supposing that they can reach a transversal velocity of the same order of magnitude, we find a cyclotron radius $r=mv/(eB)$ of the order of 0.1 mm. So the magnetic field is effective at forcing the electrons along small radius spirals (even allowing for multiple scattering cumulating the deviations).

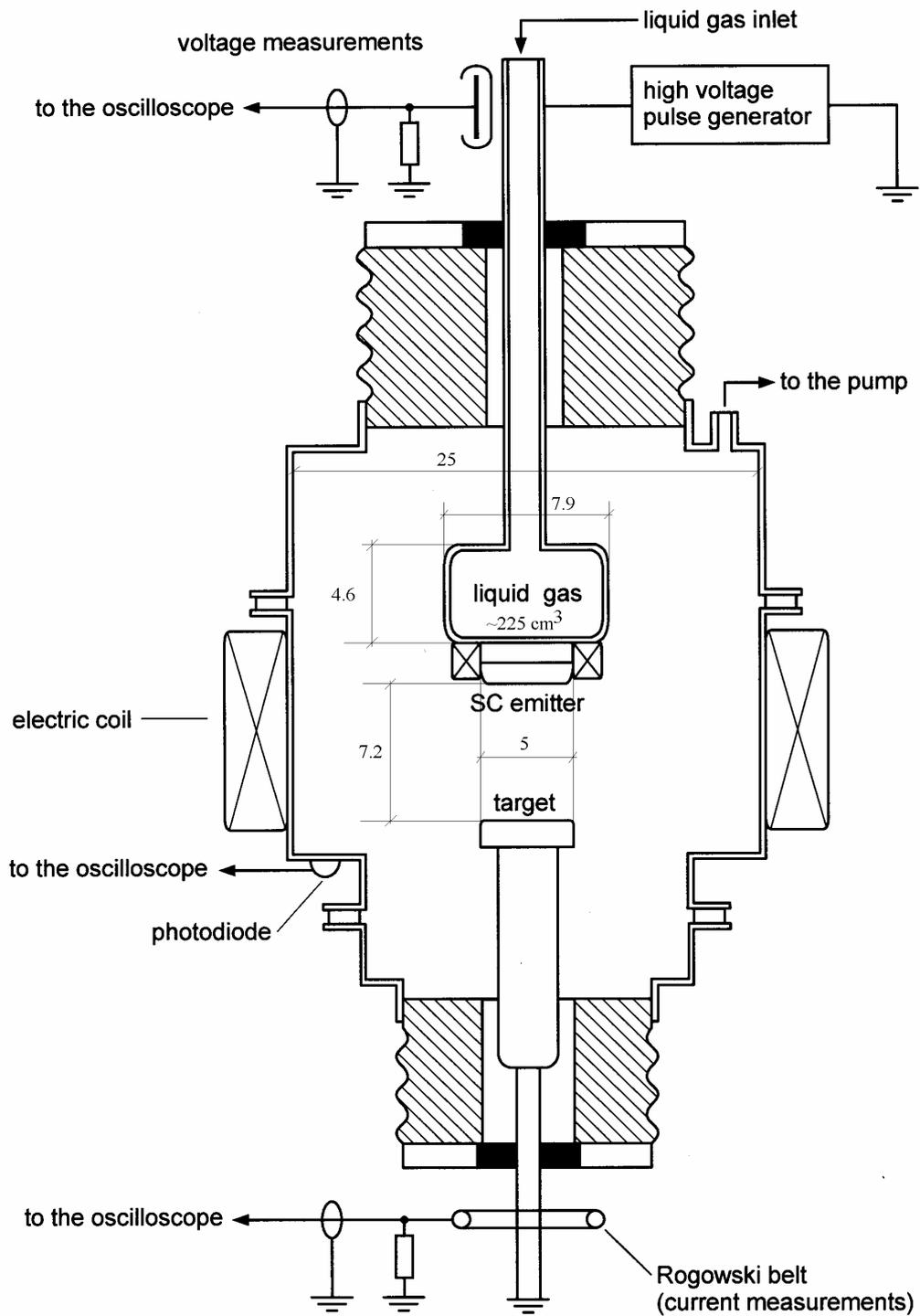

Fig. 2 – Scaling hypothesis nr. 1 (simple 50% reduction). The 25 cm chamber width is to scale on the drawing with the emitter diameter, but actually Podkletnov gave a 100 cm width in the description of the original device and so the reduced chamber width should be 50 cm.

Fig. 3 shows an example of our "Scaling hypothesis nr. 2", where pressure is increased as to allow a much shorter gap between the electrodes. The magnet on the emitter has been eliminated, renouncing to the 25% performance improvement. The big external solenoid has been eliminated, too, since lateral deviations of the discharge are much less likely in this configuration. The chamber diameter is 5 times the emitter diameter; it could be possibly further reduced. The volume of the cooling fluid reservoir must be 1/4 of the original volume (~450 cm$^3$ vs. 1800), not 1/8 as in simple volume scaling, because the emitter thickness stays the same and so its mass is decreased by a factor 4, not 8. It is probably convenient to shape the reservoir in such a way that its diameter is larger than the emitter diameter (for better heat transfer), but the wall attached to the emitter is not (to avoid a distortion of the electric field near the emitter).

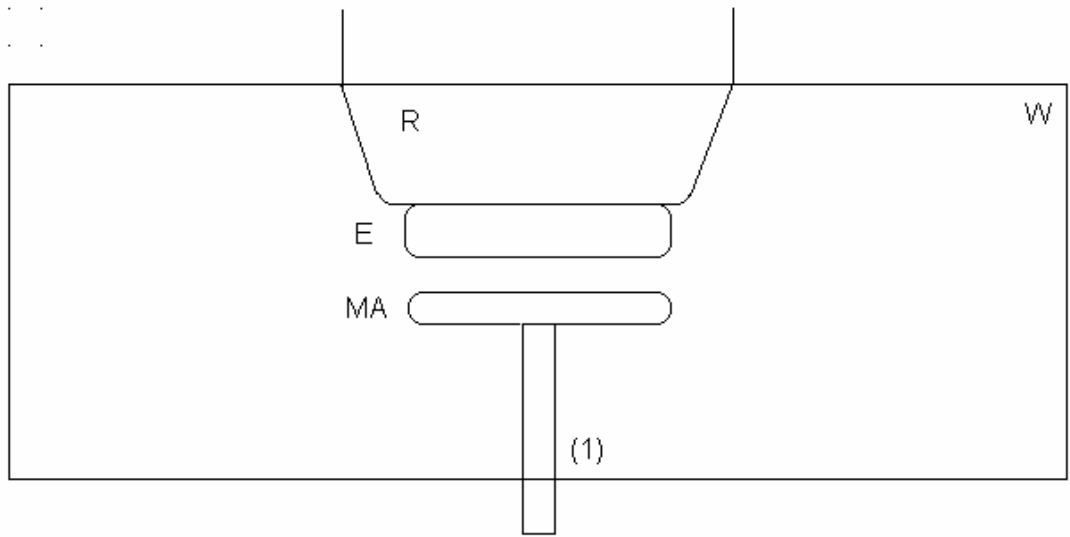

Fig. 3 – Scheme of scaling hypothesis nr. 2. The pressure in the discharge chamber (W: walls of the chamber) is higher, allowing the emitter (E; diameter 5 cm, thickness 1 cm) and the metallic anode (MA) to be closer to each other (~ 1.5 cm or less). Their distance can be adjusted by a micrometric movement mechanism on the metallic anode (1). The cooling fluid reservoir (R) must be slightly wider than the emitter. Both solenoids are eliminated.

Vacuum

Any design of the discharge chamber will require the contribution and advice of a vacuum expert. The correct use of liquid helium is not trivial. I can not give any hint on whether it will be necessary to pump new cooling fluid after each discharge, and how much. No details are known from the original paper about the evaporation rate of helium from its reservoir and it is not clear how the gas in the chamber is controlled, how the sealing needs to be and so on.

In the design of stationary gas discharges, much attention is usually paid to the problem of electrostatic charge accumulation on the chamber walls. In our case, being the discharge very short, it is unlikely that it will be affected by electrostatic charge created right at that moment; but after any discharge some ionized gas remains in the chamber, and this can lead to a gradual accumulation of static charge on the walls. So it might be necessary to periodically change the gas in the chamber. This will also depend on what the walls are made of; Podkletnov used quartz, but other choices are possible. The material of the walls will also affect any discharge deviations.

Safety

Obviously the high voltage, high frequency electromagnetic fields, cryogenics, and vacuum all have hazards. They are, however, well-documented hazards that can be handled with good engineering and standard precautions. According to Podkletnov, there are no hazards resulting from the IGG beam itself. However the main discharge is reportedly accompanied by a poorly known electromagnetic back-radiation that is dangerous for human biological tissues (see also Section 7).

Detectors

The ballistic pendulum employed by Podkletnov has some disadvantages. In particular, automatic data acquisition is difficult; it looks impossible to apply any transducers to the pendulum itself; we can at most imagine to use a photo-electric cell or to record the pendulum movements in a video, along with a clock that can be directly compared with the discharge signal. Furthermore, the pendulum needs to be well isolated from air flows and from vibrations - although checks for strict temporal coincidence with the discharge make accidental false positives very unlikely.

An advantage of the ballistic pendulum as detector is the possibility of using easily interchangeable targets with variable mass, composition, shape, cross section. The major advantage is, that the bob of the pendulum behaves, in the short time when it absorbs the radiation pulse, as a free body. Its response can be neatly predicted by the theory (see Report 1): it must acquire a velocity equal to twice the $E/p$ ratio of the incoming radiation quanta. This prediction is confirmed by the available experimental data.

The situation is different if the detector, whatever object it is, is mechanically connected to another structure. In this case, a part of the beam energy and momentum is transferred to that structure. The absorption of energy and momentum depends on the mechanical transfer function, which is in general unknown and can exhibit resonances for particular frequencies. For instance, if the detector consists of a mass connected to a spring, absorption will be maximum at a radiation frequency equal to the natural oscillation frequency of the system. If the target is very massive and rigid, like a brick wall or a thick iron plate, the absorption will be negligible.

Podkletnov observed that the effect of the radiation beam on the targets reminds that of a short-lasting pressure wave. He employed an array of pressure sensors to map the spatial width of the beam, but this was only a qualitative measurement. Podkletnov also used a microphone isolated from air, in order to detect this "pressure", which is exerted directly on the microphone membrane. The pulse was seen as instantaneous by the microphone, because the rise time and duration of the pulse ($\sim 10^{-7}$ s) are much shorter than the reciprocal of the cut-off frequency of the microphone ($\sim 10^4$ Hz $\rightarrow 10^{-4}$ s). The use of a piezoelectric sensor instead of the microphone could allow to resolve the pulse better (up to $\sim$ 1 MHz). All these techniques allow a recording and temporal analysis of the pulses, yet they give no reliable values for their absolute intensity, as mentioned. Finally, Podkletnov also reported measurements with a laser beam, which is affected by the anomalous radiation beam and can be detected with fast opto-electronics, but this measurement is definitely more complicated.

In conclusion, we think it is better to stick to the use of ballistic pendulums as detectors.

## 6. Uncertainties still present and possible ways to solve them

*Emitter structure vs. beam coherence and collimation*

Some remarks follow, about the microscopic structure of the superconducting emitter and its construction method. These are related to the "quality" of the emitted beam, in particular to its divergence angle.

*1. A melt-textured material is certainly necessary*

The melt-texture-growth (MTG) crystallization process is widely employed for the construction of ceramic superconductors and produces samples with large grains, one for each seed crystal, typically oriented with their *ab* planes parallel to the sample surface [4]. The crystal growth usually occurs from the top of the samples to the bottom, in the *c* direction, and horizontally, in the *ab* direction, at distinct moments. In general, there are some defects at the grain borders and imperfect alignment between grains and inside the grains themselves (the so-called sub-domains). A typical value of the alignment angles for high-$T_c$ levitators is "within 5 degrees or less".

Some features of Podkletnov's emitters are frequently found in commercial MTG levitators: large levitation force, large flux trapping, large critical current density. These all depend on the presence of large grains, more or less well aligned. Podkletnov wrote on page 248 of [1] that "the emitters had a structure typical for multiple-domain levitators with well crystallized and oriented grains". Later he mentioned that a large levitation force is one of the keys of success, but did not specify how large. The size of the grains and sub-domains was not specified either.

The oxygen percentage (the *x* in YBCO formula $YBa_2CuO_{6+x}$) is not specified in Podkletnov's description. From the value of $T_c$ (87-90 K) and from the lattice parameters (3.89 and 3.82 Angstrom in the *ab* plane, 11.69 in the *c* direction, note that he uses a notation with *b* and *c* exchanged), one deduces it must be a quite standard percentage, *x*=0.8 or 0.9 (see figure). This is not only fixed by the composition of the initial ingredients, but also by the control of oxygen pressure during melting. The control of oxygen doping is part of the art of melting

YBCO in a reduced oxygen environment in such a way to obtain a constant doping and so a constant $T_c$ over the samples. In our case this might be not so crucial, since the emitter is only 4 mm thick.

Podkletnov actually did not use an MTG technique, but OCMTG (oxygen controlled melt texture growth [5]. This is an improvement of MTG, but not available commercially, as far as I know. It is certainly a technique which requires consolidated previous experience in fabrication of YBCO.

*2. In order to generate an intense, narrow beam of anomalous radiation, a good alignment of the emitter crystal planes is needed, better than in commercial MTG levitators*

One of the main purposes of the peculiar three-stage fabrication process described by Podkletnov in [1], Section 2.2, is to obtain a very regular crystal structure, with well aligned *ab* planes, parallel to the emitter surface and therefore orthogonal to the discharge current. In his method, the temperature is regulated in such a way that growth is reportedly faster and isotropic; this probably leads to better alignment and smoother grains boundaries.

A regular crystal structure is crucial because it gives the emitted anomalous radiation its coherent character. The radiation is emitted at any point orthogonally to the *ab* plane at that point. If the planes are parallel all over the emitter, the radiation adds in phase and there is constructive interference. If this is not the case, the emission looks more like that of an ordinary lamp instead of a laser emission. Different portions of the emitter "shot" in different directions; we won't have a narrow and coherent beam, but a divergent beam. At a fixed position, the energy collected by a target will be small. Consider for instance that a small optical laser can have the power of few milliwatts; a lamp of the same power is clearly very weak.

The velocity of a target hit by the anomalous radiation depends only on the current density in the emitter and on the carriers density; there exists a limit, however, on the maximum energy emitted in the form of anomalous radiation at any discharge, and this in turn sets a limit on the mass of the targets: heavier targets will just be unaffected by the radiation. In Podkletnov's device, which has a clean, narrow beam and a large emitter, the maximum energy in the beam corresponds to a maximum target of ca. 20 grams. The maximum energy is inversely proportional to the emitter surface, so upon scaling it gets smaller. If in addition there is poor alignment in the crystal structure of the emitter, a reduction of a factor 10 or more can easily occur. This would mean being forced to use targets of the order of 0.1-1 grams and to stay quite close to the device.

Stimulated emission present?

The coherence of the electromagnetic radiation emitted by lasers and masers is due to the phenomenon of stimulated emission: each radiation quantum emitted by the active material is trapped for some time inside a resonant cavity, and every time it passes through the active material again it stimulates the emission of another quantum, which is coherent with the stimulating one (i.e., its wave function has the same phase). Stimulated emission is also present in systems, like certain free electron lasers, which lack a resonant cavity, so that the emission occur in a "single pass". (This latter case bears further important analogies to the IGG emitter, so we will come back to it later, see Section 7.) The question is, therefore, if in our case stimulated emission can be present. This would imply that each emitted quantum of anomalous radiation, while propagating forward in the same direction as the average Cooper pairs motion, stimulates the emission of other quanta with the same phase. The energy and momentum carried by all quanta must come in any case from the pairs, and in turn from the accelerating potential. If there is stimulated emission, the number of emitted quanta should increase exponentially through the emitter and towards the gas, with a cascade process. A coefficient can be defined, which gives the elementary probability of stimulated emission, and can be at most 1, in which case each quantum stimulates the emission of another quantum at each jump between *ab* planes.

The energetic balance described in Report 1 is unaffected by the occurrence of stimulated emission, and so are the different kinds of transitions available, their frequencies and wavelengths, which are all defined by the features of the crystal lattice and of the accelerating potential. Stimulated emission could explain, however, the small beam divergence reported by Podkletnov. In fact, let us suppose that the crystal planes in the emitter have a certain angular distribution, i.e. they are oriented on the average parallel to the emitter surface, but with small variations between one grain of the material and its neighbours. After an accurate microscopic analysis of the samples, one should be able to compile a statistical table with the percentage of planes oriented between 0 and 1 degrees from the parallel direction, and then the percentage between 1 and 2 degrees, and so on, obtaining for instance the graph and the first two columns of the table below.

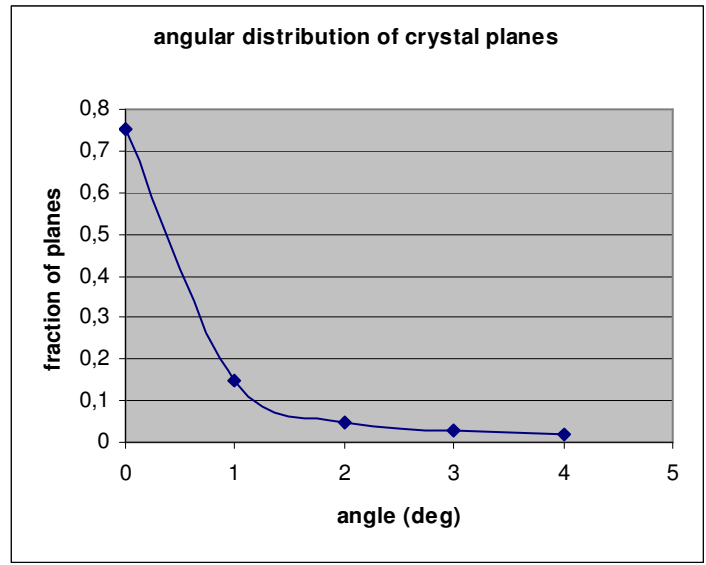

| Angle (deg) | Fraction of planes | Relative intensity $I_0:I$ for $P=10^{-4}$ | Relative intensity $I_0:I$ for $P=10^{-5}$ |
|---|---|---|---|
| 0 – 1 | 0.75 | 1 | 1 |
| 1 – 2 | 0.15 | $2.3 \cdot 10^{17}$ | 54 |
| 2 – 3 | 0.05 | $3.7 \cdot 10^{32}$ | 1782 |
| 3 – 4 | 0.03 | $1.8 \cdot 10^{41}$ | 13187 |
| 4 – 5 | 0.02 | … | … |

If the material is of high quality, deviations will be small, at most a few degrees. This would still imply, however, a beam divergence much larger than observed. According to Podkletnov, the beam does not exhibit any appreciable divergence over a distance of 150 m, i.e. the divergence angle must be smaller than approx. 0.005 degrees! If stimulated emission is present, on the other hand, the ratio between the intensity $I_0$ emitted at zero degrees and the intensity $I$ emitted at a given angle is much larger than the relative fraction of planes, because an exponential amplification occurs. This holds true even if the probability of stimulated emission is small, because the number of crystal planes crossed by the current is very large. The general formula for the relative intensity turns out to be

$$I_0:I = (1+P)^{n(1-a)}$$

where *P* is the probability of stimulated emission, *n* is the number of planes and *a* is the relative fraction of planes at a given angle. Taking *n* of the order of $10^5$, we obtain the third and fourth columns of the table above. It is clear that even for small values of the probability ($P=10^{-4}$, $P=10^{-5}$) the intensity decreases very quickly as the angle increases.

*Beam divergence vs. targets section* - A small divergence angle of the beam implies that the momenta of all virtual particles composing the beam are aligned within this angle. Note that if the divergence angle was exactly zero, the number of virtual particles absorbed by any target would not depend on the target cross section. The independence of the target velocity $v_t$ on the target cross section was observed experimentally. This looks counter-intuitive, but is a consequence of the energy-momentum conservation and of the virtual character of the anomalous radiation. (Another consequence is, that in the absence of any targets the beam does not propagate "to infinity", but it should be considered absent.) A very dense target with mass equal to the maximum movable mass can absorb all the beam energy, even if its cross section is much smaller than the beam cross section; this happens because the process in which all virtual particles in the beam are absorbed by the target is compatible with energy-momentum conservation. On the other hand, if the beam is divergent, the target cross section and the distance between target and emitter do matter, because the conservation condition also involves the target position and width (see Fig. 4). In practice, if the beam is divergent, it is better to use wider targets and place them closer to the emitter.

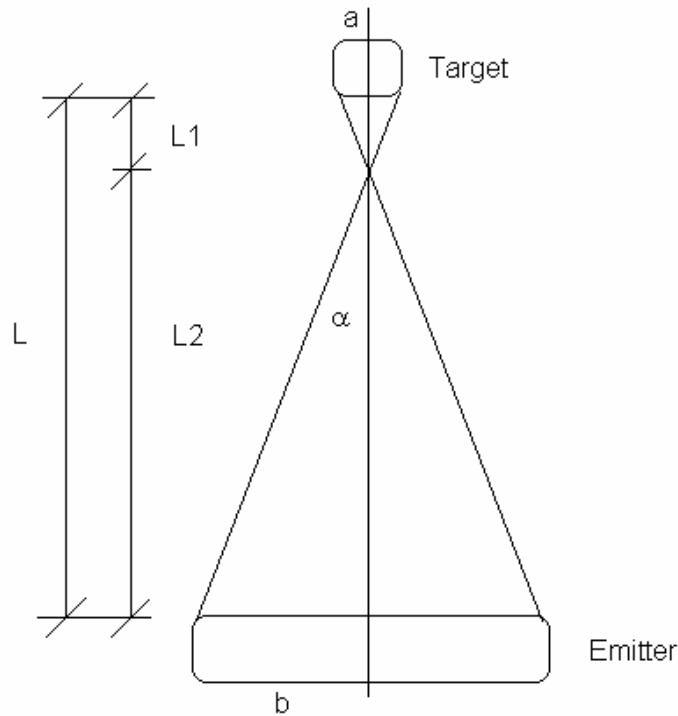

Fig. 4 – If the beam is divergent, momentum conservation implies that the target can absorb radiation quanta emitted with a maximum angle $\alpha$ with respect to the surface normal. For small angles, and $\alpha$ in radiants, one has approx. $a+b=\alpha L_1+\alpha L_2=\alpha L$ and therefore $\alpha=(a+b)/L$.

Emitter structure vs. N-layer

One peculiar feature of Podkletnov's recipe is that he performs melting two times, and after the first time he grinds the material. The first time he does not put any seeds for uniform orientation of the crystal growth; the scope of the first melting is apparently to obtain small grains with good crystal structure which are then properly oriented in the second melting.

In order to arrange the normal (rare-earth doped) layer under the superconducting layer, Podkletnov uses the material with 30-microns grains obtained in Stage 2 after grinding. He presses the two layers together at high pressure (maybe this is possible only because the grains are larger than in a powder) and then starts the OCMG treatment. In this way he obtains a unique texturing for both layers, from the top to the bottom, without separating interfaces. Perhaps, fabricating the two layers separately and then attach them is not good enough. The three stages described by him would then all seem necessary. This depends strictly on the role played by the normal layer in all the process.

According to Ref. [1], if the normal layer is absent, "the discharges tend to be less regular and the anomalous radiation is much weaker". The meaning of these observations is not very clear. In particular, it is not known whether the absence of the normal layer increases or decreases the discharge current. A simple hypothesis is that the normal layer acts as passive resistive element, limiting the current; in this case, it could be easily replaced by another passive resistor. Another obvious observation is that the normal layer avoids a direct contact between the copper cryostat and the superconducting part of the ceramic emitter. We need to look more closely at this idea, however.

When a current flows from a normal conductor (where it is carried by free electrons) to a superconductor (where it is carried by pairs, although the pairing mechanism is unknown for HTCs), the electrons get to form pairs at the boundary (Fig. 5.A). In quantum mechanical terms, we can say that electrons undergo a transition between two states, and a transition probability can be defined. If the crystal structure of the two materials is the same, without interfaces, the wave functions "overlap" much better and the transition probability is larger. On the other hand, between metal and ceramic there is a potential barrier that the electrons must overcome. So in the presence of a normal ceramic layer the passage of electrons from metal to superconducting ceramic occurs in two steps: first the free electrons jump over the potential barrier and then, when they must "condense" in pairs, they are in the right wave function already.

The current pulse in the emitter is so short, however ($\sim 10^{-7}$ s), that electrons and pairs manage to cross only a few crystal planes in that time (their drift velocity is about 0.2 m/s, so the displacement is $\sim$ 10 nm). The total emitting planes in the S layer are millions, so even if the free electrons do not form couples and start emitting right away, this does not make a difference. What happens, is that in the brief time of the discharge the normal electrons simply invade the S part for a length of $\sim$ 10 nm without forming pairs. The reasoning above then fails.

There exists another simple hypothesis. The non-superconducting part of the emitter is fixed to the copper cryostat using metal indium or Wood's metal. The contact might be not perfectly uniform, so at certain points the current flow could be easier. Then, in the ceramic layer near the metal there will be transverse currents, because after current has entered the ceramic it tends to re-distribute along the high-conductivity *ab* planes. If we want that in the S layer, where the anomalous emission occurs, the pairs move well perpendicular to the planes, we need an intermediate ceramic region where emission does not yet occur. If this intermediate region is absent, the emission will be scattered around and the beam so divergent that its effective intensity will be much lower.

## 7. Matters for further work

Electromagnetic emissions and "back-radiation"

It is important to learn more about these "parasitic" emissions, both for the sake of safety and because they can give information on the physical processes occurring in the discharge chamber and in the emitter. It is first necessary to measure the intensity and wavelength of the parasitic radiation, for instance with a field meter, supposing for a start that it is electromagnetic. Podkletnov observed that the parasitic electromagnetic radiation passes through a Faraday cage with inter-spacing of the order of 1 cm placed around all the device, therefore its wavelength should be at most that of infrared radiation, and the corresponding energy of the order of a fraction of eV per photon. Since the voltage drop per plane in the emitter is about 0.1 eV, it is possible that this electromagnetic radiation is emitted in the same microscopic process (tunnelling between crystal planes) where the anomalous radiation is also emitted. Such an electromagnetic emission could actually be one of the dissipative processes which carry away the Cooper pairs energy, since we know that the anomalous radiation quanta carry much momentum but very little energy. (By contrast, photons carry much energy but very little momentum.)

It is known that the electromagnetic "Bremsstrahlung" (braking radiation) can have a backwards peak, as happens for instance in the production of X-rays by electron bombardment of a metal anode. Emission of such a radiation from the bulk of a superconductor is in principle forbidden, however, because electromagnetic radiation can not usually penetrate a superconductor by more than few microns (the high-frequency version of the Meissner effect). Infrared emissions from ceramic superconductors were reported in the literature, but only in the case of thin films [6]. Furthermore, if the back-radiation is really originating from the emitter, it must propagate backwards passing through or around the normal part of the emitter and the metallic helium/nitrogen reservoir.

Further possible sources of electromagnetic radiation are ions recombination in the discharge chamber and the high-frequency components of the discharge current itself. In conclusion, it is clear that more experimental and theoretical investigation is needed on this front.

Comparison with a free electron laser

There are remarkable analogies between the IGG device and a free electron laser (FEL). One of the main uses of FELs is for the generation of short-wavelength electromagnetic radiation. Being this a real radiation with energy-momentum ratio *E*/*p*=*c*, its generation requires a large energy of the electrons in the beam, which usually originate from a particle accelerator. The radiation beam is narrow, in comparison to that of the IGG, and the total current of the electron beam is smaller. The most strict analogy is with the single-pass SASE FEL (see below). The single pass in a FEL requires, however, a length of several meters, because of the large electron energy needed, while in the IGG it occurs in a few millimetres. The light emitted by a FEL has a coherent character due to the coherent superposition of the waves emitted along the electron beam and possibly to the presence of stimulated emission.

In a free-electron laser, free electrons (i.e., those not bound to nuclei) from a particle accelerator or some other source are passed through an undulator (commonly called a "wiggler"), a device consisting of a linear array of electromagnets. An alternating magnetic field in the undulator bends the electrons into a spiral path around the lines of force, whereby they are accelerated to high velocities and emit energy in the form of synchrotron radiation. The intensity and wavelength of this radiation can be adjusted by modifying certain parameters of the magnetic field.

Free electron lasers can operate due to different radiation processes: "magnetic bremsstrahlung" in the undulator, Smith-Purcel and Cherenkov radiations, radiation in the laser wave. Most existing FEL devices use for the feedback forming two parallel mirrors placed at the ends of the working area. At very short wavelengths, normal-incidence mirrors of high reflectivity are unavailable. In this case, one can consider free-electron lasers based on the principle of Self-Amplified Spontaneous Emission (SASE) [7]. In a SASE FEL lasing occurs in a single pass of a relativistic, high-quality electron bunch through a long undulator magnet structure. Provided the spontaneous radiation from the first part of the undulator overlaps the electron beam, the electromagnetic radiation interacts with the electron bunch leading to a density modulation (micro-bunching) which enhances the power and coherence of radiation. In this "high gain mode", the radiation power $P(z)$ grows exponentially with the distance $z$ along the undulator.

Typical parameters of a SASE FEL are the following:

· beam energy: 233 MeV
· peak electron current: 400 A
· effective undulator length: 13.5 m
· radiation wavelength: 109 nm

A further characteristic feature of SASE FELs is the concentration of radiation power into a cone much narrower than that of wavelength integrated undulator radiation.

Solid-state device alternative to the discharge?

The essential for the IGG effect is that in a ceramic superconducting emitter having a definite structure, a supercurrent with certain features is forced (duration, intensity, carriers density, driving voltage). It is not essential that this is obtained through a gas discharge, though that might be useful for several purposes (for instance, for monitoring the current: the occurrence of a flat discharge implies that a uniform current is flowing through the surface). Can perhaps some solid-state electronic device do the same job?

It has been suggested that a solid-state thyristor could be used as switch. I have checked, however, that thyristors can not be used at the power level required by the IGG. Commercial producers offer thyristor/thyratron switched systems between ~ 10 MW and 200 MW, but only Marx generators above 200 MW. The IGG requires at least 5000 MW in the original version, and ~ 1000 MW in the 50% scaled-down version.

**8. Conclusions. Recommendations for construction of a 50% scaled-down version**

Device in general and discharge chamber

Voltage and temperature should stay the same as in the original version, i.e., *V* over 500 kV and *T* below 70 K. The emitter diameter should be 5 cm and its thickness 4 mm (superconducting layer) plus 4-6 mm (normal layer). The small coil or permanent magnet originally attached to the emitter should be omitted. It should be allowed for a micrometric adjustment of the gap length by movement of the metal electrode. We suppose at this stage that the fabrication method of the emitter and its microscopic features are exactly as described by Podkletnov.

Under the conditions above, we can safely predict that the target velocity will be the same as in the original experiment (0.35-0.40 m/s). The maximum energy available in the beam will be 1/4 of its original value and so will be the maximum target mass, thus being approximately 4 g. We recommend in general the use of a low density target (for instance, a polystyrene ball) under vacuum, positioned quite close to the emitter (e.g., no more than 2 m), with thick metal screens and possibly a wall in between.

In a 50% scaling down, the diameter of the discharge chamber passes from 100 to 50 cm. We think this size is still impractical for construction and operation purposes. We therefore propose a further reduction, to a 25 cm diameter, accompanied by a gap reduction and a pressure increase: for instance, d→d/10 (~15 cm → 1.5 cm) and p→10 p (1 Pa → 10 Pa). This step has definite physical motivations, but is less safe than the simple 50% reduction mentioned above. As a further bonus, it would allow to dispose of the large external solenoid for discharge focalization and possibly to increase the discharge current (and thus the target velocity) at the highest voltages. Two conditions must be satisfied, for this further reduction to work:

1. A fast Marx generator with a rise time of the order of 1 ns must be available. Compare the full discussion in Section 3.
2. It must be possible to control and reduce the partial pressures of $H_2O$ and $N_2$ inside the chamber below the limit when vapour condensation on the emitter occurs.

If condition 2 is satisfied and condition 1 is not, there is the alternative possibility to increase the pressure further, abandoning the exact inverse proportionality to the gap length, in order to "slow down" the discharge: for instance, p→100 p, d→d/10.

Construction of the superconducting emitter

A high quality melt-textured material is needed, with large jc, large levitation force and a good alignment of ab planes parallel to the surface. The irreversibility temperature must be above 70 K in zero field. (In case the irreversibility temperature turned out to be above 77 K, it could be possible to work at 77 K and use only nitrogen as coolant, at variance with the recommendations above.)

The normal layer can be obtained by doping the lower part of the sample during the texturing in such a way to make it non-superconducting at the operation temperature. The interface between the superconducting and normal layers does not need to be very clear-cut, but the crystal structure must be uniform from the top to the bottom.

Beam divergence: in the absence of any detailed data on the crystal structure of Podkletnov's samples and on the stimulated emission coefficient, it is impossible to predict which degree of planes alignment is necessary to obtain a beam divergence as small as in the original device. Since the beam divergence indirectly affects the intensity of the radiation hitting any single target, this factor actually represents the main uncertainty on the device performance in a replication or scaling-down.

Additional detectors and monitoring of the process

· Voltage (both on the gas gap and on the emitter) and current should be monitored through suitable high voltage probes and oscilloscopes.
· Photoelectric cells could be employed to mark the instant when the pendulum starts to move and give a trigger signal that can be compared with the other oscilloscope readings.
· Electromagnetic field meters all around the chamber and in particular on the back of the emitter and near the targets will allow a mapping of the "parasitic" electromagnetic emissions.

Outlook

In the forthcoming months we could examine the various possible design options and decide what capabilities we want to build into the experimental apparatus for conducting early experiments, and what capabilities we might want to plan for adding later.

Before that, however, Conditions 1 and 2 above should be checked. It is necessary, in my opinion, to

1. Contact commercial producers of Marx generators giving our detailed specifications and asking for a possible reduction of the rise time to ~ 1 ns.
2. Contact vacuum experts and check for the possibility of reducing the partial pressures of $H_2O$ and $N_2$ in the chamber below the level for condensation on the emitter, at total pressures from 1 to 100 Pa.

Data summary for the "safe" scaled version (compare the Data Summaries A and B of Rep. 1)

| | | |
|---|---|---|
| Total capacitance of the Marx generator in serial configuration | $C_S$ | $3 \cdot 10^{-10}$ F |
| Maximum voltage output of the Marx generator | $V$ | $2 \cdot 10^6$ V |
| Electrostatic energy in the capacitors at maximum voltage | $U_S$ | $6 \cdot 10^2$ J |
| Peak value of the discharge current | $I$ | $2.5 \cdot 10^3$ A |
| Current density | $J$ | $10^6$ A/m$^2$ |
| Pressure in the discharge chamber | $p$ | 2 Pa |
| Gap length | $d$ | 7.5-20 cm |
| Paschen $pd$ product | $pd$ | 0.1 - 0.2 Tor·cm |
| Duration of the current pulse | $\Delta t_2$ | $10^{-7}$ s |
| Effective resistance of the emitter | $R$ | 400 Ω |
| Paschen $C$ coefficient of secondary emission | $C$ | 1.6 – 2.3 |
| Free mean path of electrons in the discharge | $l$ | $10^{-3}$ m |
| Average drift velocity of electrons in the discharge | $v_d$ | $10^7$ m/s |
| Ionization rate of atoms/molecules in the discharge | | $10^{-3}$ |
| Formation time | $\Delta t_1$ | $10^{-8}$ - $10^{-7}$ s |
| Density of "Cooper" pairs in the emitter | ρ | $1.8 \cdot 10^{25}$ m$^{-3}$ |
| Drift velocity of the pairs for peak current | $v_P$ | 0.17 m/s |
| Targets velocity for peak current | $v_t$ | 0.35 – 0.40 m/s |
| Maximum targets energy for peak current | $U_{max}$ | $2.5 \cdot 10^{-4}$ J |
| Maximum targets mass | $m$ | 4.5 g |
| Energetic efficiency ($U_{max}$/energy in the capacitors) | $U_{max}/U_S$ | $10^{-6}$ |
| Targets acceleration during the pulse | $v_t/\Delta t_2$ | $10^6$ m/s$^2$ |
| Lattice spacing in the $c$ direction | $s_c$ | 1.17 nm |
| Wavelength of anomalous radiation | λ | $k \cdot s_c$, k=1,2,3,6… |
| Frequency of anomalous radiation | $f$ | $10^7 – 10^8$ Hz |
| Acceleration space of the pairs | $s_{accel}$ | $n \cdot s_c$, n=2,6 |
| Minimum emitter voltage | $V_{tot}$ | $840/(k^2 n)$ kV |
| Effective voltage per emitting plane converted into radiation energy | | $10^{-7}$ V |

| | |
|---|---|
| Rise time of the voltage pulse | 20 ns or less |
| Kind of gases present in the discharge chamber | He |
| *I*(*V*) relation | Almost linear |
| Optimal gap length *d* | Defined by the relation *pd*=exp(-*C*) |

# Chapter 3

# Evaluation of the pulse parameters, target velocity and beam energy for the IU1045 and IU1080 Marx generators (2005)

**Introduction**

In our previous analysis of the discharge process, we assumed that the Marx generator produces a voltage impulse of a certain amplitude and we stressed the importance of having a short rise-time, but we did not make reference to any specific model of Marx generator. Among those available (commercially or through research labs) two models were then identified as possible candidates, namely IU1045 and IU1080, produced by Information Unlimited. Their main specifications are given below [2]. The crucial parameter for the definition of the rise-time is the total inductance $L$ (which is unknown for Podkletnov's device). The rise-time and the ratio between the effective maximum voltage on the load $V_{max}$ and the nominal maximum voltage $V$ depend on the load, too, and can be computed according to standard circuit theory. This computation requires the knowledge of the load resistance $R$, which we can only estimate approximately, also because we do not know its physical origin (we shall address this point in detail in our next report).

We obtain in this way an estimate for $a$ (total pulse duration), $b$ (rise-time), the ratio $V/V_{max}$ and so the maximum load current $I_{max}$. Another new important outcome of this new analysis is the following: it turns out that before the breakdown ("Phase 1" - see below) the voltage on the gas gap has an high-frequency component. This can be relevant in order to define the features of the discharge.

Based on the estimated values of $a$, $b$ and $I_{max}$, we then apply our microscopic model to obtain the foreseeable performance of the scaled device as generator of anomalous radiation, summarised in the parameters $v_t$ (targets velocity) and $U_{max}$ (maximum energy available in the beam). From $v_t$ and $U_{max}$ one can compute the "maximum target mass" $m_{max}$; as we shall show later, this has the meaning that for targets with larger mass the beam acts in an inelastic way and some of its energy is dissipated.

In order to account for the uncertainty on the emitter resistance *R*, three possible values have been inserted in all calculations. We arrive in this way at Table 1, where we give targets velocity and beam energy for the IU1045 and IU1080 generators in the scaled-down IGG version (emitter with diameter 5 mm and thickness 25 cm$^2$).

For a check, we finally apply the standard formulas for *a*, *b* and $V_{max}/V$ and then the microscopic formulas to Podkletnov's device (emitter with thickness 10 mm and diameter 100 cm$^2$). In that case we know in advance the performance, but other data are unknown: notably *L*, while of $I_{max}$ we only know the magnitude order. We find a reasonable agreement and discuss the residual discrepancy (the estimate for $U_{max}$ is ca. 3 times smaller than observed) and how to eliminate it.

**1. Standard circuit theory applied to pre-breakdown and post-breakdown phases**

In this Section we want to describe the discharge process taking into account the inductance of the Marx generator and the capacitance of the gap before breakdown. There are two phases: before the breakdown and after it. The first is less important for the anomalous emission and here the gap capacitance plays a role, while after breakdown the gap just acts like a short-circuit. The breakdown is very fast in the scaled version, because the electrodes are closer, while the average ion velocity is the same as in the full version. We can consider the breakdown almost instantaneous, in comparison with the other times involved. The rise time of the current is limited by the inductance of the Marx generator, however. The requirement of a fast Marx generator is still valid, as we shall see, but not so much because the rise time of the generator has to compete with the breakdown time; rather because a fast Marx has a small inductance and this implies that the current in the load can grow larger.

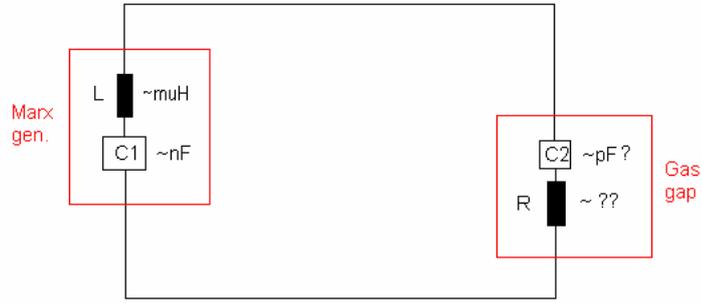

**Fig. 1** - Equivalent circuit for Phase 1 (pre-breakdown)

Phase 1 (pre-breakdown). In this first phase, the voltage pulse produced by the Marx generator is transferred to the gas gap, until breakdown occurs. The circuit comprises a large capacitance $C_1$ (~10 nF), that of the Marx, which after the quick firing/erecting process begins, say at time $t$=0, to charge the smaller gap capacitance $C_2$ (~10-100 pF) through the inductance $L$ and the resistance $R$. In the absence of the breakdown, the voltage at $C_2$ would grow until it reaches that of $C_1$, the nominal output $V$ of the Marx, or more exactly an equilibrium value slightly below that (by a fraction ~$C_2/C_1$). The circuit equation is

$$-\frac{Q_1}{C_1} - LI' + \frac{Q_2}{C_2} - RI = 0 \qquad (1.1)$$

where $Q_1$ and $Q_2$ are the charges on the capacitors 1 and 2. Taking the time derivative, we obtain the usual equation for the RLC circuit

$$LI'' + RI' + \frac{I}{C_{tot}} = 0 \qquad (1.2)$$

where $C_{tot}=(1/C_1+1/C_2)^{-1}$. The solution of this equation is exponential only if the over-damping condition $R>2\sqrt{(L/C)}$ is satisfied. However, with $L$=4.2 µH and $C$=10 pF the over-damping condition amounts to $R$>1.3 kΩ; with the same $L$ and $C$=100 pF we find $R$>410 Ω. Such high values for $R$ are very unlikely, because the estimated value of the ohmic resistance is about 400 Ω after breakdown, but is presumably much smaller (~ 1 Ω or less?) before breakdown. Therefore, the solution should be a dampened oscillation of the form

$$I(t) = I_0 e^{-t/\tau} \cos(\omega t + \varphi) \tag{1.3}$$

where the damping time $\tau$ and the oscillation frequency $\omega$ are given by

$$\tau = \frac{2L}{R}; \qquad \omega = \frac{1}{2L}\sqrt{4\frac{L}{C} - R^2} \tag{1.4}$$

The tricky point is the determination of initial conditions $I_0$ and $\varphi$ suitable for the present case. We know that $I_0=0$, but we do not know the initial value of the derivative $I'$. Putting $I(0)=0$, we find that $I_0\cos\varphi=0$; since $I_0$ cannot be zero itself, this implies in turn $\varphi=\pi/2$ or $\varphi=3\pi/2$; we choose $\varphi=3\pi/2$ in order to have a current which starts with positive values. So we obtain

$$I(t) = I_0 e^{-t/\tau} \sin(\omega t) \tag{1.5}$$

The other physical condition we can impose involves the initial charge on $C_2$, namely that $Q_2(0)=0$. To this end, we integrate (1.5) to obtain $Q(t)$ and find

$$Q_2(t) = -\frac{\tau e^{-t/\tau}[\omega\tau\cos(\omega t) + \sin(\omega t)]}{\omega^2\tau^2 + 1} + k \tag{1.6}$$

where $k$ is the integration constant. Imposing $Q_2(0)=0$ we find $k=-\tau^2\omega/(\tau^2\omega^2+1)$. A typical graph of the function (1.6) with this value of the constant $k$ is represented in Fig. 2. We see that the value of the charge must be normalized to its value for large times, which represents the equilibrium value. The potential $V_2$ on the capacitor $C_2$ is proportional to $Q_2$, so this graph also represents $V_2$ normalized to its final equilibrium value $\sim V_1$. The damping time in Fig. 2 is fixed and equal to 1 µs, which corresponds to take in eq. (1.4) $L\sim 1$ µH, as in the IU Marx generators, and $R\sim 1$ Ω. This value of $R$ is only a guess, so $\tau$ could actually be larger. The frequency obtained inserting $L\sim 1$ µH and $C_2\sim 1$ pF is of the order of 1 GHz, so the curve closest to reality is that with $\omega=100$ MHz, but we also showed the curves for lower frequency, to illustrate how the dampened oscillations are gradually converted into persistent oscillations as the frequency increases.

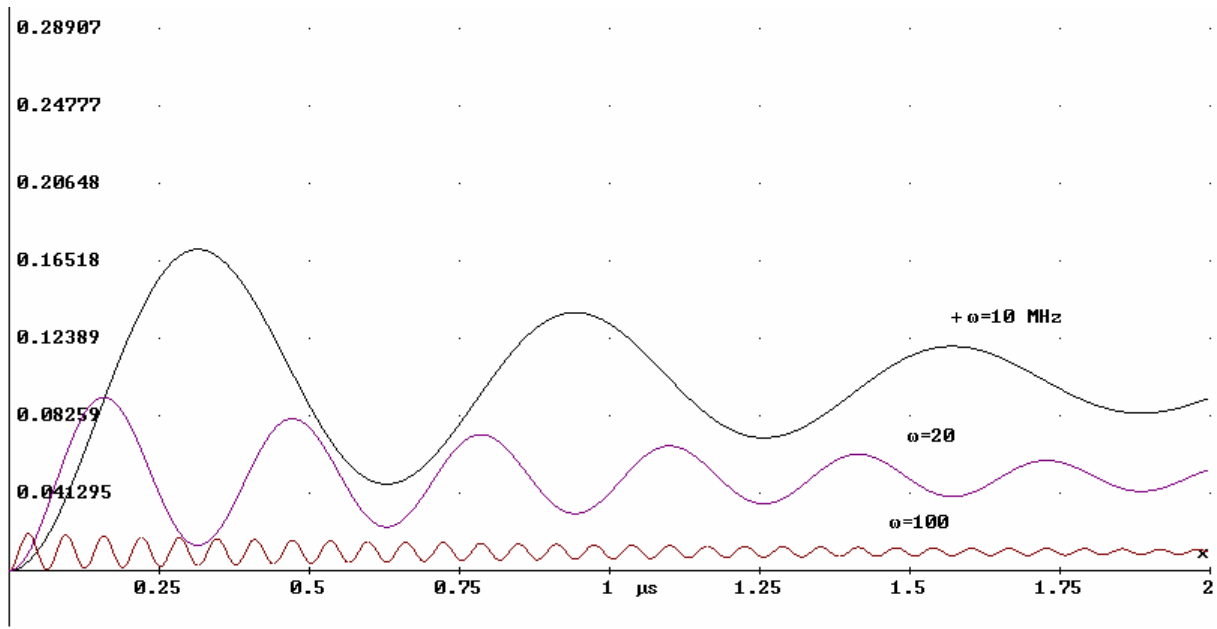

**Fig. 2**: Phase 1 (pre-breakdown). Typical voltage $V_2$ on the small gap capacitance $C_2$, for three different values of the circuit resonance frequency given by eq. (1.4) and fixed damping time $\tau=1$ μs.

<u>x-axis</u>: time (μs)

<u>y-axis</u>: voltage, to be referred to its equilibrium value for large time, visible on the right hand side.

*Discussion*

· Does the AC component of the voltage affect the formation of the discharge? The directional avalanches typical of the Townsend process can only form when the DC voltage is large enough.

· Can the AC component help expel electrons from the cathode by field effect?

· Note that the AC voltage component is only large at the beginning, then it decreases while the DC component increases.

· Much depends on the time $\tau=R/L$. The normal state resistance of the emitter in the *c* direction is of the order of 0.001-0.01 Ω and the high-frequency resistivity of the material in the superconducting state is of the same order (see next report!). This would imply a long $\tau$, possibly 100 μs or more. The S/N junctions can bear some additional resistance, however.

· In any case, it is possible that the AC voltage generates ionization near the cathode before the Townsend avalanches begin to form.

· On the other hand, if τ turned out to be short, one should also consider the finite firing time of the Marx generator. This probably cuts down the amplitude of the first oscillations.

The voltage rise on the gap is interrupted by the breakdown at some critical value $V_B$ depending on the Paschen law and the tuning of the gap. The tuning of the gap is important, to avoid a $V_B$ of just few kilovolts. At such a low voltage the electric field $E=V/d$ on the gap is still too small to expel many primary electrons from the cathode, so there would be only localized sparks, and not a full flat discharge. Fortunately, for our narrow gap this problem will be, at the same voltage, less acute than for the full version. Already at relatively small voltage, there should be abundant generation of primary electrons: say already at 50 kV, since the gap is 10 times smaller and the critical voltage in the full version is about 500 kV. It will be necessary to tune the Paschen $x$-product as to reach 50 kV, which looks feasible.

In Phase 2, the post-breakdown phase, the uncertainty on $R$ is relatively smaller. We know for sure that we are in the over-dampened case, and we can simply apply the standard formulas for the voltage transfer in an RLC circuit. That phase is also the crucial one for determining the "performance" of the device, ie the targets velocity and maximum energy.

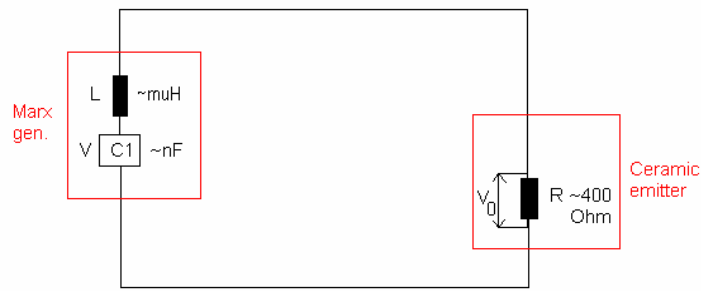

**Fig. 3** - Equivalent circuit for Phase 2 (post-breakdown)

Phase 2. At the breakdown, $C_2$ becomes a short-circuit, and the emitter is crossed by the current and exhibits a ohmic resistance. We have a series RLC circuit, where $C=C_1$. We can apply the standard equations for the rise of a pulse in this configuration ([1], Ch. 6). The initial applied voltage on $C_1$ is the nominal maximum voltage $V$ of the Marx, because the voltage drop occurred at $C_2$ in Phase 1 is very small, as noticed. The voltage $V_0$ on the load, however, can be noticeably smaller. Eq.s (6.16)-(6.17) of [1] give the time behaviour of $V_0$, in dependence on the three parameters $R$, $L$ and $C$. These formulas hold in the "over-dampened" case $\frac{R}{2L} \geq \frac{1}{\sqrt{LC}}$, ie $R \geq 2\sqrt{\frac{L}{C}}$ (for the IU generators $\sqrt{(L/C)}$ is of the order of tens of ohms, see table below). We have (compare Fig. 5)

$$V_0(t) = \frac{V}{c}\left(e^{-t/a} - e^{-t/b}\right) \tag{1.7}$$

with

$$a = \frac{2L}{R(1-c)}$$
$$b = \frac{2L}{R(1+c)} \tag{1.8}$$
$$c = \sqrt{1 - \frac{4L}{CR^2}}$$

The less well known parameter is the $R$ of the emitter. We gave an estimate in Podkletnov's case. If the rise time of the current pulse ($b$, in the following) is much shorter than the total pulse duration, we can assume that the total initial charge in the generator is equal to the peak value of the current multiplied by $a \sim RC$, or equivalently that $R=V/I$, where $V$ is the nominal peak value of the Marx voltage and $I$ is the observed peak value of the current. The current is not known exactly, we know it is of the order of 10 kA. Supposing for instance 20 kA for $V=2$ MV, we find $R=100$ Ω (this value will be confirmed in the following). Another bold extrapolation is that the resistivity of our samples will be the same, so that with surface four times smaller we will have $R=400$ Ω. We shall consider, to be safe, also the values 200 Ω and 600 Ω.

For the IU 400 and 800 kV generators we find the following values.

| $C, L$ | $R$ (Ω) | $c$ | $a$ (ns) | $b$ (ns) | $n=a/b$ | $V_{0,max}/V$ | $I_{max}$ (kA) |
|---|---|---|---|---|---|---|---|
| 15 nF, 4 µH (IU 1045) | 200 | 0.987 | 2980 | 20 | 149 | 0.97 | 3.88 |
| | 400 | 0.997 | 6000 | 10 | 600 | 0.99 | 1.98 |
| | 600 | 0.999 | 8990 | 6 | 1500 | 0.996 | 1.33 |

| 4 nF, 3 µH (IU 1080) | 200 | 0.962 | 787 | 15 | 52 | 0.94 | 1.88 |
| --- | --- | --- | --- | --- | --- | --- | --- |
| | 400 | 0.991 | 1590 | 7.5 | 211 | 0.98 | 0.98 |
| | 600 | 0.996 | 2390 | 5 | 478 | 0.99 | 0.66 |

**Table 1: pulse parameters for the IU generators**

Plotting the graph of $V_0(t)/V$ one sees that if $a>>b$, then the load voltage grows closer to the nominal potential $V$, but the effect is not dramatic.

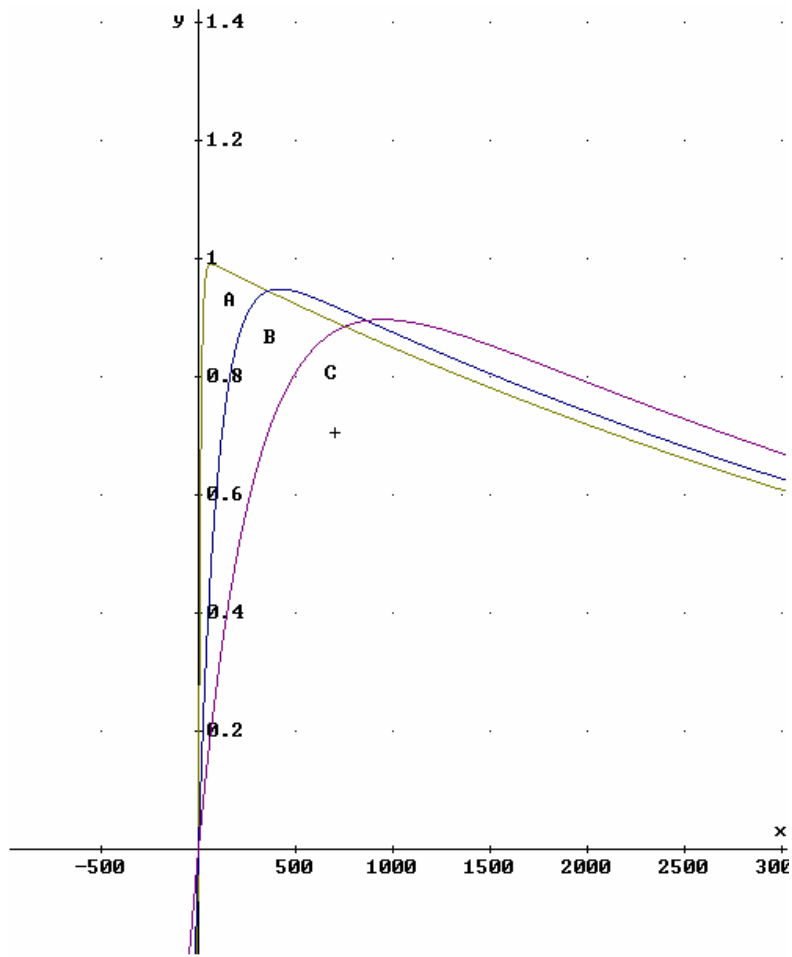

**Fig. 4** - Time dependence of the load voltage $V_0(t)$ with respect to the nominal voltage $V$, when the pulse duration is fixed to the value $a=6000$ ns, for the cases when the rise time $b$ is $b=10$ ns (A), $b=100$ ns (B), $b=300$ ns (C). <u>y-axis</u>: $V_0(t)/V$ ratio. <u>x-axis</u>: time $t$, in ns.

The equation of this curve as a function of $a$ and $b$ is $y=(e^{-x/b}/b - e^{-x/a}/a)\cdot((a/b+1)/(a/b-1))$. The factor $((a/b+1)/(a/b-1))$ is equal to $1/c$ (see below).

One can also compute the maximum of the curve, for instance as a function of the ratio $a/b=n$. The sum of the two exponentials in eq. (1.7) reaches a maximum equal to $n^{n/(1-n)}\cdot(n-1)$ and one can show that the $c$ factor is given by $(n-1)/(n+1)$; therefore, $V_{0,max}/V = n^{n/(1-n)}\cdot(n+1)$.

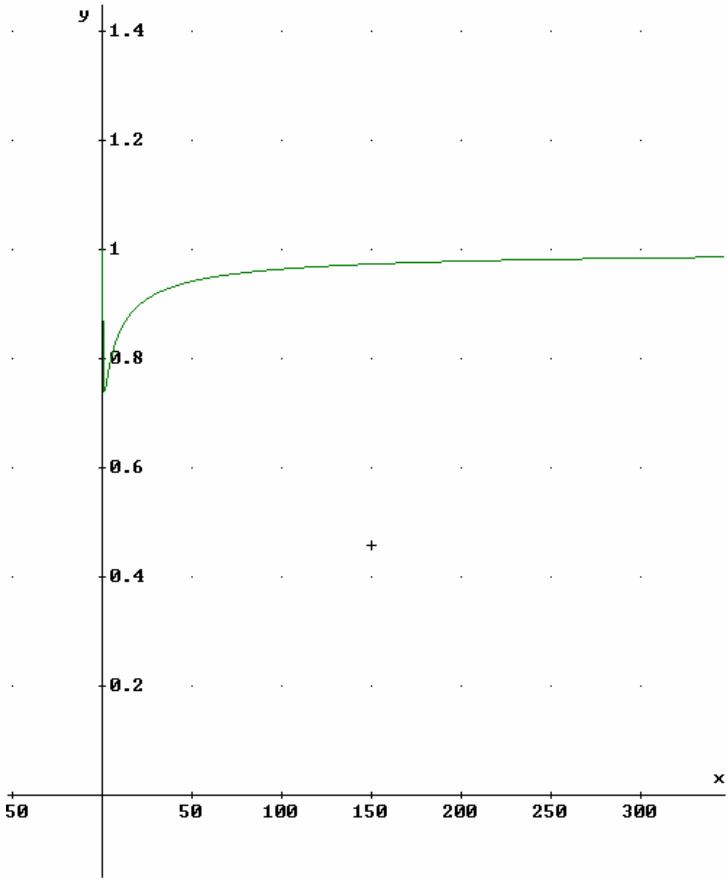

**Fig. 5** - Ratio $y=V_{0,max}/V$ between the maximum load voltage and the nominal generator voltage as a function of the ratio $n=a/b$ (<u>x-axis</u>) between the pulse duration and the rise time.

## 2. Computation of the target velocity $v_t$ and the maximum available energy $U_{max}$

We know (Report 1, Par. B.3) that the maximum energy available in the beam $U_{max}$ is given by the equation $U_{max} = IV'_{tot}\Delta t_2$, where $V'_{tot}$ is the total effective voltage and $\Delta t_2$ is the pulse duration. $V'_{tot}$ is found multiplying the number of emitting planes by the effective voltage per plane, given by $V' = hf/2e$. The number of emitting planes is given by the thickness $\delta$ of the emitter divided by the "acceleration space" $s_{acc} = ns_c$; $s_c$ is the interplane spacing, in YBCO $s_c \sim 1.17$ nm.

The frequency $f$ is itself a function of the emitter surface $S$, the current and the pairs density $\rho$, because it is given by the velocity $v$ of the pairs divided by the acceleration space:

$$f = \frac{v}{s_{acc}} = \frac{j/(2e\rho)}{s_{acc}} = \frac{I}{2e\rho S s_{acc}} \qquad (2.1)$$

Inserting this formula for $f$ in the expressions above, and writing $\Delta t_2 = a$ (the pulse duration is well approximated by the time $a$ evaluated in the previous section) we find

$$U_{max} = \frac{\delta}{n^2 s_c^2} \frac{h}{(2e)^2} \frac{I^2}{\rho S} a \qquad (2.2)$$

The energy can be equivalently computed as follows: consider that any pair emits a graviton of energy $E = hf$ when it crosses an emitting plane. The number of gravitons crossing any plane each second is $I/2e$; multiplying by the number of planes $\delta/s_{acc}$ we find

$$U_{max} = \frac{\delta}{s_{acc}} hf \frac{I}{2e} a = \frac{\delta}{s_{acc}} h \frac{I}{2e\rho S s_{acc}} \frac{I}{2e} a \qquad (2.3)$$

which is the same as (2.2).

The target velocity is half the pairs velocity multiplied by $k/n$:

$$v_t = \frac{k}{n} \frac{I}{\rho e S} \qquad (2.4)$$

The maximum mass for which the process is elastic, ie the beam energy is entirely absorbed, is given by

$$m_{max} = \frac{2U_{max}}{v_t^2} \qquad (2.5)$$

We shall show later that for larger values of the target mass, the process becomes anelastic and some of the energy is dissipated.

We want now to give a numerical estimate of the energy and the target velocity for the IU1045 and IU1080 Marx generators in the 50% scaled version. To this end we insert our previous estimate for the pairs density of approx. $\rho=1.8\cdot10^{25}$ m$^{-3}$; the electrodes area $S$ is 25 cm$^2$ and its thickness $\delta=5$ mm. We also insert the value of the fundamental constants $e$ and $h$. Finally we find

$$U_{max} = \frac{1}{n^2} 2.62 \cdot 10^{-4} I^2_{max} a \qquad (2.6)$$

$$v_t = \frac{k}{n} 0.6935 \cdot 10^{-4} I_{max} \qquad (2.7)$$

(Scaled version)

The results are summarised in Table 2 below. For the IU1045 and IU1080 generators we know exactly the inductance $L$, and so given the resistance $R$ of the emitter we are able to compute exactly the quantities necessary to estimate $v_t$ and $U_{max}$, namely $a$ and $I_{max}$. Since $R$ is only a guess based on the $I/R$ value reported by Podkletnov (see previous section), we have considered three possible values for it. For the choice of the integers $n$ and $k$ we followed similar criteria as in Podkletnov's case (see below), namely $n=k=3$ in the average.

In eq.s (2.6) and (2.7) we took into account the fact that the current varies during the pulse, decreasing exponentially from its maximum value $I_{max}$ to zero. For both formulas one finds that an adequate average value of the current is simply equal to $I_{max}/2$.

(*Proof*: For the exponential function $I(t)=I_{max}e^{-t/a}$, where $I_{max}=V_{0max}/R$, we must integrate on time, cutting, say, at $t=2a$, because at that time the potential has decreased to about 13% of its maximum value, ie below the 117 kV which allow emission with $k=2$ (this holds for the 800 kV generator; for the 400 kV generator, we can cut at $t=a$, because this corresponds to a 36% decrease, which is also below the 117 kV threshold). The integral gives in any case a factor ½. Finally, if we integrate $e^{-2t/a}$ from 0 to $2a$ we obtain approx. $a/2$. On the other hand, if we integrate to $a$, we obtain $a/2$ again, because in the square bracket with the integration limits we have $[e^{-2}-1]$ and an $a$ factor in front. For the time average of $I(t)$, we follow the same criteria, ie integrate to $2a$ for the 800 kV generator and to $a$ for the 400 kV generator. Integrating $e^{-t/a}$ (the current is not squared), in the first case to $2a$ and dividing by $2a$ to compute the average, we obtain $I_0/2$. In the second case, we obtain $-I_0[e^{-1}-1] \sim 63\%$ of $I_0 \sim I_0/2$.)

| C, L | R | a | $I_{max}$ (kA) | $v_t$ (m/s) | $U_{max}$ (mJ) | $m_{max}$ (g) |
|---|---|---|---|---|---|---|
| 15 nF, 4 µH (IU1045) | 200 Ω | 2980 ns | 3.88 | 0.26 | 1.3 | 37 |
| | 400 Ω | 6000 ns | 1.98 | 0.14 | 0.68 | 70 |
| | 600 Ω | 8990 ns | 1.33 | 0.09 | 0.46 | 106 |
| 4 nF, 3 µH (IU1080) | 200 Ω | 787 ns | 1.88 | 0.13 | 0.081 | 10 |
| | 400 Ω | 1590 ns | 0.98 | 0.068 | 0.044 | 19 |
| | 600 Ω | 2392 ns | 0.66 | 0.046 | 0.030 | 28 |
| 1.25 nF, ? (Podkletnov) | 100 Ω | 134 ns | 15 | 0.39 | 0.39 | 5 |

**Table 2: performance of the IU generators compared to Podkletnov's**

The results of the last row of the table, for Podkletnov's device, are computed inserting in the general formula appropriate values for the emitter area and thickness ($\delta$=4 mm, $S$=100 cm$^2$), while the other quantities are estimated as follows. Since the maximum current is not exactly known, but we only know it is of the order of 10 kA, we choose a value that is in agreement with the observed target velocity $v_t$=0.39 m/s, which in this case is theoretically predicted to be

$$v_t = \frac{k}{n} 1.73 \cdot 10^{-5} I_{max} \qquad (2.8)$$

(Full version)

Remember that the voltage employed by Podkletnov is very high, therefore we believe that all the microscopic transitions can be excited. The two most probable values for the acceleration space $n$ are 2 and 4; let us take 3 in the average. For the wavelength $k$, we can have at least 3 and 6 for coherent superposition; we take an average of 4.5. The average $k/n$ ratio therefore is 1.5 and in order to obtain the right velocity, the maximum current must be equal to ~15000 A. Next we estimate the effective emitter resistance $R=V_{max}/I_{max}$. The ratio between the effective voltage $V_{max}$ on the load and the nominal voltage $V$ of the Marx generator is not known, because we do not know the Marx inductance and the $a/b$ ratio. We know, however, that the condition $b \ll a$ is not satisfied, because since $R$ is of the order of 100 Ω, $a \sim RC \sim$ 100 ns and $b$ can not be much smaller than this. We take then the less favourable hypothesis, ie $V_{max}/V \sim 0.8$ (Fig. 5) or $V_{max}$~1600 kV. In this way we obtain $R=10^7$ Ω. This is remarkably close to our initial guess, so the reasoning is confirmed and this value of $R$ can be used to improve the estimate for $a$.

Then we use the values above for $I_{max}$ and $a$ to compute the energy. The appropriate formula for 4 mm thickness and 100 cm$^2$ surface is

$$U_{max} = \frac{1}{n^2} 0.524 \cdot 10^{-4} I^2_{max} a \qquad (2.9)$$

(Full version)

This value is to be compared with the observed energy $U=1/2mv_t^2$~ 1.41 mJ, with $m$=18.5 g and $v_t$=0.39 m/s. With $n$=3 (average of 2 and 4) we find from eq. (2.9) $U_{max}$=1.74·10$^{-4}$. With $n$=2 (plausible for high voltages; correspondingly, $k$=3 only, in the velocity formula, which then stays unchanged), we find $U_{max}$=3.92·10$^{-4}$, which is only about three times smaller than the observed value.

We know that the microscopic model of the emission needs to be improved by taking into account the stimulated emission. This is necessary to explain the beam collimation and also the relationship between Podkletnov's results with 4 mm and 8 mm emitters (see the next report). Even so, however, the energy momentum ratio for any emitted graviton is unchanged, and it is impossible to match exactly $v_t$ and $U_{max}$ at the same time, unless some other modification is introduced, for instance:

· Introduce a different value for the pairs density ρ
· Suppose $n$ smaller or $k$ larger
· Suppose that the frequency is slightly larger for some reason.

These are only minor adjustments, however, and at the present stage we can be satisfied with the agreement obtained.

[1] M.S Naidu and V. Kamaraju, "High voltage engineering", MacGraw-Hill, New York, 1996.

[2] The IU1045 Marx Generator specifications are as follows.

1) Erected Capacitance : 4nf
2) Total Energy : 400Joules
3) Max. O/p Voltage Possible : 400V
4) Kv per stages : 40Kv.
5) Capacitance per stage : 40nf
6) No. of stages : 10Nos.
7) Typical Rise time : Approximately 8.4 to 9.6nano seconds
8) Marx Inductance : 3micro Henry
9) Inductance of Capacitor per stage : 300nano Henry
10) Type of Spark Gap : Air Gaps ( Pre-ionized Network)
11) Typical Pulse duration : Approximately 160nano seconds in short circuit mode.
12) Overall Height : 1500mm
13) Price is $6600.00 plus freight

The IU1080 Marx Specifications are as follows.

1) Erected Capacitance : 15nf
2) Total Energy : 3200Joules
3) Max. O/p Voltage possible : 700Kv to 800Kv
4) Kv Per stage : 70Kv (80Kv max.)
5) Capacitance per stage : 150nf
6) No. of stages : 10Nos.
7) Typical Rise Time : Approximately 18nano seconds to 23nano seconds
8) Marx Inductance : 4 micro henry
9) Inductance of Capacitor per stage : 400nano Henry
10) Type of Spark Gap : Air Gaps ( precision material)
11) Typical pulse duration : 220nano seconds to 400nano
12) Overall Height : 2200mm
13) Price is $17,670.00 plus freight

# Chapter 4

## New detailed analysis of the discharge (2005)

This is our second update in 2005. In Section 1 we refine the numerical estimates given in our Report 1-2004, using more precise data for the elastic and ionization cross section. We also extend the results previously given for the original IGG to an "IGG with actual over-pressure" and to two slightly different scaled versions (Table 1).

In Sect. 2 we revise our model for the discharge mechanism. The "Townsend" mechanism suggested in our Report 1-2004 suffers from some problems due to the large energy and large free mean path of the electrons - so large that X-ray generation upon incidence on the electrodes cannot be excluded (NB for safety precautions!). In our Report 1-2004 we tried to overcome this difficulty with the hypothesis of a fine adjustment of the *pd* product to the left upwards branch of the Paschen-law diagram. We have recently established, however (Report 1-2005) that the voltage on the gas gap may undergo many oscillations at GHz frequency before the discharge. It is therefore more likely (but not sure, until the emitter resistance pre- and during breakdown is measured experimentally) that the true discharge mechanism involves a preceding formation of plasma due to this high-frequency field.

This hypothesis also helps solving problems with the limits imposed on the current and current density by the Child and Alfwen laws in the absence of stabilising plasma (Section 3).

In conclusion, we believe that all possible scenarios involved in the discharge process have been theoretically explored now, and further input can only come from the experimental tests.

The subject of the next report will be a theoretical discussion of the emitter resistance.

## 1. New estimate of electron drift velocity, collision frequency and mean free path.

Raizer [1] (page 9) gives the following general formula for the drift velocity of electrons in a gas under the action of an electric field $E$

(1) $$v_d = \frac{eE}{mf_m} \cong 2 \cdot 10^{11} \frac{E}{f_m}$$

where $f_m$ is the "collision frequency for momentum transfer", which is of the same order of the elastic collision frequency $f_c$. For the latter we can write the equation

(2) $$f_c = \frac{1}{\tau_c} = Nv\sigma_c$$

The velocity $v$ to be inserted into this equation is usually the thermal electron velocity, which for moderate electric fields is much larger than the drift velocity. In our case, however, the field is so strong that the drift velocity exceeds the thermal velocity. Let us give a first check of this, to be confirmed later in the detailed calculation. According to Raizer, typical values of the collision frequency $f_c$ at 1 tor pressure are of the order of $10^{10}$ s$^{-1}$. At 1 Pa pressure, the frequency is about 100 times smaller, so we have $f_c \approx 10^8$ s$^{-1}$. Substituting into eq. (1) with $E \approx 10^6$ V/m, we find $v_d \approx 10^9$ m/s. Of course, this implies that the velocity must be computed using relativistic dynamics, and we will find in that way a value for $v_d \approx 10^8$ m/s $< c$. On the other hand, at room temperature the electron thermal velocity is $\approx 10^5$ m/s.

So in order to treat properly our strong-field case, we want to solve simultaneously eq. (1) and eq. (2) treating $v$ and $f$ both as unknown variables, where $v$ is the drift velocity and $f=f_c=f_m$. First we determine the other two parameters, namely the density $N$ of gas molecules per cubic meter and the scattering cross section for elastic collisions $\sigma_c$. At $T=70$ K e $P=1$ Pa, $N \approx 10^{20}$ molec./m$^3$. Data for $\sigma_c$ are given by Raizer for energies up to 30-100 eV:

(a) H$_2$ and He: $\sigma_c$ decreases from $4 \cdot 10^{-16}$ cm$^2$ to $2 \cdot 10^{-16}$ cm$^2$ between 0 and 36 eV.
(b) N$_2$ and CO have a peak value of $\sigma = 28 \cdot 10^{-16}$ cm$^2$ at 4 eV, then $\sigma$ drops to about its 25% at 100 eV.
(c) For O$_2$, $\sigma_c$ keeps at the minimum values observed for N$_2$ and CO.

In conclusion, let us assume that $\sigma_c$ is of the order of $10^{-15}$ cm², ie $10^{-19}$ m², since we expect that the electron energy will be quite high. Combining eq.s (1) and (2) we find $10^{11}E/f=f/(10P)$, with $P$ expressed in Pa. From this we obtain the following expressions for $f$, $v_d$ and the mean free path $l$:

(3)
$$\begin{cases} f = \sqrt{10^{12} EP} \\ v_d = \sqrt{10^{10} \dfrac{E}{P}} \\ l = \dfrac{1}{10P} \end{cases}$$

with $P$ in Pa, $E$ in V/m, $v_d$ in m/s, $l$ in m. These are valid at T=70 K.

With the help of these equations, we can compile a table with the values of the original Podkletnov's device, those of the same device in the hypothesis pressure was under-estimated and was in fact 10 Pa, those of the proposed 50% scaled device and those of a scaled version with higher pressure. The field is the alleged threshold value (500 kV in the original IGG, on a gap between 15 and 40 cm, so we assumed for simplicity $10^6$ V/m; in the scaled version the gap is 1.5 to 4 cm).

| Device | Pressure $P$ (Pa) | Electric field $E$ (V/m) | Estim. electron collision frequency $f$ (s⁻¹) | Estim. electrons drift velocity $v_d$ (m/s) | Estim. electron free mean path $l$ (m) | Estim. electron energy (eV) |
|---|---|---|---|---|---|---|
| Original IGG | 1 | $10^6$ | $10^9$ | $10^8$ | 0.1 | $10^5$ |
| Orig. IGG w. higher real press. | 10 | $10^6$ | $3 \cdot 10^9$ | $3 \cdot 10^7$ | 0.01 | $10^4$ |
| Scaled version | 10 | $10^7$ | $10^{10}$ | $10^8$ | 0.01 | $10^5$ |
| Scaled version w. over-pressure | 100 | $10^7$ | $3 \cdot 10^{10}$ | $3 \cdot 10^7$ | 0.001 | $10^4$ |

**Table 1: gas parameters for the original IGG and scaled versions**

Note that Raizer, on page 11, gives a table of values for $l$, $f$ and $\mu$ (electron mobility, the ratio $v_d/E$) for several gases, but in a range of $E/P$ values up to 50 V/(cm tor), ie approx. 40 V/(m Pa)! Our $E/P$ values are of the order of $10^6$ V/(m Pa).

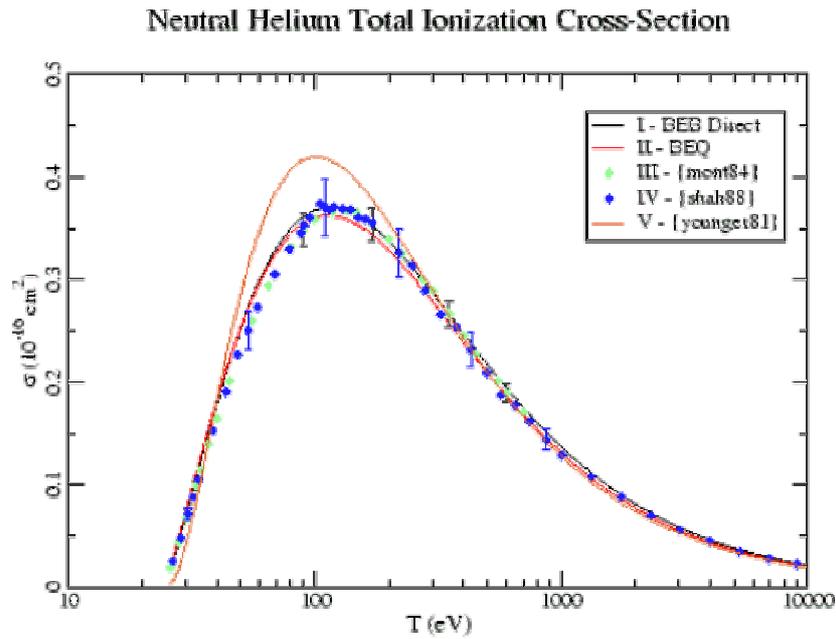

Fig. 2 – Ionization cross section for helium (from ref. [2]).

If we consider the ionization cross section by electron collision of the gas molecules, instead of the elastic cross section, the conclusions are the same: the mean free path is long and the average electron energy very large. The NIST database [2] gives data for ionization cross sections of various gas molecules. For instance, for helium the dependence of the cross section on the energy of the incident electron is plotted in Fig. 2. We see that the maximum value, $\sigma=0.35 \cdot 10^{-16}$ cm$^2$ occurs at an energy of 100 eV. Above 1000 eV, $\sigma$ is smaller than $10^{-17}$ cm$^2$, ie $10^{-21}$ m$^2$. This is two magnitude orders smaller than the elastic cross section and therefore has no influence on the drift velocity and on the mean free path.

A final check is needed of the magnitude of the Coulomb electron/ion scattering cross section $\sigma_{Coul}$, which may be relevant in ionized gases ([1], p. 13, 14). It turns out to be irrelevant in our case, however, mainly because it is inversely proportional to the squared electron energy $E_e$:

(3')
$$\sigma_{Coul} \approx \frac{10^{-14} \ln \Lambda}{E_e^2 [eV]} \approx 10^{-20} - 10^{-22} \text{ m}^{-2}$$

(ln$\Lambda$ depends on the electron energy and gas density, but the dependence is weak; typically, $\Lambda$ is of order 10).

So the values of Table 1 can be confirmed. Note that at such electron energies X-rays production is entirely possible if the electrons hit the electrodes. Actually, at no time are electrons pushed with full field strength to the electrodes, because before the discharge their motion is oscillatory (see below) and during the discharge the voltage on the gap drops quickly. Safety measures are certainly in order, however.

## 2. Difficulties for the Townsend mechanism at low *pd* values. Alternative discharge mechanism.

We see from Table 1 that the mean free electron path is quite large, compared to the gap. This means that the electrons make few collisions or no collision at all, when they go through the gap. The pressure is so low, that there are few molecules to collide with and possibly ionize, and the electric field is so strong that the ionization cross section is definitely smaller than its maximum value. In such conditions, the cascade multiplication typical of the Townsend discharges can hardly start, it seems. We are in the left upwards branch of the Paschen curve, but at very high voltage. As we stressed in Report 1-2004, assuming the left branch continues all the way up, ie Paschen law keeps valid at very high voltage, a very fine tuning of the *pd* product would be needed, in order to reach that region. Now, however, such assumption appears to be dubious, for the following reason. When the *pd* product is below its lower limit "$x_\infty$", the gas in the gap is too rarefied or the gap too narrow to sustain the discharge. Slightly above $x_\infty$, the discharge can develop if ionization is efficient; as long as the electron energy is less than 100 eV, increasing the voltage also increases the ionization cross section, so the Paschen curve goes up. But over approximately 100 eV, the ionization cross section begins to decrease, and the conditions favourable to the discharge stop there.

The idea that the voltage on the gas gap could undergo large GHz oscillations after the Marx pulse, before approaching its final value, is new. Such oscillations are only possible if the emitter resistance before breakdown is much smaller than it is during the breakdown (1 Ω or less, compared to 100 Ω). Previously I did not consider this strange possibility. So with $R$=100 Ω, the oscillations are over-dampened and absent, and the only conceivable discharge mechanism is the Townsend cascade ionization with secondary emission. How could the emitter resistance be so small before the breakdown? The question is actually, how can it be so large during the breakdown. I will address this in the next report, but in short I think it is a consequence of the emission itself.

But if there is a large high-frequency voltage on the gap for 1 micro-second or longer, the picture changes much (see below for details). The gas can become ionized and plasma-like before the discharge starts. The electrons are then swept to the anode as the DC voltage component increases. Secondary emission at the cathode could even be unnecessary. Such a mechanism would match two observed features:

(1) The longer optical duration of the discharge ($10^{-4}$ - $10^{-5}$ seconds, according to Podkletnov).
(2) The initial glow near the cathode, presumably due to positive ions accumulating there and re-combining with electrons.

The effect of an AC field in a gas is treated by L.-Jones [3] in Ch. 9 and by Raizer [1] in Ch. 7. The main concepts are simple. Some initial free electrons are supposed to be present, which oscillate back and forth under the action of the field and cause ionization. In our case, we suppose that the initiating electrons are emitted from the cathode by field effect. At very low pressure, such that collisions of the electrons during their oscillation can be disregarded, the oscillation amplitude is

$$(4) \qquad r = \frac{eE_0}{m\omega^2}$$

If the pressure, however, is such that the mean free path $l$ between collisions is shorter than this amplitude, then the effective oscillation amplitude is limited by $l$, because after each collision the oscillation starts anew. The general formula is in this case

$$(5) \qquad r = \frac{eE_0}{m\omega\sqrt{\omega^2 + f_m^2}}$$

where $f_m$ is the effective collision frequency. In our case, the amplitude computed for a field $E=10^6$ V/m is of the order of 0.1 m, ie of the same order of the mean free path. Therefore, the free electrons will make un-correlated oscillations with approximately one collision per oscillation. Almost every collision causes a ionization, especially when the electric field begins to decrease, after the first oscillations (compare Report 1-2005, Fig. 2) and the electron energy goes down near the maximum of the ionization cross section (see Fig. 2 above). For $\tau$ of the order of 1 µs (but it can even be larger!) the oscillations are ca. 1000, and being the multiplication process exponential, we obtain a large ionization and plasma formation, much more than it would be possible in a single sweep "a la Townsend" - even considering the secondary cathode emission.

Let us therefore suppose that the 1-GHz oscillations ionize a large part of the gas, and then (when the DC component grows) the positive ions are attracted to the emitter and the electrons run to the cathode, creating a current without the need of secondary emission at the cathode. Is the charge produced in the chamber by ionization alone sufficient to account for all the current? Let us assume that the effective volume of the chamber is 1 dm$^3$; at 70 K and 1 Pa this contains $10^{17}$ molecules. If they all were ionized, the charge would be 0.01 C. But the charge carried by the current pulse is 10 A in 100 ns, which means $10^{-3}$ C, so the ionization rate needed $\eta$ is only 10%.

Now let us check (like in Report 1-2004, but with the new data) if the ionization rate $\eta$ and the electron drift velocity $v$ are compatible with the required current density $j$ ($j=10^6$ A/m$^2$; $\rho$ is $10^{17}$ dm$^{-3}$ = $10^{20}$ m$^{-3}$ as above):

(6)
$$j = e\eta\rho v \approx 10^6 \Rightarrow$$
$$\Rightarrow \eta\rho v \approx 10^{25} \Rightarrow \eta v \approx 10^5$$

The minimum value of $\eta$ found above ($\eta=0.1$) gives $v=10^6$ m/s. With the maximum possible $\eta$ ($\eta=1$, in which case only 10% of the charge is swept to the electrodes), the velocity is $10^5$ m/s. A drift velocity of $10^6$ m/s is easily obtained with our field strength and free path (compare Table 1 above). Note that the voltage on the discharge chamber decreases quickly during the breakdown, but its average value is still of the order of $10^5$-$10^6$ V.

## 3. Possible limits on the current and current density.

In a conversation I recently had with an expert of plasma physics, he suggested to me to check if our 10 kA discharge current satisfies two conditions which usually set a limit on the maximum current and current density in vacuum tubes. These are the so-called Child and Alfven-Lawson limits. It is interesting to consider them for completeness, even though in our case the presence of a plasma makes those limits much less stringent. My colleague did not consider the presence of plasma when we discussed the situation, because he correctly remarked that "the applied voltage is so high and the pressure so low that plasma can not be formed; the electrons move too fast and there is too little gas to heat up". I have now come to agree with him about that, but I also have found that plasma is nevertheless formed by the GHz oscillations in the voltage.

The *Alfven-Lawson limit* on the total current is due to the magnetic self interaction of the current, and is independent on the current beam diameter. The maximum total current in vacuum is $I_{AL}=31 \cdot 10^6 \beta\gamma A/q$ A, where $A$ is the atomic mass number of the particles in the beam and $q$ is their charge in elementary units [4]. For electrons we take $A=1/1800$, and for energy $10^5$ eV we find $\beta=0.56$, $\gamma=1.2$, and so $I_{AL}=11600$ A. This is consistent with the maximum observed current.

*Child law* states that the maximum current density in vacuum between parallel plates with voltage drop $V$ (in MeV) and separation $d$ (in cm) is $j=2.34 \cdot 10^3\, V^{3/2}\, d^{-2}$ A/cm$^2$ [5]. This limit does not depend on the material of the electrodes and is due to space-charge effects: if $j$ is too large, there is an accumulation of negative charge in front of the anode. In our case (full version) we obtain a limit of $j=4.1$ A/cm$^2$ with $V=2$ MV and $d=40$ cm, or $j=29$ A/cm$^2$ with $d=15$ cm. The maximum current density measured for the full version is of the order of 100 A/cm$^2$. There appears to be a discrepancy; but if the beam of charged particles is not in high vacuum, the presence of neutralising ions makes both limits less stringent. The Alfven current and the Child maximum current density are in that case multiplied by a factor $(1-n_m)^{-1}$, where $n_m$ is the neutralization ratio, ie the ratio between the density of positive and negative ions. If the ions are generated in the gas itself, the $n_m$ ratio can differ from 1 only by a small percentage, due to space charge effects. In the limit when $n_m$ approaches 1 up to 1%, 0.1% and so on, the factor above grows to 100, 1000 etc., and so do the limits on the Alfven current and the Child maximum current density.

Another remark of my colleague: even with currents as low as 1 A, small local variations in the chemical composition near a metallic cathode lead immediately to local over-heating and thermo-ionic emission ("hot spots", usually observed at room pressure). My reply: a superconducting emitter is likely to behave much differently under this respect, namely if local over-heating occurs, the local conductivity drops sharply and the fluctuation is not self-amplified but rather self-suppressing.

This may be one of the reasons why a superconducting emitter/interface to the gas is needed. Other possible reasons are the need for highly uniform electric field and for a field emission probability larger than in a metal. All these features could be essential in the generation of flat discharges. There is much to be learnt and discovered on this point. One also should remember this before considering the possibility of replacing the discharge chamber with a more standard vacuum or gas switch or even a solid-state switch.

Finally, my colleague pointed out further possible effects of self-interaction in the electron stream carrying the discharge: electrostatic repulsion, pinch of the current, magnetic self-pressure. All these possible effects should be examined, although, again, they are more typical of electron beams in the vacuum. Since they all involve a lateral drift of the electrons, it is likely that the strong magnetic field applied by Podkletnov through his large solenoid is able to suppress them. In our scaled version without magnet, the hopes to avoid such effects are pinned to the narrower gap and larger pressure. An important remark is in order, however. According to Podkletnov, the main purpose of the big solenoid was to avoid lateral movements of the discharge as a whole, which could lead it to the walls. With the large voltage employed, the most likely cause for this are just external unbalanced electrostatic forces, instead than the mentioned self-interaction in the electron stream.

**Notes**

1. <u>Validity of Paschen law for short pulses. Applicability of the "streamer" theory.</u>

I checked the fundamental text by L.-Jones [3] about the validity of the Paschen law for short pulses. L.-Jones cites several authors who solved the basic equations for ionization growth in the case of short pulses, taking into account the possible formation of space charge. The comparison with the experimental data and the final discussion are in Sect. 8.4, in particular the data for low pressure are on page 138-140. The conclusion is that the data on ionization growth at the lower pressures ($pd$ < 150 tor cm, ie we are well in this range, with $pd$ of order 0.1) are consistent with the view that the breakdown criterion for static uniform fields is still valid. The criterion is expressed in the Townsend equation $(\omega/\alpha)[\exp(\alpha d)-1]=1$, of which the Paschen law is a consequence.

According to Naidu (see ref. in the previous reports), "it is still controversial which discharge mechanism operates in uniform field conditions over a given range of $pd$ value. It is generally assumed that for $pd$ values below 1000 tor cm and gas pressures varying from 0.01 to 300 tor (ca. 1.3 to 40000 Pa), the Townsend mechanism operates, while at higher pressures and $pd$ values the Streamer mechanism plays the dominant role in explaining the breakdown phenomena." It seems therefore that we are very far from the streamer range.

2. <u>Range of the pressure sensor.</u>

At our meeting in Munich we discussed the best choice for the range of the pressure sensor. There are two possibilities: up to 13 Pa or to 100 Pa. Our proposal for the scaled version requires a pressure of 10 Pa. If we were sure of this value within approx. a 10%, we would suggest the 13 Pa sensor, which is obviously more precise. (We mean here the precision of pressure measurement; pressure regulation is another issue, which has been discussed separately.) When we still thought that a fine tuning of the Paschen $pd$ product was necessary, we estimated the precision needed for $p$ and $d$, with reference to both electrodes distance and parallelism. This precision is quite demanding. In any case, for fine $pd$ tuning to the "$x_\infty$" value, the exact value of $x_\infty$ depends on the nature of the gas and of the electrodes, with variations up to 100%.

It is clear from all the above that we can not trust our 10 Pa prediction so much to limit the sensor range to 13 Pa. Moreover, in case we later should need a larger $pd$ value, we will not be allowed to increase the gap thickness $d$, because $d$ must keep small with respect to the emitter diameter; we only can increase the pressure. So we opted for the 100 Pa sensor. With hindsight, this choice seems even more correct, because if we can abandon the Paschen-law fine tuning of *pd*, then we have the option of increasing *p* further, as previously discussed (Report 2-2004; in that case, the scope was to slow down the discharge, which may not be actual any more; but there could be other advantages. Note that if the inverse proportionality between *p* and *d* is lost, the *E*/*p* ratio and thus the electron energy varies; compare our Table 1 above with over-pressure.)

3. Initiating electrons

We have mainly assumed until now that the initiating electrons are generated at the cathode by field effect. If this is the case, we can be assured that in the scaled version the field is about ten times stronger than in the original versions, so there should not be any shortage of initiating electrons. According to Raizer [1], p. 141, natural radioactivity causes the presence in air at any time of about 1000 ions per cubic centimetre. This figure must be scaled by pressure, however, so at 10 Pa there will be only 1 ion in 10 cubic centimetres. This might be enough to initiate the ionization avalanches, or not, depending on the ionization processes involved. Up to now, for instance, we have not considered photo-ionization, which is very efficient and can spread ionization quickly to large volumes. Since the electron energy is quite large, photo-ionization could actually play a role in our case.

(see Ch. 6 for the relativistic Child law; at 1 MeV energy this is almost equivalent to the non-relativistic approximation given in the text).

# Chapter 5

# Origin of the high-frequency resistance of the emitter. Improvements of the microscopic model (2005)

**Introduction**

According to Podkletnov, the *V*/*I* ratio during the gas discharge is approximately 100 V/A, and the discharge is apparently over-dampened. We therefore need to explain how the emitter can bear a resistance of ca. 100 Ω. We analyse several possibilities: complex impedance of superconductors for high-frequency current, radiation reaction, resistance due to the intrinsic Josephson junctions for current flow in the *c*-axis of YBCO. None of these sources of resistance can explain a 100 Ω resistance.

We then compute the possible dissipative effect of the anomalous emission. In analogy with the Drude model for ohmic conductors, we introduce a suitable scattering frequency and an "attempt frequency" for the tunnelling between superconducting crystal planes. This attempt frequency is necessary to reconcile the drift velocity of the pairs with the applied field, and turns out to be very large. We can not find any natural physical interpretation of this frequency, so we conclude that such a dissipative model is un-realistic.

Finally, we consider the possibility that the discharge is not really over-dampened, but the current is limited by the Marx inductance. This hypothesis, if confirmed, would change much the perspective of the microscopic model, because it implies that the current oscillates in both directions and that the voltage on the emitter is small. In Section 2 we also discuss the case of inelastic absorption of the anomalous radiation in the targets.

**1. High-frequency resistance of the emitter**

There are several possible causes for the high-frequency ohmic resistance of the emitter. Each one of these causes gives its own contribution to the total resistance, and it is not straightforward to tell which is the dominant contribution.

From the general knowledge about superconductors it is known that at high frequency any superconductor has an impedance with a real and imaginary part, ie a resistive and an inductive part. Both depend on frequency and are illustrated by diagrams available in the literature; I will describe this in more detail below, anyway the magnitude order of such impedance is the same as the normal-state resistivity, which for our material (see for instance the CRC Handbook [1], S. 12-91) is ~ 5 mΩ per centimetre in the $c$ direction, too small to account for the observed resistance.

**Table 1 - Normal-state resistivity of YBCO [1]**

|  | $\rho_{ab}$ (μΩcm) | | $\rho_c$ (μΩcm) |
|---|---|---|---|
|  | at 300 K | at 100 K | at 300 K |
| Single crystal | 110 | 35 | 5000 |
| Film | 200-300 | 60-100 |  |

Converting into the standard SI units $\Omega$m, we have at 100 K that $\rho_{ab} \simeq 35 \cdot 10^{-8}$ $\Omega$m, $\rho_c \simeq 5000 \cdot 10^{-8}$ $\Omega$m. For comparison, the typical resistivity of metals is about $10^{-6}$ $\Omega$cm.

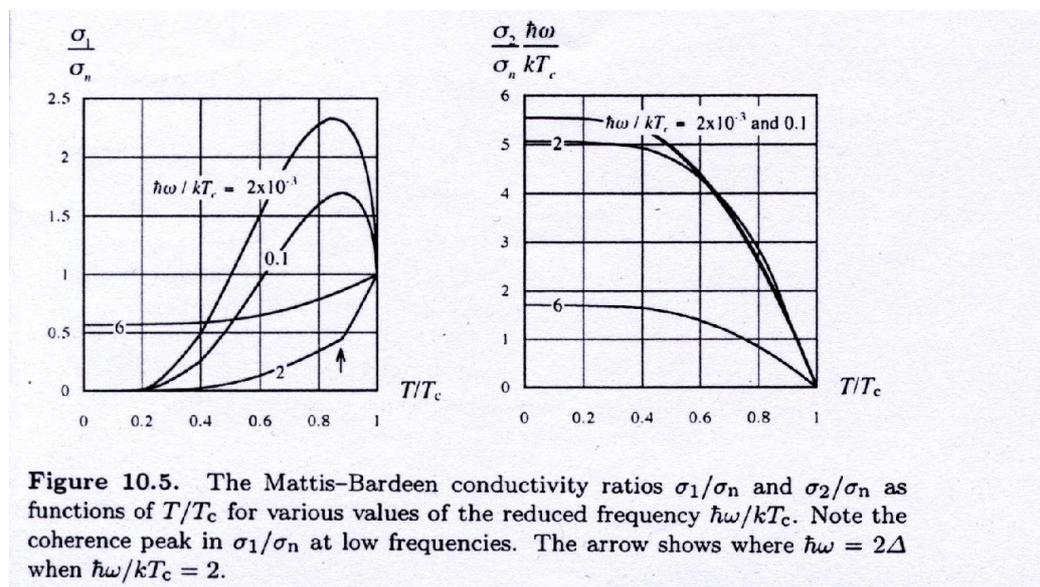

Figure 10.5. The Mattis–Bardeen conductivity ratios $\sigma_1/\sigma_n$ and $\sigma_2/\sigma_n$ as functions of $T/T_c$ for various values of the reduced frequency $\hbar\omega/kT_c$. Note the coherence peak in $\sigma_1/\sigma_n$ at low frequencies. The arrow shows where $\hbar\omega = 2\Delta$ when $\hbar\omega/kT_c = 2$.

Fig. 1 (from Waldram [2], p. 187)

In Fig. 1 we can see the two components $\sigma_1$ and $\sigma_2$ of the complex conductivity $\sigma$ (defined such that $j=\sigma E$, where $E$ is the electric field) for a conventional superconductor described by the local London theory, valid when the mean free electron path $l$ is much smaller than the coherence length $\xi$. The depicted curves are theoretical, but the agreement with experimental data obtained with Type-II superconductors is good. The real and imaginary components $\sigma_1$ and $\sigma_2$ of the complex conductivity are compared to the normal-state conductivity $\sigma_n$ for various values of the ratio $T/T_c$ (horizontal axis) and of the ratio between the frequency $\omega$ of the applied field and the "gap frequency" $\omega_c = k_B T_c/\hbar \sim 10^{13}$ in the cuprates.

We see that for intermediate temperatures (in our case, $T/T_c \sim 0.6$-$0.7$) the real component $\sigma_1$ is of the same magnitude order as $\sigma_n$, admitted that the graph is valid also for lower frequencies and the so-called "coherence peak" visible at low frequency does not grow much larger at lower frequencies. We recall that our pulses have maximum frequency components of the order of the reciprocal pulse rise-time, ie $f \sim (100\text{ ns})^{-1} \sim 10^7$ Hz. This frequency is much smaller than the gap frequencies in the microwave range typically employed in measurements of the surface resistance.

In the graph of $\sigma_2$ we can see that $\sigma_2$ is proportional to $\sigma_n k_B T_c/\hbar\omega$ and the proportionality constant is of order 1. Therefore in our conditions we have $\sigma_2 \gg \sigma_n$, in accordance with the fact that at low frequency $\sigma_2 \to \infty$. It follows, since $|\varrho|=1/|\sigma|=1/\sqrt{(\sigma_1^2+\sigma_2^2)}$, that $|\varrho| \sim 1/|\sigma_2|$ in this case. Note that $\sigma_2$ is a typical inductive impedance.

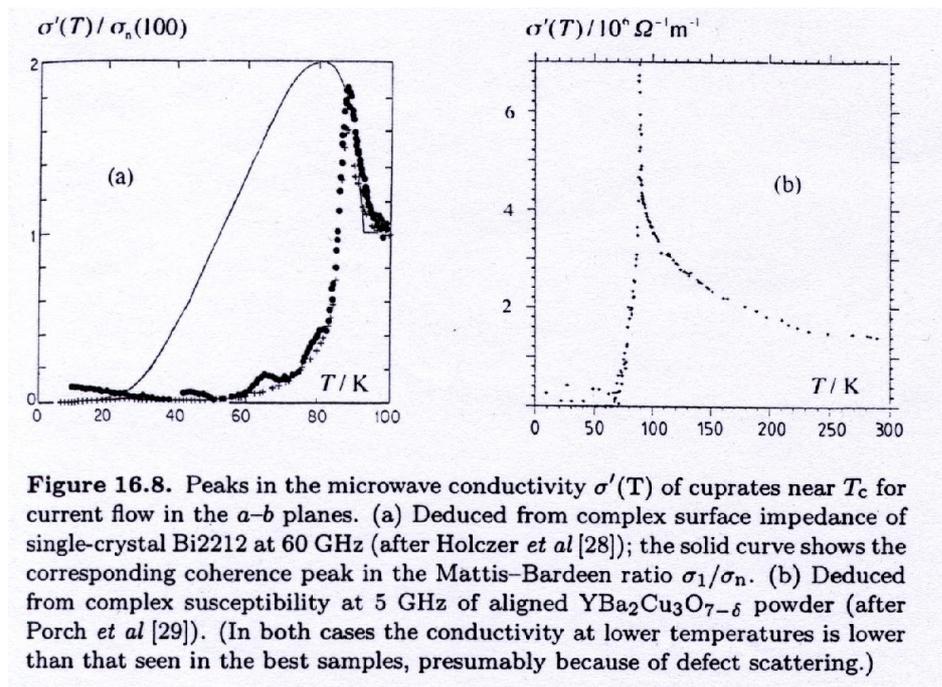

**Figure 16.8.** Peaks in the microwave conductivity $\sigma'(T)$ of cuprates near $T_c$ for current flow in the $a$-$b$ planes. (a) Deduced from complex surface impedance of single-crystal Bi2212 at 60 GHz (after Holczer *et al* [28]); the solid curve shows the corresponding coherence peak in the Mattis–Bardeen ratio $\sigma_1/\sigma_n$. (b) Deduced from complex susceptibility at 5 GHz of aligned YBa$_2$Cu$_3$O$_{7-\delta}$ powder (after Porch *et al* [29]). (In both cases the conductivity at lower temperatures is lower than that seen in the best samples, presumably because of defect scattering.)

Fig. 2 (from Waldram [2], p. 313)

Fig. (2.a) confirms, for the special case of cuprates, the behaviour of the real conductivity σ' as already seen in Fig. (1.a). We only notice a value unexpectedly close to zero around a temperature of 50 K. Also note that these values of σ' are measured at a frequency much higher than ours (60 GHz, as compared to 100 MHz).

In Fig. 2.b we see the real conductivity σ' (the same as $\sigma_1$) measured for a cuprate superconductor and referred to the *a-b* direction. This is not the case we are interested in, but let us compare it, for a check, with the data of Table 1. A resistivity $\rho \sim 35 \cdot 10^{-8}$ Ωm corresponds to a conductivity $\sigma \sim 3 \cdot 10^6$ Ω$^{-1}$m$^{-1}$ for normal material at 100 K. This is in agreement with Fig. 2.b. Also note that resistivity is independent from the frequency, as expected, since the impedance is mainly dissipative.

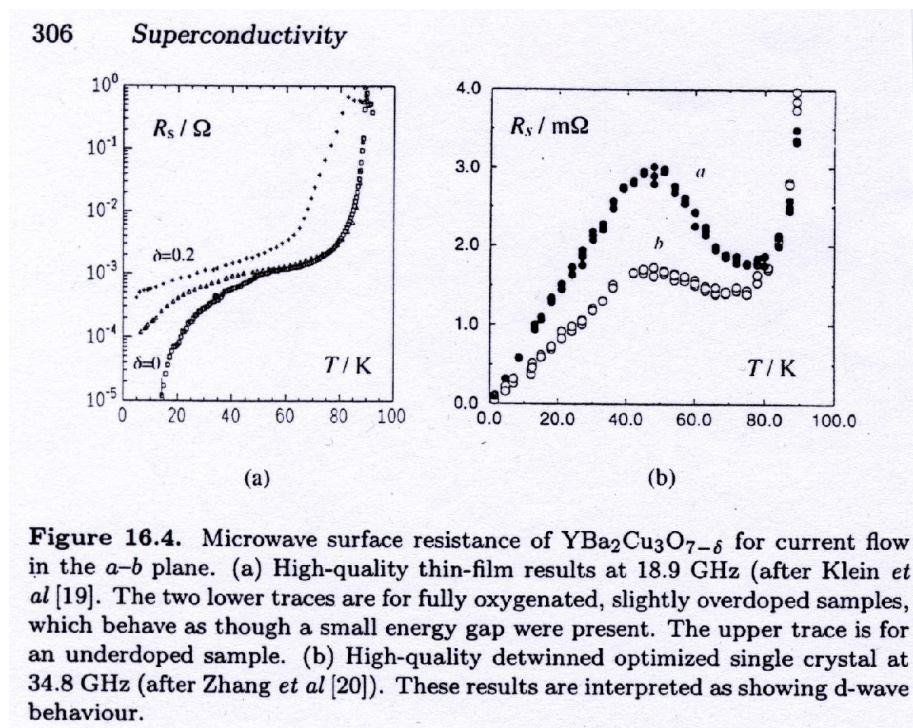

Figure 16.4. Microwave surface resistance of YBa$_2$Cu$_3$O$_{7-\delta}$ for current flow in the *a–b* plane. (a) High-quality thin-film results at 18.9 GHz (after Klein *et al* [19]. The two lower traces are for fully oxygenated, slightly overdoped samples, which behave as though a small energy gap were present. The upper trace is for an underdoped sample. (b) High-quality detwinned optimized single crystal at 34.8 GHz (after Zhang *et al* [20]). These results are interpreted as showing d-wave behaviour.

Fig. 3 (from Waldram [2], p. 306)

In Fig. 3 we see the surface resistance $R_s$ in YBCO for current flowing in the *a-b* planes, both for a film (a) and for a single crystal sample (b). $R_s$ is measured in ohms because it is defined as the resistance of a square, on the superconductor surface, when a voltage is applied to two opposite sides of the square. It is easy to show that such resistance does not depend on the length of the sides. We see from the graph that $R_s$ is of the order of milli-ohms. From this we can obtain the real part $\sigma_1$ of the conductivity through the formula ([2], pag. 305)

$$(1.1) \qquad R_s = \frac{1}{2}\omega^2 \mu_0^{\,2} \lambda^3 \sigma_1$$

where $\omega$ is the angular frequency of the applied field and $\lambda$ is the penetration length, which at high frequency is independent from $\omega$. Note that this formula has some non-obvious features, for instance the surface resistance apparently increases when the conductivity increases. Actually, for low frequencies, when $R_s$ tends to the constant $R_n$, $\lambda$ depends on $\omega$. Setting $\lambda \sim 130$ nm we find

| $\omega/2\pi$ (Hz) | $\sigma_1$ ($\Omega^{-1}\mathrm{m}^{-1}$) |
|---|---|
| 5 | $2.5 \cdot 10^8$ |
| 10 | $5 \cdot 10^7$ |
| 20 | $1.4 \cdot 10^7$ |
| 30 | $5 \cdot 10^6$ |

**Table 2 -** Real part $\sigma_1$ of the conductivity of YBCO computed from eq. (1.1) at different angular frequencies $\omega$, with $\lambda \sim 130$ nm. These data can be compared with those in Fig. 2. The agreement is good for the higher frequencies.

In conclusion, since $\sigma_2$ is proportional to $1/\hbar\omega$, at our relatively low frequencies we simply have $\sigma_2 \gg \sigma_n$ and therefore $|\rho| \sim 1/\sigma_2$. So $\rho$ is very small. Also note that all the data above are referred to measurements of the surface resistance. This is typical of measurements at GHz frequency done with microwave cavities. In our case the frequency is lower and the current flows in the bulk of the material.

Bulk measurements are reported for instance by Kunchur [3] who sent pulses of micro-seconds duration through an YBCO film. The scope, however, was to test the onset of ohmic resistance in the material due to flux slipping. This means that the current density was over $j_c$, which is not the case for our emitter ($j \sim 10^2$ A/cm$^2$, while reportedly $j_c \sim 10^4$ A/cm$^2$, as common for melt-textured YBCO). The measurements by Kunchur were made in a strong magnetic field. In this case the theory predicts ("free flux slip model") an ohmic resistance of the order of $\rho_n B/B_c$, where $\rho_n$ is the normal-state resistance and $B_c$ the critical magnetic field.

The second possible cause of resistance and dissipation is the **radiation reaction**, like in an antenna. The electromagnetic waves emitted by electrons oscillating within an antenna carry energy, which comes of course from the kinetic energy of the electrons, and therefore ultimately from the electric field causing their oscillation. The antenna does not strictly behave like an ohmic conductor, but writing the emitted power in the form $P_{rad}=1/2\ I^2 R_{rad}$ it is possible to define the so-called resistance by radiation reaction. For a linear antenna of length $d$, emitting radio waves of wavelength $\lambda$, one finds [4] that

$$(1.2) \qquad R_{rad} \approx 5\left(\frac{2\pi d}{\lambda}\right)^2 \Omega$$

This formula holds for $d\ll\lambda$, from which it clearly appears that the radiation resistance is always small (and often smaller, in practice, than the ohmic dissipation always present in the antenna). In our case, taking an emitted frequency of 100 MHz (corresponding to the maximum frequency present in the current pulse) we obtain $\lambda\sim 3$ m and $R_{rad}\sim 1$ m$\Omega$. This resistance is much smaller than the $V/I$ ratio observed in the emitter. The power emitted in radio waves when the current is $I=10^4$ A is then $P_{rad}\sim 10^5$ W, which amounts to a very small fraction of the total power dissipated ($P_{tot}\sim IV\sim 10^{10}$ W).

The formula above strictly holds for a pure harmonic motion of the electrons, but can be applied to our emitter if we imagine to split our pulse in Fourier components at several frequencies. In order to obtain a more complete picture of the electromagnetic emission, one should also consider that the emitter is not exactly an homogeneous conductor. When the pairs are accelerated through the emitter by the large voltage, they do not encounter any resistance or scattering centres like electrons in a normal metal; they behave a bit like a beam in vacuum. In such conditions, any electron beam generates electromagnetic radiation. For instance the vacuum tubes employed for microwave production, the magnetron and klystron, work this way: an electron beam is accelerated inside a resonating cavity, and in certain conditions the kinetic energy of the electrons is transferred to the electromagnetic field in the cavity, at some definite wavelengths. For this to happen, it is necessary to establish a certain phase relation between beam and field, in such a way that the beam gives energy to the radiation field, and not the reverse. This is obtained in turn with a modulation of the beam intensity (compare the free electron laser). In our case the modulation might occur because the electrons slow down before jumping from one plane to the other. (Besides the electromagnetic emission, we suppose there is the anomalous emission; while the electromagnetic emission carries almost all the energy away, the anomalous emission carries most of the momentum.)

The third possible source of ohmic resistance in the emitter are the **Josephson junctions** present in the superconductor. It is known that in sintered ceramic superconductors the grain boundaries behave like Josephson junctions. Even in single-crystal or melt-textured samples it is believed that the conduction in the $c$ direction itself occurs like a sort of tunnelling between Josephson junctions, because the pairs are localized in the $ab$ planes and need to "tunnel" between one plane and the next. There are theoretical models describing the $c$-axis conduction in this way [5,6]. The junctions would be very tiny, of nm size. (Consider that in our emitter we have the equivalent of about 1 million nano-metric junctions.) Such tiny junctions can also be fabricated artificially and are described in the literature [7]. More frequently, however, J. junctions are of micro-meter thickness.

J. junctions under finite voltage are most commonly represented by the RSJ model ("resistively shunted junction"); in this model they appear like an equivalent circuit which includes a parallel resistance (Fig. 4), so they develop a voltage when crossed by a current. In the literature one typically encounters a resistance of the order of 0.1-10 ohm, for micro-meter junctions. Another interesting feature is that J. junctions emit electromagnetic radiation when crossed by current, and this could explain some of the "parasitic" e.m. emission observed by Podkletnov (to establish a secure connection, one should be able to compare power and wavelength). Arrays of J. junctions tend to emit coherently and build standing waves if there is a resonator [8].

94    *Superconductivity*

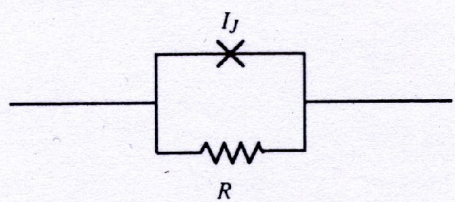

**Figure 6.1.** The resistively shunted junction (RSJ) model of a weak link. The element marked with a cross carries Josephson supercurrent in accordance with (6.3), and the normal current is described as flowing through a simple parallel resistance.

Fig. 4 (from Waldram [2], p. 94)

For Josephson junctions in conventional (non cuprate) superconductors described by the BCS theory the following relation holds between the critical Josephson current $I_J$ and the normal state resistance $R$:

(1.3) $$RI_J = V_c = \Delta_0/e$$

where $V_c$, called the critical voltage, is the equivalent in volts of the superconducting gap $\Delta_0$ and is therefore a characteristic of the junction material. For instance, for Niobium the gap is $\Delta_0 \sim 2$ meV, therefore $V_c=2$mV. $I_J$ is the maximum supercorrent flowing through the junction at zero applied voltage ($I=I_J\cos\phi$, where $\phi$ depends on the initial conditions). For a given material, $I_J$ depends on the normal state $R$, which in turn depends on the geometric and microscopic features of the junction. In Niobium junctions one can observe critical currents in a wide range, $10^{-6}<I_J<10^{-2}$ A ($I<10^{-6}$ A is regarded as noise), corresponding to $10^{-1}<R<10^{3}$ $\Omega$. For cuprate high-Tc superconductors the BCS theory is not valid. Being the gap $\Delta_0$ proportional to $T_c$, ie 10-30 times larger than in conventional superconductors, BCS would give at least RJ≈50 mV, but real data [2, p. 345] are

(1.4) $$RI_J \sim 0.1 \text{ mV} \quad (\text{YBCO, 77 K})$$

Let us see if, in view of all this, our emitter can be regarded as an array of Josephson junctions. The cross-section is 100 cm$^2$=10$^{10}$ $\mu$m$^2$. The total current during the discharge is 10$^4$ A. Let us consider a single layer, ie a couple of *ab* planes between which conduction occurs by tunnelling. We can see for instance this layer as a parallel of 10$^8$ junctions with surface 10x10 $\mu$m$^2$, each one which a current of 10$^{-4}$ A; or as a parallel of 10$^6$ junctions with surface 100x100 $\mu$m$^2$, each one which a current of 10$^{-2}$ A. For each junction, $R$=0.1 mV/$I_J$; for each layer we take the parallel, ie divide by the number of junctions. We obtain

(1.5) $$R_{\text{layer}}= 0.1 \text{ mV}/10^4 \text{ A} = 10^{-8} \text{ } \Omega$$

(The surface, in the examples above, is chosen in such a way that the critical current is in the range typical for single junctions. Alternatively, we could regard a whole couple of layers as a large Josephson junction, though this is not very realistic. In that case, we must substitute for $I_J$ the total current, so it is multiplied by $n$ and the result is the same as above.)

Then we sum over all layers. Their average distance is 3 lattice spacings, ie 3.5 nm, so in 4 mm thickness there are $1.4 \cdot 10^6$ layers. We find

(1.6) $$R_{\text{whole emitter}} = 1.4 \cdot 10^{-2} \, \Omega$$

This should be interpreted in the sense that the current in the emitter can actually be due to Josephson tunnelling if the total normal resistance of the emitter is smaller than $1.4 \cdot 10^{-2} \, \Omega$ - which is certainly true. Actually, the normal resistance is even smaller than that ($2 \cdot 10^{-5} \, \Omega$, see Table 1), and we can drive in principle a larger current through the emitter (Podkletnov mentions a critical current density of $5 \cdot 10^4$ A/cm$^2$, although this is probably not true in the c-direction).

When a finite voltage is applied, one must resort to the RSJ model for the Josephson junctions [6]. The pairs crossing the junction emit a photon of frequency $f$ given by the usual quantum relation $f = 2eV/\hbar$, where $V$ is the voltage applied. A formula holds for the maximum power available by photon emission. For instance, for a Niobium junction, suppose $I_J$=10 mA and so $R$=10 $\Omega$. The maximum power available is

(1.7) $$W = RI_J^2 = R(V_c/R)^2 = V_c^2/R$$

This power is in the micro-watt range. Eq. (1.7) must be equivalent to the general relation $W=VI$, which is valid if we assume that the entire work done by the voltage difference is transferred to the photon. This means that the maximum applicable potential is $V_c$, otherwise the pairs are broken; and in fact $2eV_c$ is just equal to the gap energy.

**2. Inelastic processes and the role of the emitter thickness**

Let us recall how the energy-momentum transfer to the target occurs, according to our microscopic model. The anomalous radiation beam is composed of $N$ gravitons with energy $E=hf$ and momentum $p=h\lambda^{-1}$. The wavelength $\lambda$ is fixed by the crystal lattice, the formula for the frequency $f$ is

(2.1) $$f = \frac{v_P}{s_{acc}} = \frac{I}{2e\rho S n s_c}$$

$N$ is given by the number of tunnelling processes; we first suppose that there is an emission in coincidence with each tunnelling. The formula for the maximum beam energy $U_{max}$ can be written as

(2.2) $$U_{max} = \frac{\delta}{s_{acc}} \frac{hf}{2e} I \Delta t_2 ,$$

from which we deduce $N$:

(2.3) $$N = \frac{1}{2es_{acc}} \delta I \Delta t_2 .$$

$N$ is very large, say $N \approx 10^{21}$.

After gravitons are absorbed in the target, there can be inelastic processes, so that a part of the energy is dissipated and "wasted", as kinetic energy of the target. Suppose first that this does not happen. Then we have

(2.4) $$\frac{1}{2} v_t = \frac{E_t}{p_t} = \frac{E_{grav}}{p_{grav}} = f\lambda$$

Therefore $v_t$ is fixed (in the sense that it is independent from $m$) and given by the known formula $v_t = \frac{k}{n} \frac{I}{e\rho S}$ ($n$=acceleration space; $k$=wavelength, both in lattice units $s_c$). It follows that for $m<m_{MAX}$ only a part of the $N$ gravitons are absorbed, ie a number $N_1$ such that

(2.5) $$\frac{N_1}{N} = \frac{\frac{1}{2} m v_t^2}{\frac{1}{2} m_{MAX} v_t^2} = \frac{m}{m_{MAX}}$$

In order to understand what happens when $m>m_{MAX}$, let us consider for instance the case $m=2m_{MAX}$. Being the total available momentum still the same, the target velocity will be half as much, if all gravitons are absorbed. It follows that the total kinetic energy of the target $E_{tot}=1/2mv_t^2$ is half the target energy for $m=m_{MAX}$. Therefore, half the energy carried by the gravitons is dissipated in anelastic processes. It is easy to show in general that if $m>m_{MAX}$, then part of the energy must be dissipated in inelastic processes.

In addition to the 4 mm emitter considered up to now, Podkletnov reports trials made with a 8 mm emitter. How are the predictions of our model affected by the emitter thickness? If the emitter resistance is ohmic, it must double, and so must the discharge time $\Delta t_2 \approx RC$, while $I$ is halved, if the applied voltage is the same. From the velocity formula

$$(2.6) \qquad v_t = \frac{k}{n}\frac{I}{e\rho S}$$

we see that $v_t$ should halve if the thickness is doubled. From the energy formula written as

$$(2.7) \qquad U_{max} = \frac{d}{s_{acc}^2}\frac{h}{(2e)^2}\frac{I}{\rho S}\Delta t_2$$

we deduce that when the thickness doubles, $U_{max}$ stays the same. But the frequency of the gravitons is halved, and so is the energy of each graviton, because $f$ is proportional to $I$: $f=I/(2e\rho S s_{acc})$. In order to keep invariant $U_{max}$, the gravitons number must double. (This matches the fact that half as much current crosses twice as much planes for twice as much time.) But the momentum of each graviton is the same, because it only depends on $\lambda$. As a consequence, the total momentum $P_{tot}$ is doubled; this is consistent with $m_{MAX}$ becoming 4 times larger.

The situation is summarized in the following table:

| Quantity | Variation | Ground |
|---|---|---|
| 1. Emitter resistance | $R \rightarrow 2R$ | Assumption of ohmic resistance |
| 2. Total current | $I \rightarrow I/2$ | Macroscopic consequence of 1 |
| 3. Capacitor discharge time | $\Delta t_2 \rightarrow 2\Delta t_2$ | Macroscopic consequence of 1 |
| 4. Target velocity | $v_t \rightarrow v_t/2$ | Microscopic conseq. of 2 and (2.6) |
| 5. Max. available energy | $U_{max} \rightarrow U_{max}$ | Microscopic conseq. of $\delta \rightarrow 2\delta$, 2, 3 |

| | | | and (2.7) |
|---|---|---|---|
| 6. Max. target mass | | $m_{MAX}$ → 4 $m_{MAX}$ | Microscopic conseq. of 4, 5 |
| 7. Frequency and energy of emitted gravitons | | $f$ → $f/2$; $E$ → $E/2$ | Microscopic conseq. of 2 and (2.1) |
| 8. Number of emitted gravitons | | $N$ → $2N$ | Microscopic conseq. of 5 and 7 |
| 9. Wawelength and momentum of emitted grav. | | $P$ → $p$; $\lambda$ → $\lambda$ | General prediction of the microscopic model |
| 10. Total momentum | | $P_{tot}$ → $2P_{tot}$ | Microscopic conseq. of 8 and 9 |

**Table 3** – Variations of macroscopic and microscopic quantities of the theoretical model when the emitter thickness is doubled.

Experimentally, these predictions are not verified. With the 8 mm emitter a target velocity $v_t$ is observed, which is about 35% larger than the target velocity for the 4 mm emitter. Part of this increase can be due to the fact that the 8 mm emitter can trap more magnetic flux. As discussed in the Report 1-2004, this amounts to a reduction of the effective surface of the emitter. The table above tells us, however, that a 50% reduction should occur. It seems therefore that the larger thickness can bring further changes, besides those considered above. In particular, if stimulated emission is present, the layers of the emitter closer to the gas host many more emission processes. In that case, the influence of the emitter thickness on the ohmic resistance is not the dominant factor.

An alternative hypothesis is the following. We have seen that for $m>m_{MAX}$ inelastic processes occur, with partial dissipation of the anomalous beam energy. We also saw that when $\delta$ is doubled, $m_{MAX}$ increases by a factor 4. Would it be possible that

(1) the targets employed by Podkletnov have $m>m_{MAX}$ with a 4 mm emitter and $m<m_{MAX}$ with a 8 mm emitter, and

(2) because of this, they can acquire a larger velocity with the 8 mm emitter?

Point 1 implies that part of the beam energy is not absorbed with the 4 mm emitter, while complete elastic absorption is possible for the 8 mm emitter. The best way to settle these questions is to make measurements with light targets and see if the relative performance of the 8 mm emitter to the 4 mm emitter remains the same.

### 3. Improvements of the microscopic model

As we have seen, a resistance of the order of 100 ohms in a superconductor can not be explained by any conventional mechanism, even at relatively high frequency (ca. 100 MHz, the highest Fourier components of the Marx pulses). It is possible to assume that such a resistance occurs when the pairs lose their momentum and energy in the anomalous emission of gravitons. Numerically this can make sense, as we show below. But there are two big problems in this view:

1. A practical problem: what if the emission does not start? The current may grow so large, that it can damage the Marx capacitors and the current and voltage meters. And how can the emission be brought about, in that case?

2. A theoretical problem: an emission-induced resistance can numerically make sense only if one admits that tunnelling of pairs between superconducting planes only succeeds after an average attempt frequency of ca. $10^{11}$ Hz (see below). We can not figure any realistic physical justification for this number. Even in the presence of anomalous emission, why should it be so difficult for the pairs to jump between planes? They usually do it much more easily, and that's why resistance is usually very low.

Let us analyze the theoretical aspects in more detail. In any ohmic conductor, ie one in which the current is proportional to the applied voltage, some microscopic process occurs where the charge carriers (electrons, in a metal) are accelerated by the electric field and acquire velocity and kinetic energy, but after a short average time $\tau$ collide with the positive ions lattice and loose all their energy, then start again, and so on ("Drude" model). Denoting by $a$ the electron acceleration under the effect of the field, we have $a=eE/m$ and the mean drift velocity is $v=a\tau$. The collision frequency $f=1/\tau$ is often introduced, too. The following relations hold:

(3.1) $$j=\sigma E=E/\rho \quad \text{(definition of } \sigma \text{ and } \rho\text{)}$$

(3.2) $$j = nev = \frac{e^2 E n \tau}{m}$$

where $n$ is the number of electrons per cubic meter. Typical values for a metal are $\rho \sim 10^{-8}$ $\Omega$m, $n \sim 10^{29}$ m$^{-3}$, $\tau \sim 10^{-13}$ s.

Now, our theoretical model tells us that the pairs are accelerated by the electric field from one superconducting *ab* plane to the other, and after each jump they loose their momentum emitting a virtual graviton, and also loose their energy, probably through a simultaneous electromagnetic emission or through some other dissipative process. (The energy carried away by the graviton is very small.) The drift velocity is $\sim 0.5$ m/s and the superconducting planes are 2 or 4 lattice spacings apart, therefore the transit time is $\sim 10^{-8}$ s.

However, in an analogy with conduction in a metal, we can not identify just this transit time as the time $\tau$ between collisions. The inter-plane voltage is $\sim 1$ V, so if the motion of the pairs between the planes was free and collisionless, the pairs would reach a velocity much larger than 0.5 m/s; it was instead of the order of $10^5$ m/s. In other words, the kinetic energy of an electron with velocity 0.5 m/s is $\sim 3 \cdot 10^{-12}$ eV, which is $3 \cdot 10^{11}$ times smaller than the energy available. It happens *as if* the pairs, in the jump between planes, would undergo $\sim 3 \cdot 10^{11}$ dissipative collisions. Let us call this number "additional frequency" $f_a$.

The total collision frequency is then $f = f_{transit} f_a \sim 3 \cdot 10^{19}$ s$^{-1}$, where $f_{transit} \sim 10^8$ as above. Let us insert this into the computation of the resistivity, also remembering that the density of pairs in YBCO is smaller than the density of conduction electrons in a metal, namely of the order of $10^{25}$ m$^{-1}$. We obtain

(3.3) $$\rho = \frac{1}{n}\frac{m}{e}f = \frac{1}{10^{25}} \cdot 4 \cdot 10^7 \cdot 3 \cdot 10^{19} \approx 10^2 \Omega m$$

(NOTE: for the pairs, $e$ and $m$ must be doubled, but we are concerned with magnitude orders here.) From $\rho$ we find the estimated resistance of Podkletnov's emitter:

$R=\rho\delta/S \sim \rho \cdot 0.01\text{m}/0.01\text{m}^2 \sim \rho$ ($\delta$ is the thickness of the emitter), which agrees well with the observed 100 $\Omega$ of the *V/I* ratio.

The physical picture differs from that of conduction in a metal, as a consequence both of the layered structure of the superconductor and of the unique wave function describing all pairs. We must imagine that the pairs make, on the average, $\sim 3 \cdot 10^{11}$ jumping attempts before they reach the next plane. (This resembles the behaviour of an alpha-particle in a radioactive nucleus, in the semi-classical approximation.) During these attempts each pair acquires the energy and momentum corresponding to a voltage of 1 V; or we can say that when the pair actually jumps, it must acquire them, but in fact the exact instant is non determined, for quantum mechanical reasons.

The good numerical coincidence above is not casual, but comes from the very structure of our microscopic model. Remember that the velocity *v* is estimated from the current through the equation *v=I/Sne*, where *S* is the cross section of the emitter. The transition frequency is given by $f_t=v/s_c$, where $s_c$ is the acceleration space. The additional frequency of attempts can then be expressed as $f_a=eV/(1/2mv^2)$, where *V* is the voltage drop over the distance $s_c$, ie $V=V_{tot}/(\delta/s_c)$ ($\delta$ is the thickness of the emitter). Taking all these relations into account, eq. (2.3) for $\rho$ can be transformed as follows:

$$(3.4) \qquad \rho = \frac{m}{e^2 n} f_t f_a = \frac{m}{e^2} \frac{1}{n} \frac{v}{s_c} \frac{eV_{tot} s_c}{mv^2 \delta/2} = \frac{1}{n} \frac{RI}{v\delta}$$

Finally, by inserting *v=I/Sne* we obtain an identity. Therefore, an ohmic-like resistive mechanism is implicit in our model and consistent with the data, but only if one admits the existence of the additional attempts frequency as given above.

Let us then consider an alternative and very different possibility. Suppose the emitter resistance is actually very small during the discharge (we already accepted it is small before breakdown). Our RLC circuit will then make un-dampened oscillations. We recall that un-dampened oscillations occur for a resistance less than approx. 27 ohms in our case (IU1080 generator, *C*=15 nF, *L*=4 $\mu$H). From the *V/I* ratio we earlier estimated *R*~100 $\Omega$, or equivalently that all the capacitors charge was released in an over-dampened oscillation with peak current $I \sim 10^4$ in a time $\Delta t_2 \sim 10^{-7}$ and therefore $R=\Delta t_2/C \sim 100$ $\Omega$. But suppose now that $R$ is much smaller, less than 1 $\Omega$.

In the un-dampened case the oscillation period is $T \sim 10^{-7}$ ($\omega=1/\sqrt{(LC)}$) and the discharge time $\tau=2L/R \sim 10^{-6}$ if $R \sim 1$ $\Omega$. Given the initial voltage of $10^6$ V, a peak current value of $10^4$ A follows from the circuit equations; we must assign a value to $L$, however, and we take that of the IU1080 generator (Podkletnov did not give the value of $L$ for his Marx generator). The magnitude order estimate for the current is straightforward: we can set $L\Delta I/\Delta t \sim 10^6$, $\Delta t \sim \Delta t_2$, because the current increase can not be faster, otherwise the induced contrary electro-motive force would exceed the applied voltage.

More precisely, we can use the same circuit equation as for Phase 1 (pre-breakdown; see Report 1-2005). In the present case the capacitance $C_2$ is absent. The emitter does have a stray capacitance, but we disregard it. The equation is simply

$$(3.5) \qquad \frac{Q}{C} + LI' + RI = 0$$

By taking the time derivative, considering that $Q'=I$, and imposing the initial condition $I(0)=0$, one obtains

$$(3.6) \qquad I(t) = I_0 e^{-t/\tau} \sin(\omega t)$$

where, if $R$ is less than $\sim 1$ $\Omega$ we have for the IU1080 generator ($C=4$ nF, $L=3$ $\mu$H)

$$(3.7) \qquad \omega \approx \frac{1}{\sqrt{LC}} \approx 10^7 \text{ Hz}$$

$$(3.8) \qquad \tau = \frac{2L}{R} \geq 10^{-6} \text{ s}$$

In order to determine $I_0$, which gives the maximum amplitude oscillation we are interested in, we must integrate $I(t)$ to find the charge $Q(t)$:

$$(3.9) \qquad Q(t) = \pm I_0 \frac{\tau e^{-t/\tau}[\omega\tau\cos(\omega t) + \sin(\omega t)]}{\omega^2\tau^2 + 1} + k$$

The ± sign needs to be checked, but is not relevant to the following. (Note that in the calculation of Report 1-2005 the factor $I_0$ was erroneously omitted, which was fortunately not essential in that case.) At the initial time we have $Q(0)=CV$, where $V$ is the maximum initial voltage of the pulse. So we find $k = CV \pm I_0 \frac{\omega\tau^2}{\omega^2\tau^2+1}$. Then we impose the condition $Q=0$ for $t=+\infty$, and find

(3.10) $$|I_0| = CV\frac{\omega^2\tau^2+1}{\omega\tau^2} \approx CV\omega \approx 4\cdot 10^4 \text{ A}$$

(Numerically, from the formula $\omega\tau \approx \frac{2}{R}\sqrt{\frac{L}{C}}$ we see that $\omega\tau \gg 1$ and therefore the fraction is simplified; we set $V=10^6$ V.) In the first oscillations the value of $I_0$ is reached almost without any dampening, because the exponential only begins to cut at times $\sim\tau$. Expressing $\omega$ as a function of $L$ and $C$ we can also write

(3.11) $$|I_0| \approx V\sqrt{\frac{C}{L}}$$

whence we see how the peak current can be made smaller, if needed.

In conclusion, it could be that all the picture of a single dampened discharge is incorrect, $R$ is actually very small, oscillations in the emitter always occur, and the current is limited only by the inductance. If this is so, compelling questions arise:

- Should Podkletnov have noticed and reported this? I asked him, but he said he did not find any direct or indirect evidence of such oscillations. They can be possibly difficult to observe, because the detectors for large currents have some limitations in frequency.

- Is the microscopic model based on a single pulse crossing the emitter still valid? Should one not expect emission in both directions if the current oscillates back and forth? By analogy with energy-momentum conservation in other mechanical and electromagnetic phenomena, I would say it is possible that the model still correctly predicts the total energy and momentum given to the targets. Multiple current "sweeps" over the emitter could even enhance the probability of anomalous emission, both spontaneous and stimulated.

On the practical side, the computations above tell us that if *R* is small, the peak current is $I=V\sqrt{(C/L)}$, so it can be controlled by a series inductance, if needed. *L* should not be too large, because it also affects rise-time and duration of the discharge.

Also, an alternative scheme was recently considered of "sandwiching" the emitter between two metal plates and sending a pulse switched simply by an air gap. This would allow to eliminate the low-pressure discharge chamber. We wonder: would the air gap switch close the circuit for a time long enough that the oscillations can occur? The dampening time of the oscillations is $\tau=2L/R$. Note that the presence of a normal part in the emitter and its resistance can now play some role, if we are not supposing any more that the resistance of the SC part is so large.

# Chapter 6

# The emitter modelled as a stack of intrinsic Josephson junctions (IJJs) (2006)

**1. Introduction**

In the 2005 Reports our theoretical model has been much changed and improved, as far as the current in the emitter is concerned. We passed from the original idea of a single over-dampened pulse of duration 100 ns, with an effective emitter resistance of the order of 100 Ohms, to the idea of a discharge with oscillating current. This picture is much more plausible from the point of view of superconductivity. The frequency is thought to be $\omega \approx 1/\sqrt{(L_L C_L)} \approx 1$ MHz, where $L_L$ and $C_L$ are those of the Marx generator. The maximum current $I_0$ is limited by the Marx inductance $L_L$. The ohmic resistance $R_E$ of the emitter is of the order of $10^{-4}$ Ohms, the emitter voltage $I_n R_E$ is just about 1 mV ($I_n$ is the normal current in the emitter). The oscillation is under-dampened, with decay time $\tau = 2L/R_L \approx 10^{-4}$ s defined by the load resistance of the external circuit $R_L$; the oscillation merit factor Q, expressing the number of oscillations in the decay time, is given by $Q = \omega\tau \approx 100$.

This picture should be regarded as a theoretical guess, because it is not supported by Podkletnov's data. Those data are contradictory: (i) Podkletnov does not mention any oscillating current, though he was probably unable to detect the oscillations with his equipment. (ii) On the other hand, he denies that a large load resistance was present, such to obtain over-dampening. (iii) From our experience with the Marx generator, it appears implausible that any generator can support 10 kA - 2000 kV un-dampened oscillations without damage.

As written in my Report 3-2005, the emitter can be represented as a stack of intrinsic Josephson junctions (IJJs), in agreement with all modern studies of the conduction of cuprates in the c direction [1]. Alternatively, one can describe the material as a homogeneous superconductor with complex conductivity $\sigma = \sigma_1 + i\sigma_2$; the conclusions are compatible with the Josephson junctions model, but the information available in the literature about $\sigma_1$ and $\sigma_2$ in the cuprates in dependence on T and other parameters is scarce (one should extrapolate from Bardeen theory [2]). The IJJs model is furthermore better suited to describe the electromagnetic emission of the material, and we shall see that the electromagnetic emission is important because partly related to the anomalous emission.

The parameters of the IJJs will be discussed in detail below. For now we observe, as confirmed by numerical simulations, that being the emitter inductance and capacitance $L_E$ and $C_E$ much smaller than those of the external circuit, they do not substantially affect the oscillation frequency. The simulations also show, as can be expected since the material "must" anyway conduct with excellent values of $\sigma$ at MHz frequency, that the normal current in the junctions adjusts itself to a value $I_n$ ($<<I_s$) such that the voltage-per-plane corresponds to an AC Josephson frequency equal to the external frequency. This AC Josephson frequency is in turn the same of the Cooper pairs interplane tunnelling, and so the same of the anomalous emission.

Being the anomalous emission a virtual process, its energy/momentum ratio is different from the E/p ratio (ie the $f\lambda$ product) of the e.m. emission. We shall discuss later the tricky problem of momentum conservation in the anomalous emission and the problem of the emission direction. We shall in addition show that the main factors relevant for high anomalous emission performance are:

- high oscillation frequency
- several oscillations in the damping time
- large current.

## 2. Inductance, capacitance, plasma frequency, dampening parameter

Let us find the inductance of the emitter as a series of $\approx 10^7$ Josephson junctions. The inductance of a single junction is $L \approx \phi_0/I_J$, with $\phi_0=h/2e\approx 2\cdot 10^{-15}$ Wb and $I_J$ critical current (at least $10^4$ A in our case). We find $L \approx 10^{-19}$. With $10^7$ crystal planes, the total inductance is: $L_E = 10^{-12}$, to be compared with $L_L\approx 10^{-6}$ of the Marx generator.

Then we find the capacitance of the emitter as a series of junctions. For a couple of crystal planes (cross-section $S\approx 20$ cm$^2$, distance $d\approx 2$ nm, relative dielectric constant $\varepsilon$ of the order of 10), we have $C = \varepsilon_0\varepsilon S/d \approx 10\cdot 10^{-11}\cdot 20\cdot 10^{-4}/2\cdot 10^{-9} \approx 10^{-4}$. Dividing by $10^7$, the total capacitance is found to be $C_E\approx 10^{-11}$, to be compared with $C_L\approx 10^{-8}$ of the Marx generator.

Therefore the proper frequency of the Josephson junctions, also called plasma frequency $f_P$, is $f_P = 1/2\pi\sqrt{(L_L C_L)} \approx 10^{11}$. It is natural to expect that this proper oscillation does not influence the behaviour of the system under the effect of an external forcing frequency of the order of 1 MHz. Note that $f_P$ is the same for the emitter and for any single junction, because the capacitances and inductances in series scale in the opposite way. The formula for $f_P$ can be easily re-written as follows, with reference to the single junction [1]:

$$f_P = \left(\frac{I_J}{2\pi\phi_0 C}\right)^{1/2} = \left(\frac{j_J d}{2\pi\phi_0 \varepsilon\varepsilon_0}\right)^{1/2}$$

The McCumber parameter of a junction $\beta_c$ is defined by $\sqrt{\beta_c}=2\pi f_P RC$. This is connected to the hysteresis of the I-V curve of the junction, because $\sqrt{\beta_c}=(4/\pi)I_J/I_r$, where $I_r$ is the so-called return current. For $\beta_c<1$ we have over-damped, non-hysteretic junctions; for $\beta_c>1$ we have under-damped, hysteretic junctions. With the data above, one finds for the single junction $\sqrt{\beta_c}\approx 10^8 R$. Therefore our junctions are strongly over-damped, because R is less than $10^{-10}$ Ω for the single junction.

## 3. Normal resistance of the intrinsic Josephson junctions in YBCO and "resistive shunts"

The intrinsic Josephson effect has been observed in YBCO as clearly as in BSCCO. Kawae et al. [3] give evidence of I-V curves in YBCO with multiple branches and hysteresis, very similar to those reported by Kleiner and Muller for BSCCO [1]. This can only be seen, however, in very small samples, with area about 0.25 μm². In larger samples, grain borders or other defects act as low-resistance shunts. The total resistance is essentially determined by these shunts, and so depends on the micro-structure and not just on the material. The junctions become non-hysteretic, because the McCumber parameter $\beta_c$ is proportional to *R*, and a small $\beta_c$ means no hysteresis. For this reason, the presence of the single junctions can not be seen in the I-V curves of "large" samples (not to mention our giant melt-textured discs!). From the practical point of view, all this does not disturb much, except that it is impossible to know in advance the resistance of our material, it depends on the micro-structure. The CRC data [8] is only an indication: ρ=5·$10^{-5}$ Ωm at room temperature, implying $R_E=10^{-4}$ Ω for our emitter.

According to CRC, there is only a small variation in the normal resistance between room temperature and 100 K.

Ref.s [3] and [4] also allow to estimate the normal resistance of the employed samples. For a stack of 80 junctions, ref. [3] gives at 4.2 K a critical current of 0.1 mA (40 kA/cm$^2$); the return voltage is 0.2 V and the slope of the I-V characteristic in the single normal branch is 800-1000 Ω. The $I_cR$ product is therefore 2-3 mV per junction; the material has $T_c$=43 K, so the BCS prediction is about 10 mV. The resistivity computed from the data above is $3 \cdot 10^{-3}$ Ωm. In ref. [4] the junctions have size 0.65×0.85 mm$^2$ and R=2 μΩ per crystal plane at the peak resistance value (84K; $T_c$=93 K). This gives ρ=$1.1 \cdot 10^{-3}$ Ωm.

The measured resistance of our emitter, including contacts, is $3 \cdot 10^{-4}$ Ω at room temperature and $5\text{-}12 \cdot 10^{-6}$ Ω at 77 K. Now, for noble metals the resistivity varies by a factor 5-10 between 300 K and 80 K; for iron more than 10 times. Alloys with noble metals show smaller variations, typically a factor 2-3. At 77 K the resistance of YBCO alone is zero, therefore the residual resistance is that of the metal layer and of the contact. On the contrary, the $3 \cdot 10^{-4}$ Ω at room temperature are essentially due to the YBCO, in agreement with the CRC data above. (I think that data on the normal resistance of YBCO below $T_c$ are extrapolated or obtained after application of a magnetic field; compare for instance the data given in Ref. [4].)

The aim of works like [3] is to see the features of the microscopic junctions (their resistance, capacitance, impedance, McCumber parameter), in view of possible applications for fast electronics, microwave collective emission or detection etcetera. In our case we will be satisfied to know that the intrinsic Josephson junctions are active and syncronized. Seeing their individual signature is not essential, actually impossible in a bulk sample. I believe they have to be syncronized, because otherwise they could not sustain the oscillating current; they would instead pass into a resistive state and the oscillating current would be normal current. The voltage on the emitter would then be of the order of $10^{-4}$ A $\cdot 10^{-4}$ Ω ≈ 1 V.

**4. First estimate of the total anomalous radiation energy $U_{max}$ and of the dissipation in the emitter.**

I have found a simple way to make an order-of-magnitude estimate of $U_{max}$, which agrees with the results of the detailed simulations (see below). Assume that the voltage-per-plane in the emitter is given by the Josephson relation $V=hf/2e=\phi_0 f$. Here $f$ is the frequency of the *external* circuit. This is necessary in order that the external current flows in the emitter as supercurrent (except for the small normal component $I_n$ which gives the finite voltage). The numerical simulations confirm this coincidence of external frequency and Josephson frequency.

Take the circuit values discussed in January 2006 for a 10 stages series generator.

Measured circuit parameters

| | | |
|---|---|---|
| L | Load (Marx+circuit) inductance | 6 µH |
| C | Load (Marx) capacity | 15 nF |
| $R_L$ | Load (circuit) resistance | 0.1 Ω |
| $V_0$ | Max pulse voltage | 100 kV |
| $\omega=2\pi f$ | Oscillation frequency | 3.3 MHz |
| $I_0$ | Max pulse current | 5 kA |
| $\tau$ | Dampening time (to 36% of max amplitude) | $1.2 \cdot 10^{-4}$ s |
| $Q=\omega\tau$ | Merit factor (nr. of oscill.s in time $2\pi\tau$) | 400 |
| $U_{tot}$ | Electrostatic energy stored in the capacitors | 75 J |

The voltage over a single junction is $V = 2\cdot 10^{-15}\cdot\omega/2\pi = 2\cdot 10^{-15}\cdot 5\cdot 10^5 \approx 10^{-9}$ V. The total number of junctions in 1 cm thickness is $\approx 10^7$ (each is 1.17 nm). Thus the total voltage on the emitter is $10^{-2}$ V. The IV product in the emitter, also called DC-power $P_{DC}$ is $P_{DC}=IV=5\cdot 10^3\cdot 10^{-2} = 50$ W. In the time $\tau$ this makes available in the emitter an energy of 6 mJ.

The electromagnetic emission generated in the AC Josephson effect has an energetic efficiency which is typically of the order of 10% [5]. We suppose that the anomalous emission is associated with the electromagnetic emission, ie a graviton is emitted together with the photon at each Cooper pair tunnelling, at the same frequency and with an emission probability of the same magnitude order. So the total energy $U_{max}$ of the anomalous radiation is ≈ 10% of 6 mJ = 0.6 mJ. This is quite small, but consider that the total energy is only 75 J with the parameters above, compared for instance with 2500 J of Podkletnov's device at 2000 kV (but only 156 J at 500 kV!). With a target of mass 5 g, the energy 0.6 mJ gives a velocity of ca. 0.5 m/s.

Approximately the 80% of the IV product, say 5 mJ, is wasted for the emission and dissipated as heat. This does not cause any serious temperature increase. My latest estimate for the thermal capacity of the emitter is 200 mJ/K; the temperature increase of the bulk is therefore negligible.

[This thermal capacity can be obtained as follows. Tilley [6] gives a graph of the specific heat of YBCO as a function of temperature, in units mJ/gK (Fig. 1). The density of YBCO is not easy to find in the literature, probably because it depends on the cell parameters, which are variable. Taking for instance a=0.38 nm, b=0.39, c=1.17 nm (Waldram [7], $YBCO_{7-x}$, x=0.4), one finds a unit cell volume of 1.7E-28 $m^3$. The unit cell (Fig. 2) contains 1 Y atom, 2 Ba atoms, 3 Ca atoms, for a total mass of 560 a.m.u. This gives a density of 6300 kg/$m^3$. Therefore the thermal capacity of the emitter, supposed it has volume 20 $cm^3$, is 200 mJ/K. 10 mJ/K$cm^3$ corresponds to 1.5 mJ/Kg.]

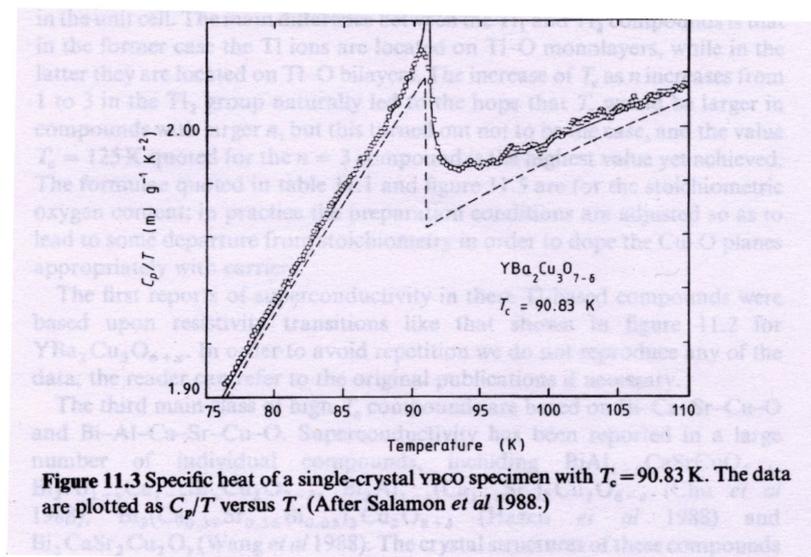

Figure 11.3 Specific heat of a single-crystal YBCO specimen with $T_c$ = 90.83 K. The data are plotted as $C_p/T$ versus $T$. (After Salamon et al 1988.)

Fig. 1 – From Tilley [6]

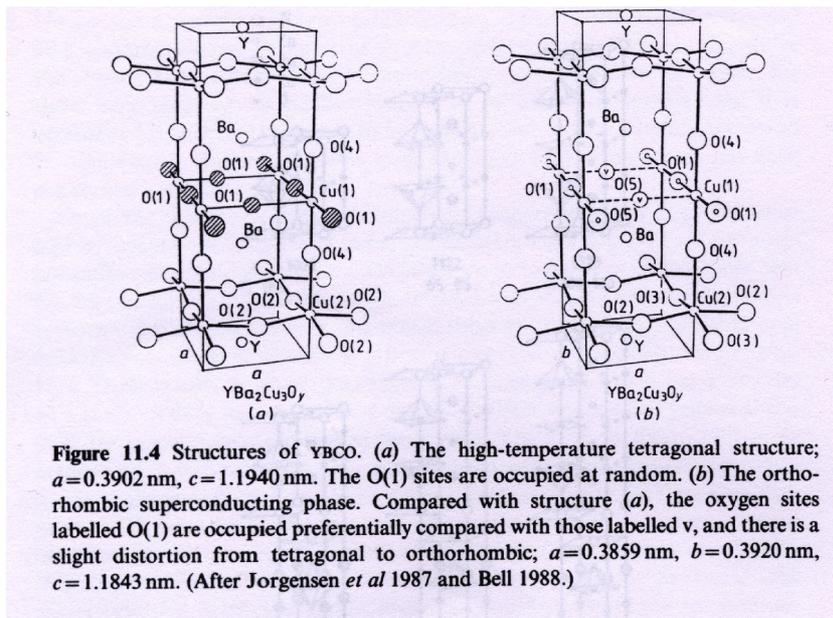

**Figure 11.4** Structures of YBCO. (a) The high-temperature tetragonal structure; $a=0.3902$ nm, $c=1.1940$ nm. The O(1) sites are occupied at random. (b) The orthorhombic superconducting phase. Compared with structure (a), the oxygen sites labelled O(1) are occupied preferentially compared with those labelled v, and there is a slight distortion from tetragonal to orthorhombic; $a=0.3859$ nm, $b=0.3920$ nm, $c=1.1843$ nm. (After Jorgensen et al 1987 and Bell 1988.)

Fig. 2 – From Tilley [6]

Note that the IV product on the emitter does not depend on the external load resistance $R_L$. If we can reduce $R_L$, we will increase $\tau$ and so proportionally increase the target energy. There are two practical problems with a small $R_L$, however: increased dissipation at the S/N contacts and possible damage to the capacitors of the Marx generator due to the large number of oscillations.

For comparison, the classical radio emission from the whole circuit seen as an antenna, with frequency $\omega=3.3$ MHz, current 5 kA, emitting length 10 cm, would be 150 W, ie 18 mJ in the time $\tau$. In any case, we do not expect any anomalous emission from an homogeneous superconductor without Josephson currents (or should we, provided there are sufficiently strong wave function gradients? This is an open question). For Podkletnov's device, which has a large metallic helium reservoir, at 2000 kV, the radio emission is much larger. This could be the reported "back radiation".

The normal current $I_n$ in the emitter can be computed through the relation $V=I_n R_E$. The resistance $R_E$ can roughly be guessed from the data in the literature. For instance, if $R_E \approx 10^{-4}$ Ω (CRC) one finds $I_n \approx 10^2$ A; if $R_E \approx 10^{-3}$ Ω [4], $I_n \approx 10$ A. The pure ohmic heating is therefore irrelevant. The figure for $I_n$ gives the *total* normal current. The density of normal current varies locally, as more current flows in the shunts with smaller resistance.

Estimated emitter parameters

Data with a * are not the ones estimated above, but exact data from numerical simulations (see below).

| | | |
|---|---|---|
| $R_E$ | Emitter normal resistance (measured at 77 K) | $10^{-4}$ Ω |
| $C_E$ | Emitter capacitance | $10^{-11}$ F |
| $L_E$ | Emitter inductance | $10^{-12}$ H |
| $f_P$ | Emitter proper oscillation frequency (plasma freq.) | $10^{11}$ Hz |
| $\beta_c$ | McCumber parameter of emitter junctions | $10^{-2}$ |
| $V_E$ | Emitter voltage* | $10^{-3}$ V |
| $P_{DC}=I_0 V_E$ | DC Josephson power in the emitter | 50 W |
| $I_n$ | Normal current in the emitter* (w. $R_E=10^{-5}$) | $10^2$ A |
| $U_{max}=10\%\tau P_{DC}$ | Target energy from anomalous radiation | 0.6 mJ |
| $c_E$ | Thermal capacity of the emitter | 200 mJ/K |
| $c_c$ | Thermal capacity of copper contact region | 1.92 mJ/K |
| $\rho$ | Emitter normal resistivity | $5 \cdot 10^{-6}$ Ωm |
| $\rho_c$ | Emitter contact resistivity | $2 \cdot 10^{-4}$ Ωcm$^2$ |
| $I_c$ | Emitter critical current | > 50 kA |
| $R_c$ | Emitter contact resistance at 77 K | $5\text{-}12 \cdot 10^{-6}$ Ω |
| S | Emitter surface | 20 cm$^2$ |
| d | Thickness of intrinsic J. junctions | 1.17 nm |
| $\varepsilon$ | Relative dielectric constant of IIJs | 10 |
| N | Number of junctions in emitter thickness 1 cm | $10^7$ |

## 5. Synchronization of the emission. Critical current. Magnetic field.

Emission of laser-like, coherent radiation from intrinsic Josephson junctions has been observed [9] when the superconductor is enclosed in a microwave cavity, which serves to impose a definite common oscillation frequency to the junctions. In our case the common frequency is set by the external circuit. The superconductor just follows the external oscillation. The general response of a superconductor to an AC voltage in the KHz-MHz range is to exhibit a small impedance, with small resistive and inductive components (related to the $\sigma_1$ and $\sigma_2$ mentioned in previous reports). For cuprates this is still true, independently from their intrinsic-Josephson structure. I am assuming that in large samples with resistive shunts the intrinsic Josephson structure, un-observable in the I-V curves, is nevertheless active for coherent electromagnetic and anomalous emission.

- P. Barbara, A.B. Cawthorne, S.V. Shitov and C.J. Lobb, Stimulated emission and amplification in Josephson junction arrays, Phys. Rev. Lett. 82 (1999) 1963. A similar experiment is reported by Vasilic et al., Appl. Phys. Lett. 78 (2001) 1137 (not acquired).

- Vasilic, B., Barbara, P., Shitov, S.V., Lobb, C.J., Constant-voltage resonant steps in underdamped Josephson-junctionarrays and possibilities for optimal millimeter-wave power output, IEEE Transactions on Applied Superconductivity, 11 (2001) 1188-1190 (not acquired).

- B. Vasilic, P. Barbara, S.V. Shitov, and C.J. Lobb, Direct observation of a threshold for coherent radiation in unshunted Josephson-junction arrays with ground planes Phys. Rev. B 65, 180503 (2002).

- G. Filatrella, B. Straughn, P. Barbara, Emission of radiation from square arrays of stacked JJs, J. Appl. Phys. 90 (2001) 1137. Only a proposed design plus theoretical model. Includes the case of intrinsic junctions.

In all the works above, the array size is comparable or larger than the free-space radiation wavelength $\lambda=2$ mm. The emission frequency corresponds to a high-Q resonance in the structure formed by the array and the resonator ground plane. (The power coupled to the detector is actually transmitted through a non-linear transmission line, so lambda is not exactly that of free space. The detector is itself made of junctions, and is very close.)

At MHz frequency, $\lambda$ is clearly much larger than the system's size. In our case, lambda is comparable to the system size, but the radiation is only virtual.

The transition of the junctions array to a coherent state, "analogous to the transition undergone by lasers", was predicted by Bonifacio et al. on the basis of the formal analogy between JJs arrays and free electron lasers [Lett. Nuovo Cim. 34 (1982) 520, not acquired].

Earlier, such a quantum coupling mechanism was predicted by Tilley [D.R. Tilley, Superradiance in arrays of superconducting weak link, Phys. Lett. 33A (1970) 205-206]. Jain [5], on the contrary, describes classical synchronization of over-dampened JJs with resistive shunts and low efficiency, about 1%. Note that our junctions are over-dampened because $\omega_c RC \ll 1$, due to resistive shunts and low resistance. Our junctions are driven from an external AC current, however, so they keep oscillating in spite of being over-dampened.

According to Barbara et al. and Tilley, "the output power scales as the square of the number of active junctions". I could not understand how, for the same number of emitted photons, coherence can change the transported energy. Maybe the point is that the coupling with the receiver changes. This implies that if the energy received by the detector is smaller, the rest remains in the emitter, where it is probably dissipated. The general concept relevant here is that of "super-radiance"; see for instance S. Barnett, P. Radmore, "Methods of theoretical optics", Oxford 1997.

I made an extensive literature search and study about the onset of synchronization in arrays of artificial and intrinsic Josephson junctions. This confirmed that an external load causes synchronization. In many experiments and simulations, the external load is just an RLC circuit as in our case. My simulations with few junctions clearly exhibit synchronization. Note that in our case, not only has the external circuit a definite proper frequency, but the initial conditions are such that the circuit oscillates from the beginning, while in several other experiments and simulations the junctions are DC biased and coupled to a resonant circuit which is initially passive.

In general, a magnetic field in the ab direction should increase the inter-plane coupling (Kleiner et al., [1]). It does so, however, at the price of decreasing $I_c$. Other coupling mechanisms should be more effective in our case, in particular the external driving frequency and the normal current ("quasi-particles" current). The numerical simulations predict that, paradoxically, a lower $I_c$ (provided still larger than $I_0$) increases the emitter voltage and thus the DC Josephson power. Therefore it can be helpful to apply an uniform magnetic field to the emitter. The dependence of $I_c$ on the field is not strong (see below, see below, with field in ab planes) and $V_E$ is inversely proportional to $I_c$, so the power gain is not expected to be dramatic. Also we can not define the degree of uniformity needed. On the other hand, all Podkletnov's trials included a magnetic field, either applied directly to the emitter by a coil or magnet, or propagating (with an inevitable radial component, in this case) from the large solenoid which concentrates the discharge. It would be therefore prudent to include it, if not too demanding. It can not be excluded that the field plays a symmetry breaking role in the microscopic transitions related to the Josephson and anomalous emission.

In the computations by Rogovin and Scully [10] the B field appears explicitly. Sometimes it couples e.m. normal modes with different polarizations (see also the work by Almaas and Stroud). In principle, certain modes of the Josephson junctions would be decoupled from the radiation in the absence of a static magnetic field.

- Acebron et al., The Kuramoto model: a simple paradigm for synchronization phenomena, Rev. Mod. Phys. 77 (2005)

Review article. Considers several synchronization problems, among them the one we are concerned with: Josephson-junction arrays connected in series through a load exhibit "all-to-all" (that is global) coupling (K. Wiesenfeld et al., Phys. Rev. Lett. 76 (1996) 404). Gives a schematic circuit with ideal JJs in series coupled through a resistance-inductance-capacitance load; in parallel to both is a bias current generator I_B. The model eq.s are

(hbar*C_j/2e)phi_j'' + (hbar*C_j/2er_j)phi_j' + I_j*sin(phi_j) + Q' = I_B;   j=1,...,N   (1)

for the junctions and

LQ'' + RQ' + Q/C = (hbar/2e) Sum_1...N(phi_k')   (2)

Note that the junction capacitance is taken into account. The first three terms on the l.h.s. in (1) are resp. the JJ currents In,c, In,r and Is. The term on the the r.h.s. of (2) is the total voltage on the JJs array.

Physical and numerical experiments on these equations were done by H. Sakaguchi and K. Watanabe, J. Phys. Soc. Jpn. 69 (2000) 3545. (Not acquired.)

- B. Daniels et al., Phys. Rev. E 67 (2003) 026216. (Not acquired) RSJ equations describing a ladder array of over-dampened (zero capacitance) JJs with different and disordered Ic. It is not exactly our case, since intrinsic JJs are supposed to be almost equal.

- G. Filatrella et al., High-Q cavity-induced synchronization in oscillator arrays, Phys. Rev. E 61 (2000) 2513.

Model for a large number of JJs coupled to a cavity; attempt at an explanation of the experiment by Barbara et al. for 2D arrays. The synchronization behavior was reproduced. Junctions are under-dampened, with non-zero C. A bias current is taken such that each junction is in the hysterretic regime. Depending on the intial conditions, the junctions may work in each of two possible states, with zero or non-zero voltage. In the latter case, the phases vary with time and the junctions are called "active".

- Filatrella, G., Pedersen, N.F., The mechanism of synchronization of Josephson arrays coupled to a cavity, Physica C, Vol. 372-376, SUPPL. PART 1, pp. 11-13, Aug. 2002. (Not acquired.)

Here they find that the transition from a state where the junctions are essentially oscillating at the unperturbed frequencies to one where they oscillate at the same frequency occurs above a threshold number of active junctions, in agreement with the experimental results by Barbara et al.

- G. Filatrella, N. Falsig Pedersen, C.J. Lobb and P. Barbara, Synchronization of underdamped Josephson-junction arrays, European Physical Journal B 34 (2003) 3-8.

In this latter model there is no threshold.

In general, the model employed in the papers above includes a global, "classical" coupling (external oscillating circuit), while Barbara et al. in their PRL article stressed the fact that the coupling is local and typically quantum-mechanical, with stimulated emission. For this reason, probably, is the threshold behaviour not properly reproduced. I am even a little surprised that they can reproduce the SIRS without considering the effect of the radiation exchange in the cavity (see for this the model by Stroud, Almaas and Harbaugh). But in fact, Shapiro steps, similar to SIRS, can be produced applying a periodic potential to the junctions.

In our simulations we send an oscillating current into the JJs instead of the DC bias used here. We are almost sure from the beginning that the frequency is the same in all junctions, while the authors here suppose that R is the same (for us, not necessarily, because variations are compensated by variations in In), and let instead the critical current Ic vary (our simulation also supports this; the voltage on the junctions depends on Ic; in any case we have I<Ic).

In this model, for a high-Q load the JJs locked at the resonance frequency drive the load much harder than junctions which are not locked (and so off-resonance). Consequently, the locked junctions interact far more strongly with the array.

Dependence of $I_c$ on the magnetic field

- S. Sanfilippo et al., Physica C 282-287 (1997) 2313

Field along *ab* planes. Measured at 77 K. Bulk textured mono-domain samples, top seeded MT. Bars w. 3 mm length, 0.4 mm$^2$ cross section.
$J_{c,c}$=6500 A/cm$^2$ for B=0 T; 5000 A/cm$^2$ for B=1 T; 3000 A/cm$^2$ for B=2 T.

- H. Ishii et al., Physica C 225 (1994) 91

Melt-grown YBCO fiber crystal, fiber diameter 245 µm, length 3.4 mm.
$J_{c,c}$=20000 A/cm$^2$ for B=0.01 T; 10000 A/cm$^2$ for B=0.1 T; 6000 A/cm$^2$ for B=1 T.

Information on field dependence in BSCCO is given in Kleiner et al. [1]. Fig. 15: field dependence of $I_c$ in the lowest branch of an IJJ. After some oscillations, about $4H_0$=0.1 T the critical current remains at ca. 75% of its maximum value. Kleiner and Muller [1]: see Fig. 11.

## 6. Contact resistance and heating.

In experiments on intrinsic Josephson junctions there is usually a transport current along the c axis, fed in from a generator through special contacts on the top and bottom of the samples. At the contacts, most of the external current is converted into super-current. The same should happen in our case, but our current is large and there is the problem of contacts over-heating. (The current is well below $I_c$, but this is so because melt-textured materials have especially large $I_c$.) If the material is driven normal near the contacts, all the mechanism of superconducting conduction and Josephson tunnelling is lost. The material can be driven normal also because contact is not uniform and local current density exceeds $J_c$.

We have seen that dissipation and heating in the bulk of the emitter can be disregarded. We must then check heating at the contacts. In a paper by Takeya et al. [1] the heat diffusion in BSCCO is taken into account. A heat diffusion length $l$ can be defined, both in the ab and c directions. Heat delivered at one point spreads over a volume of approximate size $l^2_{ab} \cdot l_c$. Knowing the specific heat of the material, one can compute the temperature increase of that volume. We are interested only in $l_c$, since heat is generated at planar contacts.

Next we need a guess for the surface resistance of the contacts $R_c$. Take for instance $R_c=10^{-5}$ $\Omega$, ie $\rho_c=2 \cdot 10^{-4}$ $\Omega cm^2$; then for larger or smaller values all scales in proportion. Heat generated at the contacts is of the order of the total energy (75 J) multiplied by the ratio $R_c/R_{load}$, where $R_{load}=0.1$ $\Omega$ as above. We find a dissipation of 0.0075 J. The temperature increase would then be negligible.

The heat diffusion length is given by the formula $l=2\sqrt{(Kt/c)}$, where K is the heat conductivity, c the specific heat, t the duration of the pulse. For BSCCO, Takeya et al. give K=0.25 W/mK, c=2 kJ/Km$^3$. For YBCO, K=15 W/mK along ab ([7], p. 254), c=10 kJ/Km$^3$ (see above). With a pulse duration t=0.5 μs, we find l=0.77 mm and the interested volume is $1.54 \cdot 10^{-6}$ m$^3$; its thermal capacity is 15.4 mJ/K. The thermal capacity of the copper layer, however, is almost ten times bigger, so that it takes most of the heat. For copper, K=560 W/mK at 80 K, c=4 J/Kcm$^3$; this gives a diffusion length l=0.24 mm with pulse duration $t=10^{-4}$ s; the interested volume is $4.8 \cdot 10^{-7}$ m and the thermal capacity 1.92 J/K. (One should also take into account the indium layer between YBCO and copper.)

So, supposed a surface resistivity of the order of $10^{-4}$ $\Omega cm^2$ can be obtained, there would be room for a reduction of the external resistance $R_{load}$, admitted this is possible in practice (if the external resistance is too small, there are many oscillations and the Marx capacitors could be damaged). If $R_{load}$ is smaller, then $\tau$ is larger and the total energy $U_{max}$ in the target increases in proportion (see detailed table in Sect. 7). Dissipation in the bulk of the emitter and in the contacts also increases in proportion to $\tau$.

**7. Simulation of a Josephson junction inserted in a RLC circuit, in the RSJ model**

According to the RSJ model (resistively-shunted junction), a Josephson junction can be represented as circuit element by a non-linear element obeying the Josephson effect equations below, plus an ohmic resistance R in parallel. In the purely Josephson element flows only supercurrent while in the resistance flows normal current. We have seen that when the Josephson junction is placed in an external oscillating circuit with large C and L, it should not influence the external current. This is confirmed by the simulation below and is true also for many junctions in series. Therefore we first simulate one single junction and then we shall consider the synchronization of several junctions.

The two fundamental equations of the Josephson effect are

$$I_s = I_J \sin\phi \qquad (1)$$

where $I_s$ is the supercurrent in the junction, $I_J$ is the critical current and $\phi$ the phase difference over the link, and

$$\phi' = \frac{2e}{\hbar}V \qquad (2)$$

where the prime denotes time derivative and V is the voltage applied to the junction. According to the RSJ model, $V=RI_n$, where $I_n$ is the normal current flowing in the normal resistance R of the junction, parallel to $I_s$.

Only $I_n$ generates a voltage in the emitter, but both $I_n$ and $I_s$ flow in the external capacitance and inductance ($I_s$ after conversion to normal) and discharge the capacitor.

Denote $a = \frac{2e}{\hbar}R$ and rewrite (2) and the second derivative of (1) as follows

$$\begin{cases} \phi' = aI_n \\ I_s'' = aI_J(I_n'\cos\phi - aI_n^2\sin\phi) \end{cases} \qquad (3)$$

These are the first two eq.s of a system, whose unknowns are the functions of time $\phi(t)$, $I_s(t)$, $I_n(t)$.

Write the derivative of the Kirchoff equation over the loop including the external load ($L_L$, $C_L$, $R_L$) and the junction

$$\frac{1}{C_L}(I_s + I_n) + L_L(I_s'' + I_n'') + R_L(I_s' + I_n') + RI_n' = 0 \qquad (4)$$

This is going to be the third equation of the system. Divide by $L_L$ and note that the proper frequency of the external circuit is $\omega = 1/\sqrt{L_L C_L}$. Disregard the last term because R is about $10^{10}$ times smaller than $R_L$. Replace $I_s''$ with the 2$^{nd}$ eq. in (3), where $I_J$ is denoted g. Finally define $b = R_L/L_L$. We find

$$\omega^2(I_s + I_n) + ag(I_n'\cos\phi - aI_n^2\sin\phi) + I_n'' + b(I_s' + I_n') = 0 \qquad (5)$$

Isolating $I_n''$, we obtain the final complete non-linear system, where the currents are denoted simply by s and n:

$$\begin{cases} \phi' = an \\ s'' = ag(n'\cos\phi - an^2\sin\phi) \\ n'' = -\omega^2(s+n) - ag(n'\cos\phi - an^2\sin\phi) - b(s'+n') \end{cases} \qquad (6)$$

Summarizing, the symbols and magnitude orders of the parameters are, with quasi-real parameters (more precise data are considered below for several specific cases)

| | |
|---|---|
| $n = I_n(t)$ | $R = R_E/N = 10^{-11}$ |
| $s = I_s(t)$ | $\omega = 3.3 \cdot 10^6$ |
| $a = \dfrac{2e}{\hbar}R = 3 \cdot 10^4$ | $g = I_J = 5 \cdot 10^4$ |
| $b = \dfrac{R_L}{L_L} = 1.5 \cdot 10^4$ | |

The initial conditions at time t=0 (when the Marx spark gaps close) are the following:

$\phi(0) = 0$

$I_n(0) = 0$

$I_s(0) = 0$

$I_s'(0) = 0$

$I_n'(0) = I_0 \omega$

These are the same as for the RLC circuit alone. At the time t=0 the external circuit begins to oscillate, starting from a state in which the capacitor is full (in practice, the spark gaps of the Marx are triggered in a short time). The initial value for $I'_n(0)$ is standard for an RLC circuit (see our previous Reports). $I_0$ is the maximum external current, which depends on V, $C_L$ and $L_L$ as

$$I_0 \approx V \sqrt{\frac{C_L}{L_L}}$$

Note that $I_s'$ is initially zero due to eq. (1) and (2), since V (emitter voltage, in that case) is initially zero. It is interesting to note that in spite of this, $I_s$ rapidly grows and becomes almost equal to $I_0$ in the emitter, where $I_n$ stays small (see below).

With these initial conditions the equation system (6) can be solved numerically through the Runge-Kutta method (see the Mathematica code below, including two junctions in order to show synchronization). The result is clear: for $I_0 < I_J$ (which is usually the case) all functions oscillate with the external frequency. [For $I_0 = I_J$, the phase makes a complete oscillation in the period of the circuit oscillation. For $I_0 > I_J$, the phase does not even reach the value $\phi = 1$, and then reverses.]

```
NDSolve[{
f1'[t] == a*n1[t],
f2'[t] == a*n2[t],
s1''[t] == a*g*(n1'[t]*Cos[f1[t]]-a*n1[t]^2*Sin[f1[t]]),
s2''[t] == a*g*(n2'[t]*Cos[f2[t]]-a*n2[t]^2*Sin[f2[t]]),
n1''[t] ==
 -w^2*(s1[t]+n1[t])
 -a*g*(n1'[t]*Cos[f1[t]]-a*n1[t]^2*Sin[f1[t]])
 -b*(s1'[t]+n1'[t]),
n2''[t] ==
```

```
  -w^2*(s1[t]+n1[t])
  -a*g*(n1'[t]*Cos[f1[t]]-a*n1[t]^2*Sin[f1[t]])
  -b*(s1'[t]+n1'[t])
  +a*g*(n1'[t]*Cos[f1[t]]-a*n1[t]^2*Sin[f1[t]])
  -a*g*(n2'[t]*Cos[f2[t]]-a*n2[t]^2*Sin[f2[t]]),
f1[0] == 0, f2[0] == 0, s1[0] == 0, s2[0] == 0, n1[0] == 0, n2[0] == 0,
s1'[0] == 0, s2'[0] == 0, n1'[0] == d*w, n2'[0] == d*w},
{f1, f2, s1, s2, n1, n2}, {t, 0, 6.283/w}]
```

Plotting instruction:

```
Plot[Evaluate[f1[t] /. %...], {t, 0, 0.000001}]
```

NB: in the expression for $n_2$'', the 2. and 4. row cancel each other, but are left here for clarity.

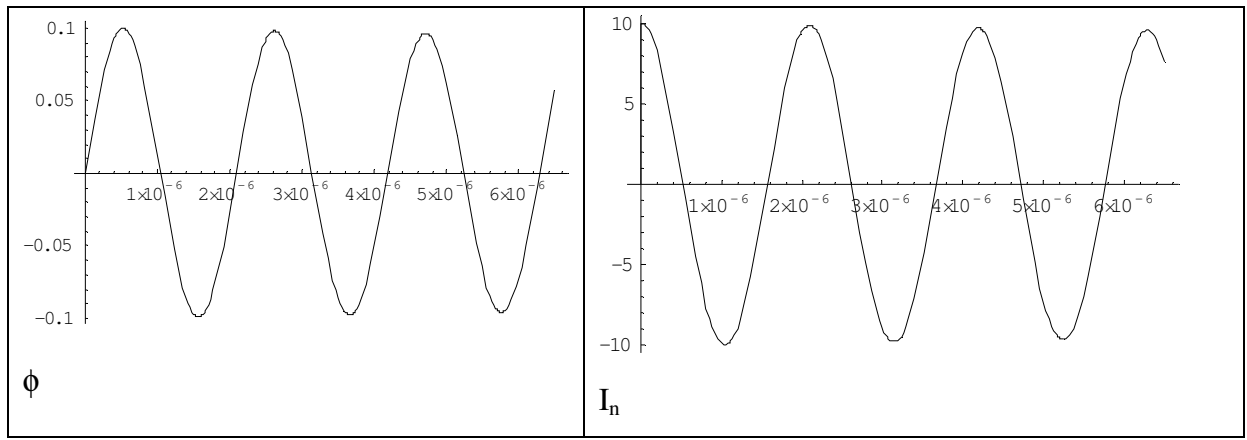

A refined version of the RSJ model includes a junction capacitance $C_J$ in parallel to the resistance [11]. For high frequency, the capacitive channel can become important. We have seen that $C_J \approx 10^{-4}$ F, so the impedance of the C channel at $\omega \approx 1$ MHz is of the order of $10^{-2}$, much larger than $R_E \approx 10^{-11}$. It should therefore be legitimate to disregard $C_J$. For a check, we included a capacitive channel in the numerical simulation. The results are at first sight strange and unrealistic, because in this case the capacitance of the Josephson junction affects the circuit behaviour much more than the (smaller) external capacitance $C_L$; but this is an artefact, because eventually we want to simulate a large number of junctions in series, and in that case their total capacitance will be small, so $C_L$ will actually dominate and the C-channels of the junction carry very little current.

Let us write the equation for 2 junctions in series:

$$s_1 = g \sin \phi_1$$
$$s_2 = g \sin \phi_2$$
$$\phi_1' = a n_1$$
$$\phi_2' = a n_2$$
$$s_1'' = ac(n_1' \cos \phi_1 - a n_1^2 \sin \phi_1)$$
$$s_2'' = ac(n_2' \cos \phi_2 - a n_2^2 \sin \phi_2)$$
$$\frac{1}{C_L}(n_1 + s_1) + L_L(n_1'' + s_1'') + R_L(n_1' + s_1') = 0$$

In the last equation one isolates $n_1''$ and replaces $s_1''$.

Finally, for $n_2''$: we note that $n_1+s_1=n_2+s_2$ → $n_1''+s_1''=n_2''+s_2''$ → $n_2''=\ldots$ We have 6 eq.s with unknown $\phi_1$, $s_1$, $n_1$, $f_2$, $s_2$, $n_2$. The initial condition is the same, as is easily obtained differentiating the equation $n_1+s_1=n_2+s_2$.

For 3 junctions: the equations for $\phi_3'$ and $s_3''$ are simple. Then current conservation gives $n_3''=n_1''+s_1''-s_3''$. This also holds for the first derivatives, and for the initial condition, which is just the same.

In this way we check directly the synchronization, at least for 3 junctions, and regarding a whole crystal layer (with surface of the order of square centimetres!) as a single junction. We see in fact that all phases oscillate in exactly the same way, and the same is true for the normal current and the voltage. The synchronization occurs for all values of the parameters in the range above, and also for higher frequency (larger than 10 MHz).

The simulations allow to compute the emitter voltage $V_E$ by multiplying $I_n$ and $R_E$. This voltage turns out to be smaller (typically 10 times smaller; see below, with real parameters) than the simple estimate based on the relation $V=(h/2e)f$. This relation holds rigorously for a constant voltage, while in our case we have $V=RI_n$, and $I_n$ oscillates (R normal resistance of the junction, $I_n$ normal current).

In addition, the emitter voltage depends on the critical Josephson current $I_J$, and is larger when $I_J$ is smaller (inversely proportional, see below). This could not have been predicted without the simulations, but with hindsight the reason is clear. The supercurrent is fixed (approx. equal to the total current) and $I_s=I_J\sin\phi$. If $I_J$ is larger, then the oscillations of $\phi$ are smaller, and so $\phi'$ is smaller too; and V is proportional to $\phi'$. A possible way to depress $I_J$ is to apply a magnetic field. So the magnetic field is not needed for synchronization (Sec. 5), but improves the DC power and the target energy.

It is not easy to understand intuitively how an oscillating $I_s$ is obtained when the voltage itself oscillates. Mathematically, the point is that $\phi$ does not evolve linearly in time, but oscillates in turn, therefore $I_s$ is not perfectly harmonic while $I_0$ is harmonic, and the difference $I_n=I_0-I_s$ oscillates.

The simulations show that after t=0 the normal current, starting from zero, rises quickly and then begins to oscillate from its maximum. The Josephson junctions are very quick ($10^{-11}$-$10^{-12}$ s) to adapt to the least energy configuration, where most external current is converted into super-current.

Some simulations were run to look for the dependence of the normal current upon $I_J$. It turns out that there is an inverse proportionality. For instance, with $a_1=a_2=3\cdot10^4$ one finds

| Critical current (kA) | Normal current (A) |
|---|---|
| 20 | 25 |
| 40 | 12.5 |
| 80 | 6.1 |
| 160 | 3.1 |

Simulations with real parameters

**EP 500 kV**  (C=1.25 nF, L=15 µH)

Data known or estimated:

- $\omega = 7.3$ MHz
- $I_0 = 4560$ A (max current amplitude)
- $R_E = \frac{1}{4} \cdot 10^{-4}$ Ω  (1/4 of our estimated resistance, because area is 4 times larger; anyway, simulations show that changes in $R_E$ do not affect the emitted power)
- $I_J = 4 \cdot 10^5$ A (critical current; based on 5 kA/cm$^2$, like for our samples, although EP mentions 50 kA/cm$^2$)

The simulation gives

- $I_n = 11$ A (normal current in the emitter)
- $V_E = 2.75 \cdot 10^{-4}$ V (voltage on the emitter)
- $P_{DC} = I_0 \cdot V_E = 1.25$ W (DC Josephson power)

The target energy measured by EP is 0.45 mJ. This is referred to an emitter with thickness 0.8 cm, while our estimate is for thickness 1 cm. With the power estimated above, this energy can be obtained in a time $\tau \approx 3.6 \cdot 10^{-4}$ s, ie ca. 400 oscillations and load resistance $R_L = 0.1$ Ω (EP mentiones an N-layer resistance of "less than 0.5 Ω"). So it appears to be plausible, but note that we have not applied any Josephson DC-AC efficiency conversion factor. We may hypothesize that due to strong stimulated emission (as confirmed by the strong beam directionality), this factor is very close to 1. Also for our emitters below I consider that the DC Josephson power is entirely converted into target energy.

**Our generator with 100 kV** (C=15 nF, L=6 μH)

- $\omega = 3.3$ MHz
- $I_0 = 5$ kA
- $R_E = 10^{-4}$ Ω
- $I_J = 10^5$ A (critical current; based on 5 kA/cm$^2$, as from literature)

The simulation gives

- $I_n = 5.5$ A (normal current in the emitter)

- $V_E = 5.5 \cdot 10^{-4}$ V (voltage on the emitter)
- $P_{DC} = I_0 \cdot V_E = 2.75$ W (DC Josephson power)

**Our generator with 50 kV**

- $\omega = 3.3$ MHz
- $I_0 = 2.5$ kA
- $R_E = 10^{-4}$ Ω
- $I_J = 10^5$ A (critical current)

The simulation gives

- $I_n = 2.75$ A (normal current in the emitter)
- $V_E = 2.75 \cdot 10^{-4}$ V (voltage on the emitter)
- $P_{DC} = I_0 \cdot V_E = 0.69$ W (DC Josephson power)

These data are not bad. Consider the following load resistances and numbers of oscillations:

|  | $R_L$ load resistance (Ω) | $\tau = 2L/R_L$ damp. time (μs) | $Q = \omega\tau$ oscillations | $U_{max}$ target energy (mJ) | Target veloc. (m/s) for m=2 g |
|---|---|---|---|---|---|
| $I_0 = 5$ kA, P=2.75 W | 1 | 12 | 40 | 0.033 | 0.18 |
|  | 0.5 | 24 | 80 | 0.066 | 0.26 |
|  | 0.1 | 120 | 400 | 0.33 | 0.57 |
| $I_0 = 2.5$ kA, P=0.69 W | 1 | 12 | 40 | 0.008 | 0.09 |
|  | 0.5 | 24 | 80 | 0.016 | 0.13 |
|  | 0.1 | 120 | 400 | 0.083 | 0.29 |

(Formula for the target velocity: $v_t = \sqrt{2U_{max}/m}$, where one can also put $U_{max}$ in mJ and m in grams)

The question is: is it better, to avoid stress to the capacitors, to have more current and less oscillations, or less current and more oscillations? Thinking of an elastic material, like a spring, it seems to me that if it is far from its strain limit, it can make many oscillations without damage, while near the strain limit even few oscillations can damage it.

**8. Wavelength of the anomalous emission and momentum conservation**

Concerning the wavelength of anomalous emission and momentum conservation, the stacked IJJs model implies big changes with respect to the old model. With a normal current of just tens of A, the voltage on the emitter turns out to be very small (mV instead of kV!). The corresponding electric field is not sufficient to accelerate the pairs enough to give them the momentum they should pass over to the gravitons. How can then emission occur? It is possible that the pairs tend to recoil after the emission, but the super-current carries them ahead and transfers the recoil to the whole emitter. This process is more difficult to analyze theoretically, because it involves not single pairs but the current as a whole.

In the quantum picture of superconductivity, all pairs are described by a single, "rigid" wave function, and individual behaviour of the pairs makes no sense, strictly speaking, although Cooper pairs are often visualized as particles for simplicity.

(Note that every electromagnetic emission from an atom causes the atom to recoil, but usually the recoil velocity is very small because the momentum carried by photons is very small and the atom mass is large; the recoil is only measurable for gamma emission by nuclei - the Mossbauer effect.)

In other words, the large voltage generated by the Marx is not directly available as large electric field in the emitter, but it drives the load current, and thus the supercurrent, and the pairs can not "recoil out of the supercurrent", because the wave function is rigidly collective.

If there is no driving electric field, it is necessary to make some conjecture about the emission direction. One could think that the emission in the direction of the pairs motion is still favoured, although the velocity of this motion is much smaller than the recoil velocity. The stimulated emission then further selects the direction orthogonal to the emitter. For the emission versus we can make different hypotheses:

- There are actually both forward and backward emission.
- The versus of the first oscillation is decisive, but I could not explain why (maybe it triggers stimulated emission?).
- Conduction in the gas is not isotropic (negative carriers being much quicker than positive carriers), and while this does not affect stationary or 50 Hz currents, it is important in our case. In this case the presence of the gas could be important.

Another change in the new model concerns the emission frequency and wavelength. The emission frequency is now thought to be equal to the Josephson frequency, and so in turn to the oscillation frequency of the external circuit. This can visualized in two ways.

Quantum mechanically, we can think the graviton emission as associated to the Josephson emission, taking place when the pairs tunnel through the junctions. (See [11]; not every tunnelling gives a photon or graviton, the so-called DC-AC conversion efficiency can vary from 1% [5] to 17-30% in the presence of stimulated emission [9].)

Semi-classically, we think of the condensate as undergoing density and gradient changes during the AC Josephson oscillation of the supercurrent with frequency $\omega$. These changes amount to oscillations of the local amplified $\Lambda_{eff}$, and this in turn acts as an external source for emission of gravitons with energy $E=h\omega/2\pi$.

The Josephson frequency is smaller, by at least a factor 20, than the frequency previously estimated as the reciprocal of the "transit time" $s_c/v$ of the accelerated Cooper pairs.

If the frequency of the anomalous virtual gravitons is smaller, their wavelength must be larger than previously supposed. This is because the product $f\lambda$ is fixed and must be equal to $1/2v_t$, where $v_t$ is the target velocity. (The energy/momentum ratio in the target is $E/p=1/2v_t$ and in the gravitons it is $E/p=f\lambda$.) The new estimate for $\lambda$ is about 100 times the inter-plane spacing.

This is in contrast with our previous idea that $\lambda$ was equal to the lattice spacing, or a small integer multiple of it. That idea was suggested by a requisite of spatial coherence between subsequent emissions: we thought that emissions from different *ab* planes could add in phase only if the radiation waves made an integer number of oscillations between two planes. Now it seems that the large frequency required by this condition is not available in the emitter. So let us see if it is possible to relax the requirement of spatial coherence. [Our device is different from a free electron laser, where coherence arises in this classical way, but there is no stimulated emission in quantum two-level systems.]

Stimulated emission requires equal frequency throughout the whole emitter, which is in fact present, and equal to the Josephson frequency. The observed small divergence of the beam is most probably caused by stimulated emission. Although spatial coherence may enhance the total amplitude through in-phase amplitude addition, papers on laser-like emission from Josephson junctions arrays only mention stimulated emission as the main amplification factor (not directly, but through the external cavity). In conclusion, it seems that the condition of spatial coherence may indeed be relaxed.

Note the since stimulated emission also gives phase coherence, there is coherence into any single cascade of gravitons originated by the same seed graviton, but not between one cascade and the others. This is different from what happens in a laser, where the resonant cavity makes it possible in principle for one single photon to generate all the others (disregarding the necessary cavity losses); here the subsequent emission are on the same single run.

If $\lambda$ is not equal to the inter-plane spacing, what else fixes it? There are two possibilities. The first hypothesis is that being the emission a virtual intermediate process, its momentum is automatically self-adjusted to the target momentum, and then $\lambda=p/h$. In others words, the gravitons are created in an excited state of the gravitational vacuum only for a very short time, and picked out of the whole virtual momentum spectrum just with momentum equal to the final momentum. The probability function of the whole process contains a $\delta(p-p')$, where p is the final momentum and p' the intermediate momentum.

The second hypothesis (incompatible with the data, however, and superseded by our considerations above on the "rigid" wave function) is that the total momentum has to be compatible with the electric field on the emitter, acting for several oscillations. Consider the total number N of emitted virtual gravitons and their momentum, connected to $\lambda$. Suppose the total momentum is supplied by the electric field on the emitter. The electric force in one oscillation is $F=QE=I\Delta t \cdot \Delta V/\Delta x$. The momentum is $p=F\Delta t$. Taking $E=10^{-3}$ V/cm, $\Delta t=2\cdot 10^{-5}$ s, $I=10^4$ A, we find $p=4\cdot 10^{-7}$ kg·m/s. In 400 oscillations, $p_{tot}=10^{-5}$. With $U_{max}=0.6$ mJ and angular frequency $\omega=3.3$ MHz, N turns out to be $N=2\cdot 10^{24}$. Setting $p_{tot}=Nh/\lambda$, we find $\lambda=10^{-4}$ m, while the E/p ratio gives $\lambda=10^{-7}$ m.

This would imply that: (a) It is impossible to define lambda in this way; (b) the momentum is not made locally available in the emitter, but in the whole circuit, and transferred by the current. The circuit "recoils", but which part exactly?

## 9. Things are not better with a peaking capacitor!?

We discussed this possibility earlier. The connection I have in mind is the following. The Marx has a capacitance $C_L=15$ nF and an inductance $L_L=6$ µH. The peaking capacitor in this example has $C_P=0.1$ nF and $L=0.5$ µH, proper frequency $\omega=141$ MHz, and a (small) resistance $R_P$ to be defined below.

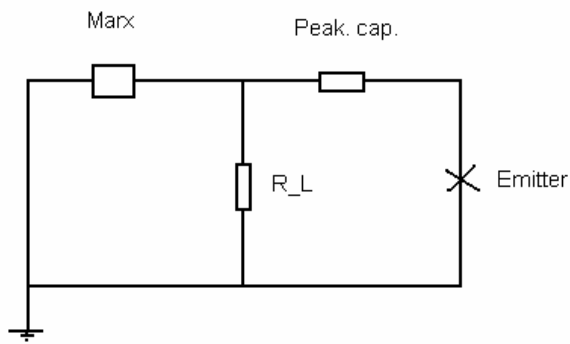

The idea would be to put a large load resistance $R_L$ so that the Marx discharge is underdamped but the peaking capacitor is charged and then oscillates with higher frequency than the Marx, and makes many oscillations, because the resistance $R_P$ is small.

But the problem is that in this connection, if my calculations are right, the presence of the $R_L$ in some way causes also the oscillations in the peaking capacitor to be quickly dampened. So much, actually, that if for instance you put $R_L=100$ Ω in order to have the Marx discharge under-damped, than the oscillation in the peaking capacitor is under-damped, too. You then put for instance $R_L=1\ \Omega$, so that the Marx oscillates but not many times; the Marx voltage is set to 100 kV and current 5 kA. Then you get:

- With $R_P=1\ \Omega$, the current in the peaking capacitor oscillates with its proper frequency (corresponding to period 44 ns), amplitude ca. 100 A, dampening time ca. 1 µs (Fig. 1). The corresponding Josephson DC power in the emitter is 22 W, without taking into account the reduction factor dependent on $I_J$. Thus the energy delivered to the target in 1 µs is small, definitely smaller than using the Marx directly with the same $R_L=1\ \Omega$! (26 W, but for a longer time.) Here $d=5000 \cdot 3.3 \cdot 10^6$ as usual.

- With $R_P=0.1\ \Omega$, the dampening time is ca. twice as much, and with $R_P=0.01\ \Omega$, it does not change any more (Fig. 2), confirming that it depends on $R_L$ more than on $R_P$, for small $R_P$. (This can also be seen increasing $R_L$: with $R_L=10\ \Omega$ and $R_P=1\ \Omega$, the oscillation in the peaking capacitor is almost over-dampened (Fig. 3).) In that case y'(0) must be adapted to the present maximum current.

- With $C_P=10$ pF and $L_P=0.1$ µH the frequency increases to ca. 1 GHz; but the current is only 15 A and the dampening time only 0.1 µs, so the total energy is still small (Fig. 4).

In all figures, the vertical axis is current in the peaking capacitor, in A.

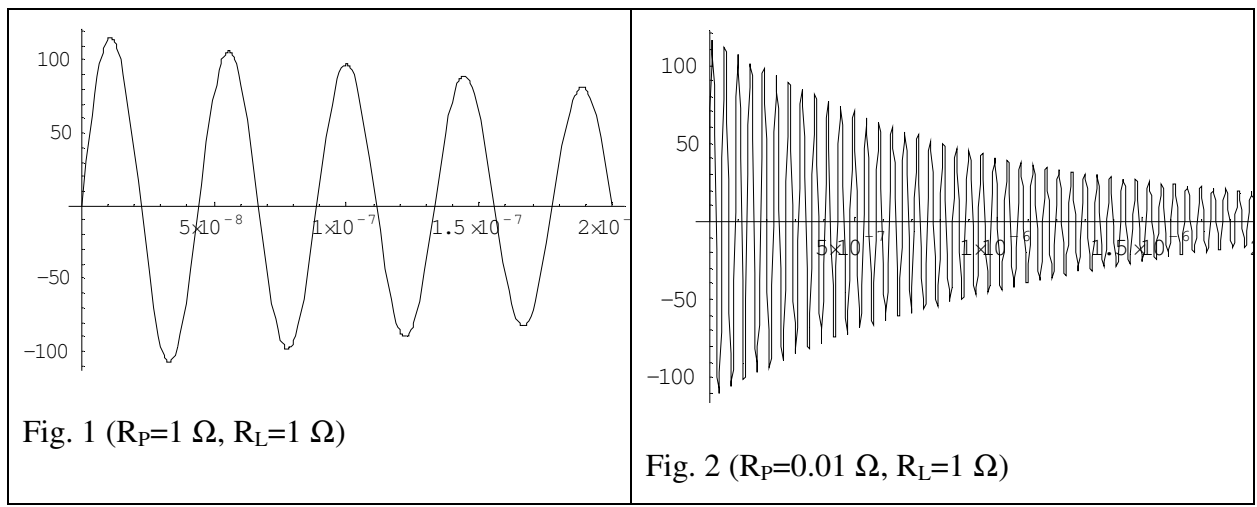

Fig. 1 ($R_P=1\ \Omega$, $R_L=1\ \Omega$)

Fig. 2 ($R_P=0.01\ \Omega$, $R_L=1\ \Omega$)

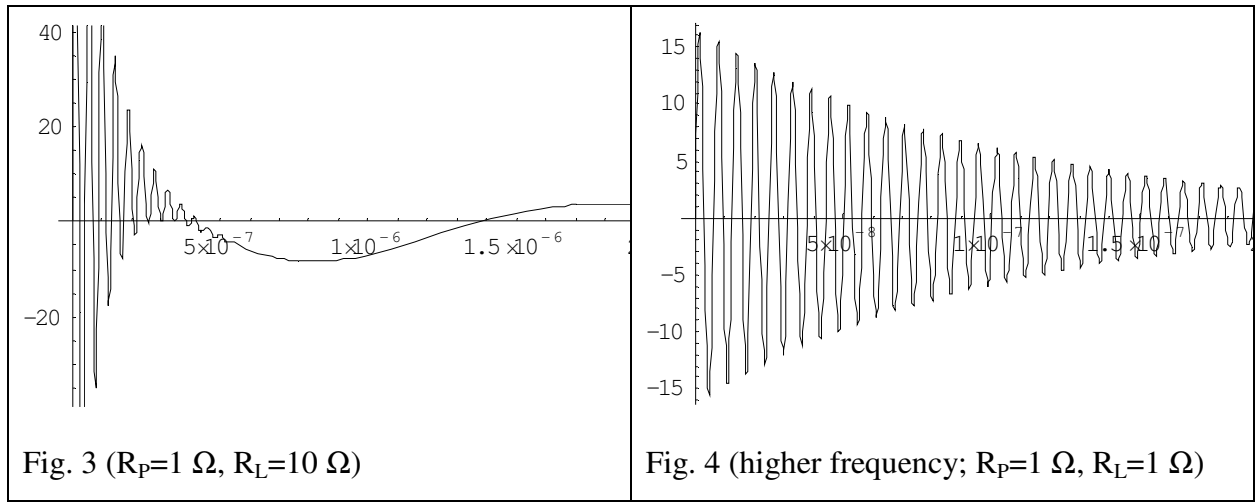

Fig. 3 ($R_P$=1 Ω, $R_L$=10 Ω)  Fig. 4 (higher frequency; $R_P$=1 Ω, $R_L$=1 Ω)

In conclusion, in this configuration at least, things do not work as desired. In some way, one should be able to disconnect the peaking capacitor from the Marx and from $R_L$ immediately after it has been charged, but that does not seem easy to do.

Along a completely different line, I am wondering if high-frequency (100 MHz) current generators are commercially available, which can give a current of several kA. What do they use for radio stations?

[Circuit equations:
$a_1(x+y)+b_1(x''+y'')+c_1 x'=0$
$a_2 y+b_2 y''+c_2 y'-c_1 x'=0$
$a_1=1/C_L$, $a_2=1/C_P$, $b_1=L_L$, $b_2=L_P$, $c_1=R_L$, $c_2=R_P$, $x=I_L$, $y=I_P$
Initial conditions $x=y=x'=0$, $y'=\omega I_0$]

Table: External circuit parameters for different tentative configurations.

These are examples of possible Marx configurations, among which we eventually chose that of Section 4 (with lower-voltage version and various RL as in Section 7).

| Type of generator | $L_L$ (μH) | $C_L$ (nF) | $R_L$ (Ω) | V (kV) | ω (Hz) | $I_0$ (A) | τ (s) | ωτ | $E_{tot}$ (J) |
|---|---|---|---|---|---|---|---|---|---|

| Configuration | | | | | | | | | |
|---|---|---|---|---|---|---|---|---|---|
| 1-stage, current gen. | 1 | 150 | 0,1 | 60 | 2,58E+06 | 2,32E+04 | 2,00E-05 | 5,16E+01 | 270 |
| 4-stages | 4 | 37,5 | 0,1 | 240 | 2,58E+06 | 2,32E+04 | 8,00E-05 | 2,07E+02 | 1080 |
| 10 stages | 10 | 15 | 0,1 | 600 | 2,58E+06 | 2,32E+04 | 2,00E-04 | 5,16E+02 | 2700 |
| Peaking capacitor | 0,5 | 3,2 | 0,1 | 675 | 2,50E+07 | 5,40E+04 | 1,00E-05 | 2,50E+02 | 729 |
| | | | | | | | | | |
| 1-stage, current gen. | 1 | 150 | 0,01 | 60 | 2,58E+06 | 2,32E+04 | 2,00E-04 | 5,16E+02 | 270 |
| 4-stages | 4 | 37,5 | 0,01 | 240 | 2,58E+06 | 2,32E+04 | 8,00E-04 | 2,07E+03 | 1080 |
| 10 stages | 10 | 15 | 0,01 | 600 | 2,58E+06 | 2,32E+04 | 2,00E-03 | 5,16E+03 | 2700 |
| Peaking capacitor | 0,5 | 3,2 | 0,01 | 300 | 2,50E+07 | 2,40E+04 | 1,00E-04 | 2,50E+03 | 144 |
| | | | | | | | | | |
| 4-stages parallel | 0,25 | 600 | 0,005 | 36 | 2,58E+06 | 5,58E+04 | 1,00E-04 | 2,58E+02 | 388,8 |
| Real (10 stages series) | 10,6 | 15 | 1 | 310 | 2,51E+06 | 1,17E+04 | 2,12E-05 | 5,32E+01 | 720,75 |
| Real (10 stages series) | 6 | 15 | 0,1 | 200 | 3,33E+06 | 1,00E+04 | 1,20E-04 | 4,00E+02 | 300 |
| Real (3 stages parallel) | 1,5 | 450 | 0,1 | 36 | 1,22E+06 | 1,97E+04 | 3,00E-05 | 3,65E+01 | 291,6 |
| | | | | | | | | | |
| EP 2000 kV | 15 | 1,25 | 0,1 | 2000 | 7,30E+06 | 1,83E+04 | 3,00E-04 | 2,19E+03 | 2500 |
| EP 500 kV | 15 | 1,25 | 0,1 | 500 | 7,30E+06 | 4,56E+03 | 3,00E-04 | 2,19E+03 | 156,25 |

- "4-stages" is made of four capacitors.
- "1-stage, current generator" is a single capacitor.
- "EP" is Podkletnov's generator, with $L$=15 $\mu$H as from your estimate.
- $E_{tot}$ is the total electrostatic energy initially stored in the capacitors.
- "Real": other possibile onfigurations.

## 10. Final remarks

<u>Our current "gambles"</u>

After 3 years of theoretical and practical experience, we have now come to a point where we attempt "replication" with radical modifications of the original device. It is a big gamble, with high payoff in terms of simplification, control and understanding of the basic mechanism. The main modifications are:

- The emitter surface is 4 times smaller than Podkletnov's.
- There is no gas discharge chamber.
- The emitter has no normal layer.
- Voltage is below 500 kV.
- The emitter has a standard melt-textured structure.

The role of the normal layer in Podkletnov's design was probably twofold:

1. Provide one efficient S-N contact (the other being provided by the gas).
2. Provide a small load resistance.

According to Podkletnov, "the conductivity of both S and N layers was large, more than 1.5 S/m", ie the resistivity was less than 0.6 ohms*m. This piece of information is not very useful, actually, because we know that the normal state resistivity of YBCO is much smaller, ca. $5*10^{-3}$ ohms*m even in the c direction. The cited data is probably just an upper limit.

Podkletnov wrote me recently, about the total circuit resistance, that it could mainly reside in the part made of steel (the current flows to the emitter through the nitrogen/helium reservoir and its tube). But I computed that the resistance of that part should be very small, although it is made of steel and not of copper. The resistivity of iron at 80 K is less than $10^{-8}$ ohms*m (CRC Handbook, 12-45). Consider the tube, say with diameter 2 cm, thickness 1 mm and length 50 cm. We so have R=$10^{-4}$ ohms.

If the resistance of steel and copper was the only resistance in the circuit, the dampening time $\tau$ of the oscillations would be extremely long ($\tau$=2L/R). Podkletnov noticed that without the N-layer the discharges were irregular. The reason might be that the oscillations are not dampened, so the DC component of the voltage is very small. It is this DC component that causes the front of the plasma to move to the cathode.

If the resistivity of the N-layer is not far from the limit mentioned by Podkletnov, say between 0.1 and 1 ohms*m, then its total resistance is 0.1-1 ohms (because surface is ca. 0.01 $m^2$ and thickness of the order of 0.01 m). Supposing the N-layer only has a passive resistance role, it would clearly be easy to replace it with another else. Note that the resistive part of the circuit also dissipates the power and heats up.

The SN contacts, with N metal or else, always bear resistance and cause small voltage drops. With 100 ohms and 1000 kV in the emitter, the contacts were negligible. But now one could think that for some reason the N-layer provides a better contact, with respect to a direct metal contact.